\documentclass{aa}  

\usepackage{graphicx}
\usepackage{txfonts}
\usepackage{lipsum}
\usepackage{subcaption}         
\usepackage{lscape}           
\usepackage{placeins}           
\usepackage{hyperref}

\newcommand{\krakow}{Krak\'ow{}}

\begin{document}

   \title{Observation of the Crab Nebula with the Single-Mirror Small-Size Telescope stereoscopic system at low altitude}
   \authorrunning{SST-1M Collaboration}
   \titlerunning{Observation of the Crab Nebula with the SST-1M}

\author{
    C.~Alispach\inst{\ref{inst1}}\and
    A.~Araudo\inst{\ref{inst2}}\and
    M.~Balbo\inst{\ref{inst1}}\and
    V.~Beshley\inst{\ref{inst3}}\and
    J.~Bla\v{z}ek\inst{\ref{inst2}}\and
    J.~Borkowski\inst{\ref{inst4}}\and
    S.~Boula\inst{\ref{inst5}}\and
    T.~Bulik\inst{\ref{inst6}}\and
    F.~Cadoux\inst{\ref{inst1}}\and
    S.~Casanova\inst{\ref{inst5}}\and
    A.~Christov\inst{\ref{inst2}}\and
    J.~Chudoba\inst{\ref{inst2}}\and
    L.~Chytka\inst{\ref{inst7}}\and
    P.~\v{C}echvala\inst{\ref{inst2}}\and
    P.~D\v{e}dic\inst{\ref{inst2}}\and
    D.~della Volpe\inst{\ref{inst1}}\and
    Y.~Favre\inst{\ref{inst1}}\and
    M.~Garczarczyk\inst{\ref{inst8}}\and
    L.~Gibaud\inst{\ref{inst9}}\and
    T.~Gieras\inst{\ref{inst5}}\and
    E.~G{\l}owacki\inst{\ref{inst9}}\and
    P.~Hamal\inst{\ref{inst7}}\and
    M.~Heller\inst{\ref{inst1}}\and
    M.~Hrabovsk\'y\inst{\ref{inst7}}\and
    P.~Jane\v{c}ek\inst{\ref{inst2}}\and
    M.~Jel\'inek\inst{\ref{inst10}}\and
    V.~J\'ilek\inst{\ref{inst7}}\and
    J.~Jury\v{s}ek\inst{\ref{inst2}}\thanks{Corresponding authors; emails: jurysek@fzu.cz, tavernier@fzu.cz, novotnyv@fzu.cz}\and
    V.~Karas\inst{\ref{inst11}}\and
    B.~Lacave\inst{\ref{inst1}}\and
    E.~Lyard\inst{\ref{inst12}}\and
    E.~Mach\inst{\ref{inst5}}\and
    D.~Mand\'at\inst{\ref{inst2}}\and
    W.~Marek\inst{\ref{inst5}}\and
    S.~Michal\inst{\ref{inst7}}\and
    J.~Micha{\l}owski\inst{\ref{inst5}}\and
    M.~Miro\'n\inst{\ref{inst9}}\and
    R.~Moderski\inst{\ref{inst4}}\and
    T.~Montaruli\inst{\ref{inst1}}\and
    A.~Muraczewski\inst{\ref{inst4}}\and
    S.~R.~Muthyala\inst{\ref{inst2}}\and
    A.~L.~Müller\inst{\ref{inst2}}\and
    A.~Nagai\inst{\ref{inst1}}\and
    K.~Nalewajski\inst{\ref{inst5}}\and
    D.~Neise\inst{\ref{inst13}}\and
    J.~Niemiec\inst{\ref{inst5}}\and
    M.~Niko{\l}ajuk\inst{\ref{inst9}}\and
    V.~Novotn\'y\inst{\ref{inst2}, \ref{inst14} \star}\and
    M.~Ostrowski\inst{\ref{inst15}}\and
    M.~Palatka\inst{\ref{inst2}}\and
    M.~Pech\inst{\ref{inst2}}\and
    M.~Prouza\inst{\ref{inst2}}\and
    P.~Schovanek\inst{\ref{inst2}}\and
    V.~Sliusar\inst{\ref{inst12}}\and
    {\L}.~Stawarz\inst{\ref{inst15}}\and
    R.~Sternberger\inst{\ref{inst8}}\and
    M.~Stodulska\inst{\ref{inst1}}\and
    J.~\'{S}wierblewski\inst{\ref{inst5}}\and
    P.~\'{S}wierk\inst{\ref{inst5}}\and
    J.~\v{S}trobl\inst{\ref{inst10}}\and
    T.~Tavernier\inst{\ref{inst2} \star}\and
    P.~Tr\'avn\'i\v{c}ek\inst{\ref{inst2}}\and
    I.~Troyano Pujadas\inst{\ref{inst1}}\and
    J.~V\'icha\inst{\ref{inst2}}\and
    R.~Walter\inst{\ref{inst12}}\and
    K.~Zi{\c e}tara\inst{\ref{inst15}}
    }

\institute{
D\'epartement de Physique Nucl\'eaire, Facult\'e de Sciences, Universit\'e de Gen\`eve, 24 Quai Ernest Ansermet, CH-1205 Gen\`eve, Switzerland\label{inst1}
\and 
FZU - Institute of Physics of the Czech Academy of Sciences, Na Slovance 1999/2, Prague 8, Czech Republic\label{inst2}
\and
Pidstryhach Institute for Applied Problems of Mechanics and Mathematics, National Academy of Sciences of Ukraine, 3-b Naukova St., 79060, Lviv, Ukraine\label{inst3}
\and
Nicolaus Copernicus Astronomical Center, Polish Academy of Sciences, ul. Bartycka 18, 00-716 Warsaw, Poland\label{inst4}
\and
Institute of Nuclear Physics Polish Academy of Sciences, PL-31342 \krakow, Poland\label{inst5}
\and
Astronomical Observatory, University of Warsaw, Al. Ujazdowskie 4, 00-478 Warsaw, Poland\label{inst6}
\and
Palack\'y University Olomouc, Faculty of Science, 17. listopadu 50, Olomouc, Czech Republic\label{inst7}
\and
Deutsches Elektronen-Synchrotron (DESY) Platanenallee 6, D-15738 Zeuthen, Germany\label{inst8}
\and
Faculty of Physics, University of Bia{\l}ystok, ul. K. Cio{\l}kowskiego 1L, 15-245 Bia{\l}ystok, Poland\label{inst9}
\and
Astronomical Institute of the Czech Academy of Sciences, Fri\v{c}ova~298, CZ-25165 Ond\v{r}ejov, Czech Republic\label{inst10}
\and
Astronomical Institute of the Czech Academy of Sciences, Bo\v{c}n\'i~II 1401, CZ-14100 Prague, Czech Republic\label{inst11}
\and
D\'epartement d'Astronomie, Facult\'e de Science, Universit\'e de Gen\`eve, Chemin d'Ecogia 16, CH-1290 Versoix, Switzerland\label{inst12}
\and
ETH Zurich, Institute for Particle Physics and Astrophysics, Otto-Stern-Weg 5, 8093 Zurich, Switzerland\label{inst13}
\and
Institute of Particle and Nuclear Physics, Faculty of Mathematics and Physics, Charles University, V Hole\v sovi\v ck\' ach 2, Prague 8, Czech~Republic\label{inst14}
\and
Astronomical Observatory, Jagiellonian University, ul. Orla 171, 30-244 \krakow, Poland\label{inst15}
}

   \date{Received ... / Accepted ...}
 
  \abstract{
  The Single-Mirror Small-Size Telescope (SST-1M) stereoscopic system is composed of two Imaging Atmospheric Cherenkov Telescopes (IACTs) designed to deliver optimal performance for gamma-ray astronomy in the multi-TeV energy range. It features a 4-m diameter tessellated mirror dish and an innovative SiPM-based camera. Its optical system features a 4-m diameter spherical mirror dish based on the Davies-Cotton design, maintaining a good image quality over a large field of view (FoV), while minimizing optical aberrations. In 2022, two SST-1M telescopes were installed at the Ond\v{r}ejov Observatory, Czech Republic at an altitude of 510 meters above sea level, collecting data for commissioning and astronomical observations since then.
  
  We present the first SST-1M observations of the Crab Nebula, conducted between September 2023 and March 2024 in both mono and stereoscopic modes. During this observation period, 46 hours for the SST-1M-1 and 52 hours for the SST-1M-2 were collected (of which 33 hours were in stereoscopic mode). In this work, we used the Crab Nebula observation to validate the expected performance of the instrument, as evaluated by Monte Carlo (MC) simulations that were carefully tuned to account for instrumental and atmospheric effects. We determined that the energy threshold at the analysis level for the zenith angles below $30^\circ$ is 1 TeV for mono mode and 1.3 TeV for stereo mode. The energy and angular resolutions were approximately 20\% and $0.18^\circ$ for mono mode and 10\% and $0.10^\circ$ for stereo mode, respectively.
  
  We present an off-axis performance assessment of the instrument and a detailed study of the systematic uncertainties. The full simulation results for the telescope and its camera are compared to the data for the first time, enabling a deeper understanding of the SST-1M array performance.
  }


   \maketitle

\section{Introduction} \label{sec:intro}

Observations of very-high-energy (VHE) cosmic gamma-ray sources are crucial for understanding the extreme processes that accelerate particles to PeV energies in our Galaxy \citep[e.g.,][]{2019IJMPD..2830022G, 2024NatAs...8..425D} and beyond \citep{Bose2022-ks}. However, detecting VHE gamma rays presents significant challenges. Since photon fluxes from cosmic sources follow negative power laws, which are also reduced by propagation in the interstellar medium, satellite-based experiments such as  Fermi-Large Area Telescope \citep{2009ApJ...697.1071A} do not have enough collection area in their detection time frame to measure them. Ground-based experiments, such as imaging atmospheric Cherenkov telescopes (IACTs) and water Cherenkov arrays, offer a much larger collection areas than space-borne detectors by exploiting the showers of secondary particles produced when gamma-ray photons interact with atmospheric nuclei. These secondary particles can be detected over a large area on the ground and discriminated by the cosmic ray-induced background. In brief, IACTs detect the Cherenkov light emitted shortly after charged relativistic secondary particles in the showers pass through a medium. These are captured by the optical systems of IACTs and focused on their nanosecond-sensitive cameras. Alternatively, other experiments detect the light induced by shower charged particles directly in the water of their tanks \citep[see e.g.,][for a review of gamma-ray detection techniques]{DavidJFegan_1997, 2015CRPhy..16..610D, 2022EPJST.231....3B, 2022Galax..10...21S}. Beyond TeV energies,  the flux sensitivity of IACTs is also limited by the number of telescopes spread on a given area and by their optical field of view (FoV), due to the large angular size of the high-energy showers at moderate impact distances. Additionally, VHE gamma-ray sources in the Galaxy are often extended \citep[see e.g.,][]{2024ApJS..271...25C, HESS-Collaboration2018-ca}, further limiting their observability with the current generation of IACTs, which tend to have a rapidly decreasing gamma-ray acceptance with increasing offset angle on a scale of $\sim1^\circ$ \citep[e.g.,][]{2016APh....72...76A}.

The Single-Mirror Small-Size Telescope (SST-1M) was developed to probe VHE gamma rays while addressing the usual IACT limitations, such as a small FoV and limited duty cycle due to moonlight. Its optical system features a 4-m diameter spherical mirror dish based on the Davies-Cotton design \citep{1957SoEn....1...16D}, maintaining a good image quality over a large FoV, while minimizing optical aberrations \citep{sst1m_hw_paper}. That is crucial for observations of extended sources, follow-ups of poorly localized transients\footnote{In the case of a large direction uncertainty on the order of several tens of degrees, a larger FoV means fewer pointings are needed to cover the uncertainty region.} and a good reconstruction of VHE gamma-ray showers generating images that can extend over several degrees in the camera. The aperture of the Cherenkov camera is composed of a 3 mm thick Borofloat window, which integrates a narrow-band optical filter composed of dielectric layers and an anti-reflective coating \citep{sst1m_hw_paper}. This enables the transmission of Cherenkov light to the pixel sensors while reducing the NSB. The camera comprises 1296 silicon photomultipliers (SiPMs) \citep{CameraPaperHeller2017}, enabling safe operation in the presence of high night sky background (NSB) without human intervention, effectively improving robustness against high light rates and duty cycle compared to cameras equipped with photomultiplier pixel \citep{2013JInst...8P6008A}. The degradation of the SiPM performances under high NSB has been studied in \citep{nagai_sipm_2019} for the SiPM of the camera and thus are copped for in the analysis. Moreover, SiPMs are insensitive to the magnetic field and no high-voltage power supply is needed.

Currently, two SST-1M telescope prototypes are operating at the Ond\v{r}ejov Observatory in the Czech Republic, at an altitude of $510\,\rm{m\,a.s.l.}$ The telescopes\footnote{Hereafter referred to as SST-1M-1 ($long=14.782924^\circ$, $lat=49.911800^\circ$) and SST-1M-2 ($long=14.782078^\circ$, $lat=49.913084^\circ$).} are separated by 152.5 meters, and their timestamps are synchronized to nanosecond precision using the White Rabbit timing network \citep{white_rabbit}. They operate in a stereoscopic mode, allowing for precise shower reconstruction when observed from different directions. Both telescopes are currently in the commissioning phase, focusing on optimizing operations and gaining a detailed understanding of all components of the stereoscopic system. A comprehensive description of their design and operation is provided in \citet{sst1m_hw_paper}.

The Crab Nebula, is a nearby ($\sim 2\,\rm{kpc}$) remnant of a supernova observed in 1054 A.D., hosting a pulsar with a rotation period of $33\,\rm{ms}$. It was the first detected VHE gamma-ray source \citep{1989ApJ...342..379W}, and exhibits a spectrum extending up to PeV energies \citep{2021Sci...373..425L}. While signs of variability or flares were reported at GeV energies \citep{2011Sci...331..736T, doi:10.1126/science.1199705}, the Crab Nebula maintains a stable flux at VHE, despite years of monitoring with various instruments \citep[e.g.,][]{2014A&A...562L...4H}. It is therefore usually referred to as the ``standard candle" of VHE astronomy and is used to assess the performance of VHE gamma-ray observatories. In this study, we evaluate the mono and stereo performance of a pair of the SST-1M telescopes through observations of the Crab Nebula conducted between September 2023 and March 2024. 

In Section~\ref{sec:calib}, we describe the calibration procedure and the Monte Carlo (MC) model of the instrument. Section~\ref{sec:mc} details the MC simulations. In Section~\ref{sec:pipeline}, we explain the event reconstruction and the data analysis pipeline. Section~\ref{sec:performance} is dedicated to the mono and stereo performance of SST-1M. In Section~\ref{sec:crab}, we present the Crab Nebula data sample, focusing on the validation of the MC model and presenting the results on the spectrum and the skymap. Section~\ref{sec:systematic} provides a detailed discussion of systematic uncertainties. Finally, we conclude the paper with a summary and final remarks in Section~\ref{sec:summary}.

\section{Calibration and MC model of the telescopes}
\label{sec:calib}

IACTs collect Cherenkov light produced by secondary charged particles in atmospheric air showers. As already explained in Section~\ref{sec:intro}, by imaging this light in the cameras of IACTs, the direction and energy of the primary particle can be reconstructed. The data reduction process, from the photosensor signals into calibrated images, requires a precise characterization of several factors: optical efficiency, baseline, dark count rate, night sky background, gain, and crosstalk. These parameters are used to construct a telescope model for MC simulations, which, together with the precise model of the atmosphere at the observational site, simulate air shower development and Cherenkov photon detection (Section~\ref{sec:mc}). These MC simulations provide a necessary dataset to train machine learning (ML) algorithms used to estimate the energy and direction of incoming particles and discriminate gamma rays from hadrons (Section~\ref{sec:pipeline}). The MC simulation is obtained using \texttt{CORSIKA}~\citep{1998cmcc.book.....H} for detailed simulation of extensive air showers initiated by gamma rays and cosmic rays, and \texttt{sim\_telarray}~\citep{BERNLOHR2008149}, which simulates the ray tracing, geometry and response of the detector, as well as the attenuation of Cherenkov light in the atmosphere.

\subsection {Calibrations of the SiPM response}

The telescopes are remotely operated via a graphical user interface, which allows users to execute calibration procedures and monitor the telescope subsystems during operation. The dark run is performed before the observation after the thermal stabilization of the telescope camera, with the camera lid closed. Another dark run is performed after the end of the observation. The data are used to obtain pixel gain, dark count rate, optical crosstalk, and electronic noise. 
Analogue signals from the pixels are routed to the crate-based electronics called Digicam, where they are digitized using 250 MHz Flash Analog-to-Digital Converters. The calibration process, based on \cite{alispach_large_2020} and further developed, provides the necessary conversion coefficients from analog-to-digital converter (ADC) counts to unbiased photoelectrons (p.e.). Within this procedure, the baseline, namely, the average measured signal in 1024 preceding time bins is computed in the Digicam and subtracted. 
\subsubsection{Photoelectron spectrum}
\label{sec:photoelectron_spec}

We defined the gain, $g$ [ADC/p.e.], of a pixel, coupled with its readout electronics, as the average number of integrated ADC counts recorded per single avalanche event in the SiPM. During observations, optical crosstalk may occur when a secondary avalanche is triggered within a secondary SiPM micro-cell by photons emitted during the discharge of a neighboring primary cell \citep{nagai_sipm_2019}. This effect occurs only between micro-cells within the same SiPM pixel and, if it is not corrected, leads to an overestimation of the measured p.e. count. The conversion factor from ADC counts to p.e. is given by $g^* = \frac{1}{\nu g}$ [p.e./ADC], where $\nu$ is the averaged p.e. produced per primary avalanche. These parameters are routinely monitored before and after each night of observation through dedicated dark runs as shown in \citet{sst1m_hw_paper}.
During dark runs, spontaneous avalanches are triggered by thermal electrons with a Poisson probability, $P_{\lambda}$. The total number of produced p.e. per primary avalanche, accounting for crosstalk, is well described by a Borel distribution of parameter $\nu^*$ \citep{2012NIMPA.695..247V}. These events exhibit the same characteristics as photon-triggered avalanches, allowing dark runs to be used for evaluating the SiPM's response.
For each pixel, the p.e. spectrum is constructed by stacking the maximum charge from the integrated signal in $n_s = 7$ consecutive samples from each waveform. The ADC distribution of dark events is built from approximately 30'000 randomly triggered waveforms of $n_T=50$ samples. The resulting histogram is then fitted using the probability density function of the ADC counts ($n_\mathrm{ADC}$), namely, the p.e. spectrum: $\mathrm{pes}(n_\mathrm{ADC}$), built by summing each $n$ p.e. component as

\begin{equation}
\mathrm{pes}(n_\mathrm{ADC}) =
P_{\lambda}(0) A_\mathrm{el}
+ (1-P_{\lambda}(0))A_\mathrm{pe},
\label{eq:mes}
\end{equation}

where $A_\mathrm{el}$ describe the ADC distribution for waveforms without dark events described by a Gaussian distribution, $G$, with respective mean $\mu_\mathrm{el}$ and standard deviation $\sigma_\mathrm{el}$ : $A_\mathrm{el} = \mathrm{G}\left( \mu_\mathrm{el},\sigma_\mathrm{el}\right)(n_\mathrm{ADC}) $. $A_\mathrm{pe}$ represents the sum of each $n>0$ p.e., weighted with the  point mass function, following the Borel distribution,
\begin{equation}
B_{\nu^*,n} = \frac{e^{-n\nu^*}(n\nu^*)^{n-1} }{n!}.
\label{eq:borel}
\end{equation}
Then,
\begin{equation}
A_\mathrm{pe} = \sum_{n=1}^{\infty} B_{\nu^*,n}  \mathrm{G}\left(n_0 + n g,\sqrt{\sigma_\mathrm{el}^2 + n  \sigma_\mathrm{pe}^2 }\right) (n_\mathrm{ADC}
),\,
\label{eq:mes2}
\end{equation}

where $\sigma_\mathrm{pe}$ represent the standard deviation of ADC counts distribution per single avalanche charge.
The averaged p.e. produced per primary avalanche, $\nu$, used in the analysis is corrected for the rate of dark event in the integration window: $\nu = \frac{1}{1-\nu^*} - \lambda \frac{ n_s}{n_T} $. $n_0 =-\nu \lambda g \frac{n_T}{n_s}$ is the true level of the electronic baseline.

\begin{figure}[!t]
    \centering
    \includegraphics[width=1\linewidth]{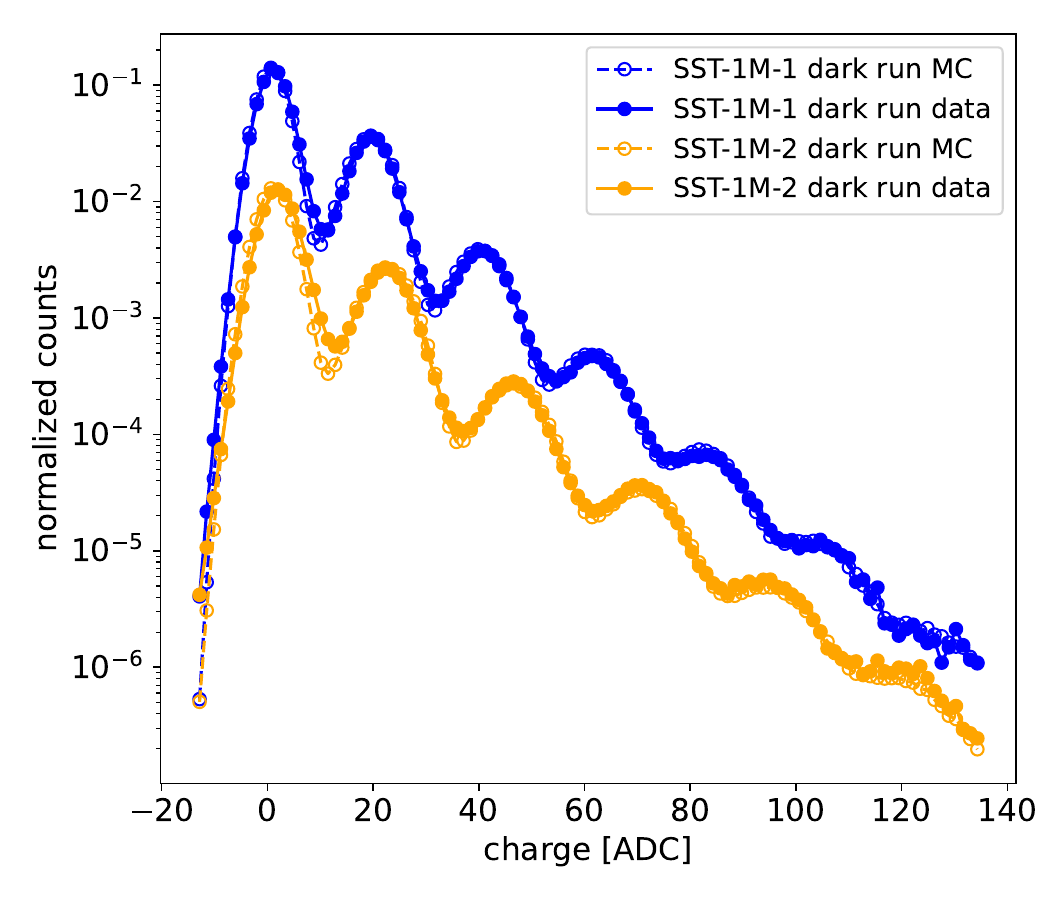}
    \caption{Comparison between data and MC of the normalized (to area) ADC charge distributions extracted from the dark runs for the two telescopes. To avoid overlaps of the two curves, the SST-1M-2 distribution is multiplied by 0.1. 
    Each count is calculated from the 50 samples ($200\,\rm{ns}$) long waveform in the same manner as done for the triggered data. See the explanation in the text. An asymmetric distribution of the electronic noise (more significant in SST-1M-2) can explain the slight disagreement between the first and the second peak.
    }
    \label{fig:dark_run}
\end{figure}

This fitting procedure gives an estimation of the SiPM gain and crosstalk probability with relevant uncertainties for each camera pixel.
In Figure~\ref{fig:dark_run}, we show the averaged distribution of dark ADC counts for all camera pixels.

To reliably describe the response of the SiPM camera to incoming photons, we tuned the \texttt{sim\_telarray} MC model parameters to comply with the measured dark run data.
In the \texttt{sim\_telarray}, the response parameters of individual pixels are drawn from a single population with parameters defined by the MC model.
The tuning of the model includes the optical crosstalk probability, electronic noise, and the amplitude of a p.e. pulse in the waveform, which is the parameter used by \texttt{sim\_telarray} to transform the detected p.e. into ADC counts.
To mimic the thermal rate of avalanches (dark noise) in SiPMs, an NSB value in the simulations was set to 1.8 and 1.5 MHz for SST-1M-1 and SST-1M-2, respectively.
The single p.e. pulse shape was determined from the data, stacking more than 50'000 individual p.e. pulses collected during dark runs. To incorporate the above-described crosstalk in simulations, we use the single-photoelectron spectrum generated from a Poisson branching process for the optical crosstalk smeared by the gain fluctuation of the SiPM. The resulting distribution is a (smeared) Borel distribution. The model have been found to be in good agreement with the experimental probability distributions for dark counts of SiPMs \citep{2012NIMPA.695..247V}.

A comparison of the measured and MC-simulated dark runs is shown in Fig.~\ref{fig:dark_run}, where we present a normalized distribution of waveform-extracted charges of all camera pixels across 11'972 and 100'000 events for the data and simulations, respectively.
The first peak, centered slightly above $0\,\rm{ADC}$ counts, corresponds to a complex interplay between the baseline subtraction, search for the pulse maximum in the noise-only waveform and the electronic noise.
The slight disagreement in its shape for SST-1M-2 between the first and second peak can be attributed to the asymmetric distribution of the electronic noise, more important in SST-1M-2. This effect is currently not simulated and will be the subject of further studies.
Nevertheless, the high-level analysis is not affected as the measured signal is composed of stacked pulses coming from the second peak and further peaks, which have been aptly modeled for both telescopes.

\subsubsection{Voltage drop induced by NSB}
\label{sec.voltage_drop}

The SST-1M cameras are equipped with SiPM sensors, which offer several advantages over traditional photomultiplier tubes (PMTs). Unlike PMTs, SiPMs can operate under high and continuous illumination without experiencing aging effects or damage. For IACTs, continuous light primarily originates from the NSB, which includes diffuse starlight, artificial light sources, moonlight, and ground reflections. Continuous exposure to light causes a steady current to flow through the SiPM bias circuitry. To prevent this high current from causing damage, a bias resistor ($R_\mathrm{bias}$) is connected in series with the SiPM. The current flowing through $R_\mathrm{bias}$ induces a drop of the biasing voltage, which translates naturally into a reduction of the gain (g), the photon detection efficiency (PDE), and optical crosstalk probability. The frontend electronics of the SST-1M is DC coupled so the direct current resulting from the feedback loop can be evaluated by measuring the pixel's baseline shift. The behavior of SiPM parameters under such conditions has been fully described and experimentally tested for the camera of SST-1M-1 in \citet{nagai_sipm_2019}.

\begin{figure*}[!t]
\centering
\begin{tabular}{cc}
\includegraphics[width=.49\textwidth]{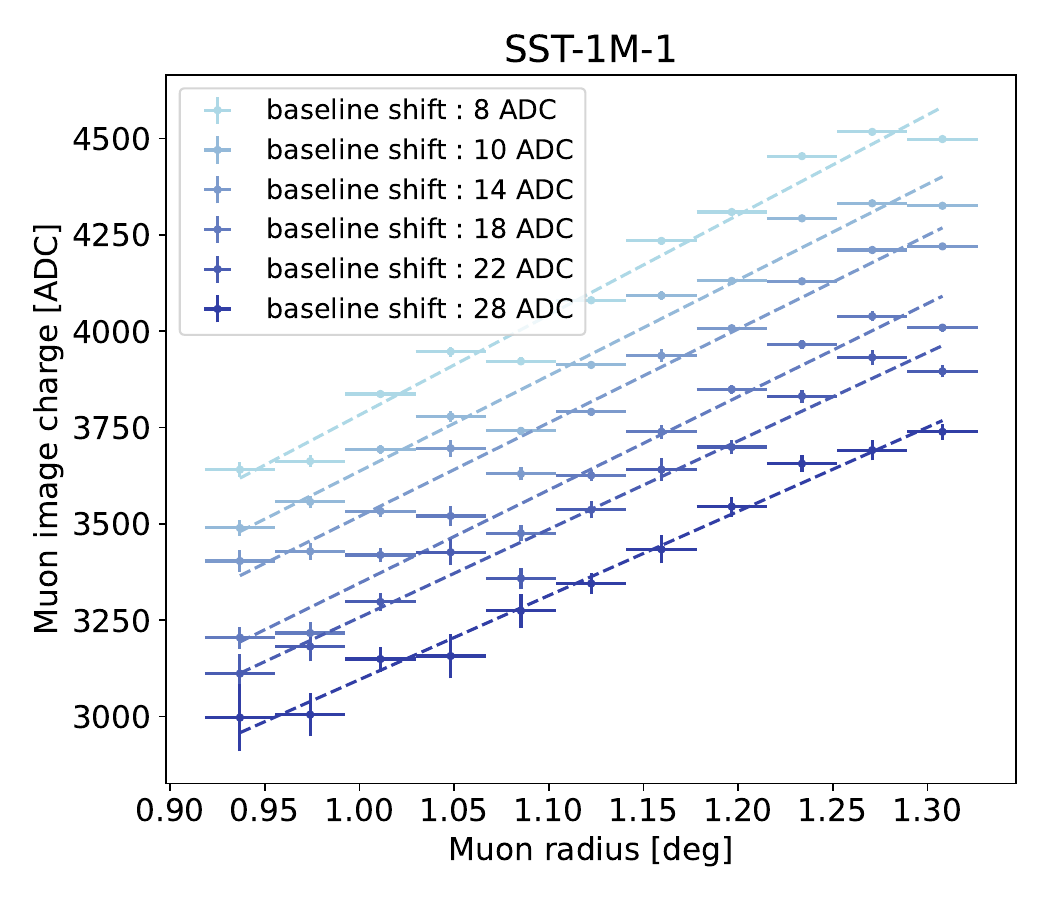} & \includegraphics[width=.49\textwidth]{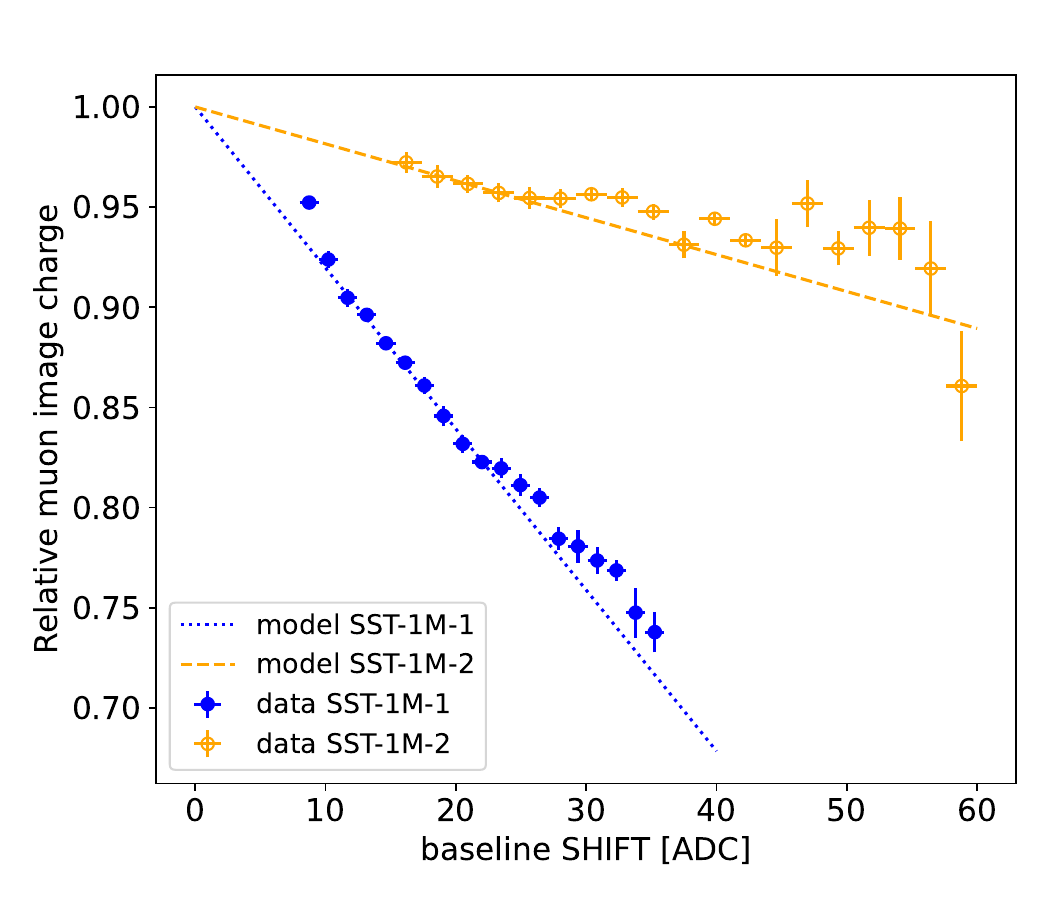}
\end{tabular}
\caption{\textit{Left}:  Relation between muon ring radius and ADC counts under various NSB conditions for SST-1M-1. \textit{Right}: Behavior of measured muon ring charge as a function of the baseline shift induced by the NSB. Blue dotted and orange dashed lines represent the expected behavior given in \cite{nagai_sipm_2019} for SST-1M-1 and SST-1M-2, respectively. Filled blue (SST-1M-1) and empty orange (SST-1M-2) markers represent the fitted estimation of the relative ADC counts for $1.15^{\circ}$ radius muon images.}
\label{fig:QvsNSB}
\end{figure*}

The two camera prototypes in operation feature different $R_\mathrm{bias}$ ($R_\mathrm{bias} = 10\ $k$\Omega$ for the camera of SST-1M-1 and $R_\mathrm{bias} = 2.4\ $k$\Omega$ for SST-1M-2) as it was observed after operation of SST-1M-1 that lowering $R_\mathrm{bias}$ would decrease the impact of their order to lower the effects of NSB for the second prototype. This results in significantly different behaviors of the two telescopes under the same NSB levels. The baseline shifts induced by the NSB are monitored during observations using $100\ $Hz clocked interleaved pedestal events. 
The impact of the voltage drop induced by the NSB on the reconstructed charge can be evaluated using Cherenkov light produced by cosmic muons under different NSB conditions. Figure~\ref{fig:QvsNSB} shows the behavior of the average ADC counts in cleaned muon images with the baseline shift for both cameras (see Section~\ref{subsec:eff} for a detailed description of the muon analysis), demonstrating a good agreement between the voltage drop impact on the muon image charge data and the expected electronic behavior that was modeled according to \citet{nagai_sipm_2019}. This good agreement confirmed that the model can be applied in the data processing, during the calibration step, to correct the image intensity (sum of the integrated pixel charges) as a function of the voltage drop.
The voltage drop is derived by measuring the shift between the baselines measured during the dark runs and the one estimated every second during the observation. The extracted intensity is then corrected according to the model to a level that would be the one measured in the absence of NSB.

\subsection{Optical efficiency calibration} \label{subsec:eff}

The optical efficiency of an IACT refers to the effective fraction of Cherenkov light transmitted through the entire optical system before reaching the photon sensors. It is influenced by mirror reflectance, the NSB filtering window transmittance, the reflectance of the lightguides coupled to the SiPMs, the shadowing of the mirror by the telescope structure, and the SiPM PDE \citep{CameraPaperHeller2017}. Since the early 1990s \citep{fleury_cerenkov_1991}, Cherenkov light produced by atmospheric muons has been recognized as an excellent test beam for the estimation of optical efficiency \citep{gaug_using_2019-2}. 
Atmospheric muons are produced in hadronic showers at high altitudes and can be detected if they pass near one of the telescopes. Cherenkov light is emitted along an emission cone following the muon trajectory. If this trajectory aligns with the telescope FoV, the Cherenkov cone of light will sweep large portion of the mirror and be projected on the focal plane as a ring, with a radius that corresponds to the Cherenkov angle.
The characteristic shape and consistency of these images make them an ideal tool for calibration and testing the accuracy of the MC model in the simulations.

Muon events are collected during routine observations and processed using a dedicated analysis pipeline.
Based on the small timing spread in photon arrivals and stable delay between the trigger and the signal, charge extractions are done using a unique fixed integration window of 28 ns for all pixels.
The integrated images are cleaned using the tailcuts algorithm \citep{2001APh....15....1L}. The brightest pixels in the camera can be identified when they are above the so-called image threshold, set at 5 p.e.. Clusters of pixels, called islands, are built by adding neighboring pixels if their intensity is above the boundary threshold set at 4 p.e. Here, islands with two or more neighboring pixels were retained.
A circle was fitted to the surviving pixels. We estimated the muon charge by summing the charge in all pixels with centers within 0.15 degrees of the fitted circle. Taking into account both the pixel geometry and the optical PSF, this led to the collection of approximately 90\% of the muon signal. Only events with a fitted circle radius between 0.8 and 1.4 degrees were selected for the analysis.
To assess the completeness of the muon ring, the signal within the ring region was summed across 12 regions, each covering a $30^\circ$ angular section. Only images that exhibited a signal above $7\,\rm{p.e.}$ in at least 10 out of the 12 regions were considered for further analysis. 
Additionally, the signal outside the selected pixels was also taken into account. Images with an integrated charge exceeding $20\,\rm{p.e.}$ outside the selected pixels were excluded.
These criteria have been designed to filter out shower images and noise fluctuations, ensuring that only muon images with well-defined ring structures are used in the calibration studies. During typical observations, the muon image rates after selection are approximately $0.1\,\rm{Hz}$ and $0.2\,\rm{Hz}$ for SST-1M-1 and SST-1M-2, respectively.

\begin{figure}[!t]
    \centering
    \includegraphics[width=0.99\linewidth]{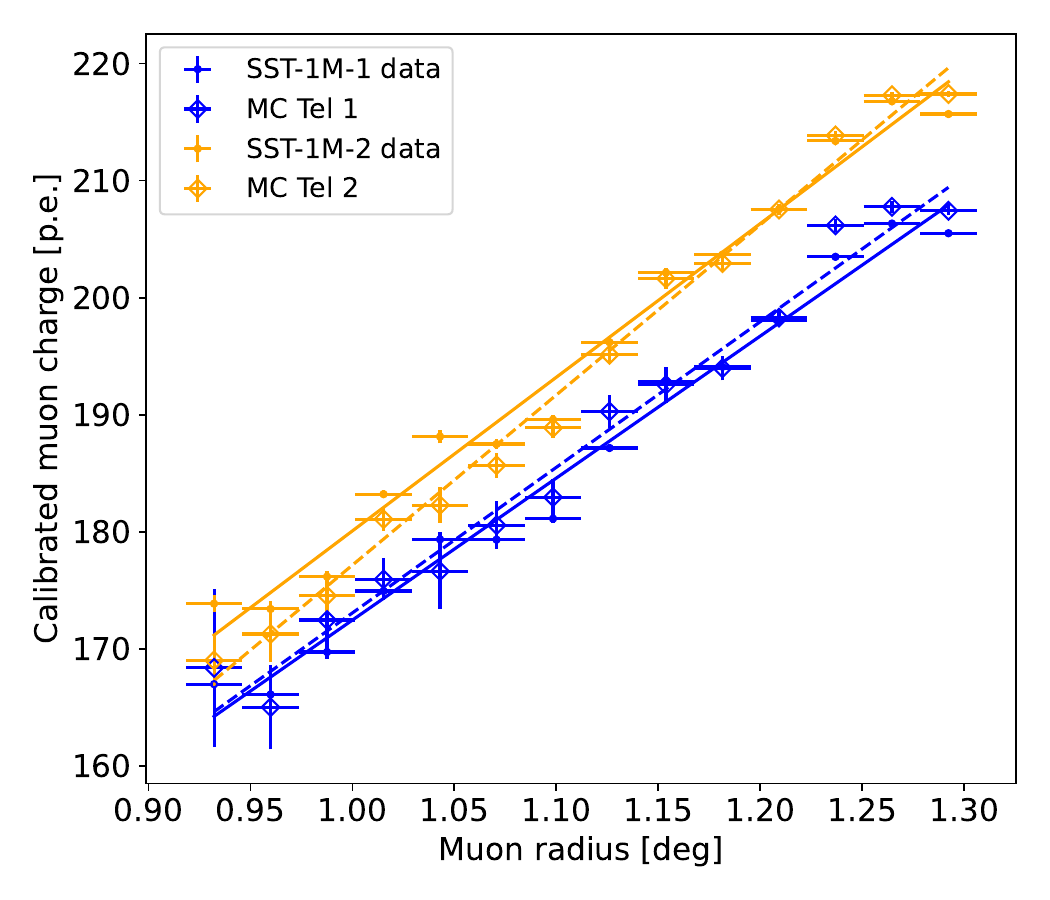}
    \caption{Charge of muon images as a function of the fitted ring radius for SST-1M-1 and SST-1M-2. The solid markers represent observational data collected from September 2023 to April 2024, during the first stereo Crab observation campaign (blue for SST-1M-1 and orange for SST-1M-2). Measured data agree with the MC simulation, shown with open markers and dashed lines. Notable differences between SST-1M-1 and SST-1M-2 arise from differences in their optical designs.}
    \label{fig:muon_RvsI}
\end{figure}

\begin{figure}[!t]
    \centering
    \includegraphics[width=0.99\linewidth]{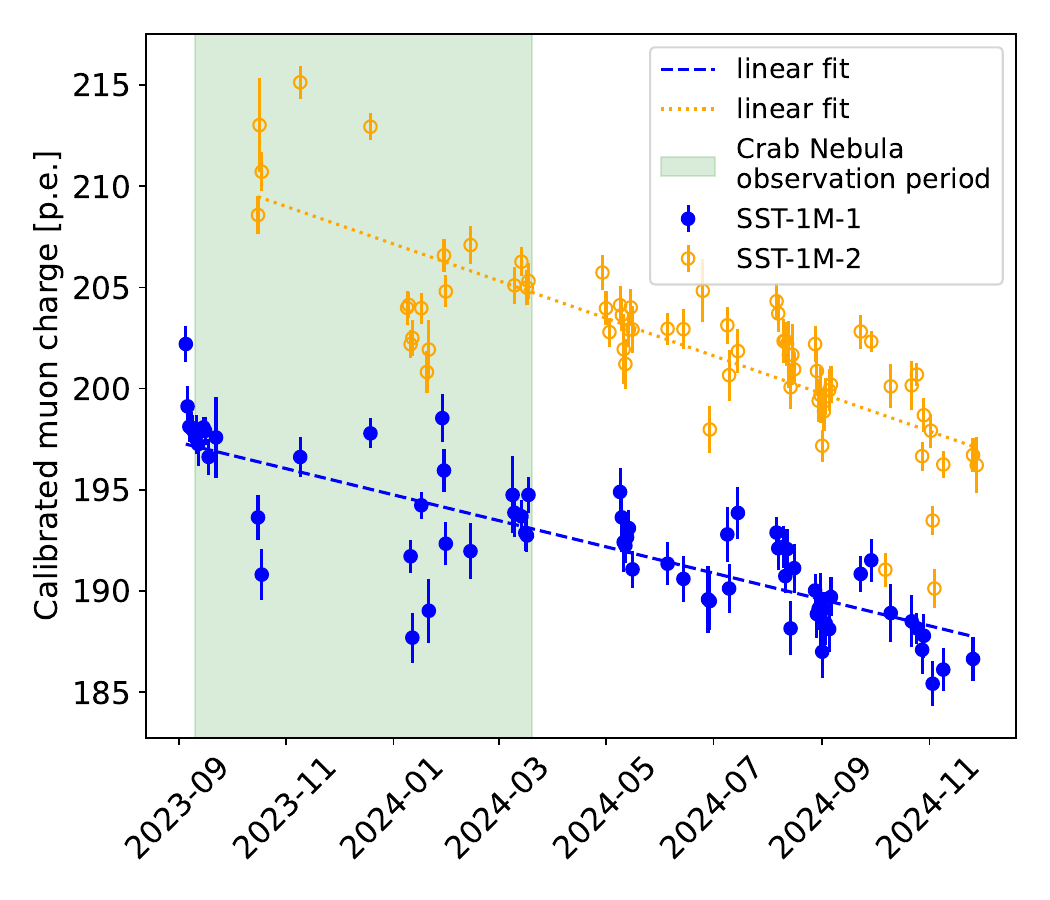}
    \caption{Evolution with time of the charge in the muon ring pixels for fixed radius of $r=1.2^\circ$ for SST-1M-1 (blue points) and SST-1M-2 (orange points). Each data point corresponds to one single observation night. The green-shaded area indicates the time period of the Crab Nebula observation campaign presented in this work. The linear fits (dashed for SST-1M-1, dotted for SST-1M-2) enable the  estimation of the relative decrease in optical efficiency over time.}
    \label{fig:muon_IvT}
\end{figure}

A dedicated MC simulation of muon events with parameters listed in Section~\ref{sec:mc} was used to tune the optical efficiency of the instrument model. Figure \ref{fig:muon_RvsI} shows the relation between the muon ring radius and its integrated intensity. The measured values for both telescopes in the time period of the Crab Nebula observation campaign (see Section~\ref{sec:crab}) are compared with the one obtained with MC simulations, showing the consistency between observational data and the instrument model. The optical efficiency of the telescopes is not constant over time. Gradual degradation of mirror reflectivity, camera window transmittance, and the performance of camera electronic components contribute to a slow but steady decline in the reconstructed muon intensity.
Figure \ref{fig:muon_IvT} shows the evolution of the optical efficiency with time for both telescopes. During the time period of the Crab Nebula observation campaign presented in this work, the degradation of optical efficiency is estimated to be approximately 2\%.

\subsection{Optical point spread function}

The optical point spread function (PSF) is measured with a dedicated CCD camera installed in the center of the mirror dish and a dedicated PSF screen installed on the camera lid \citep{sst1m_hw_paper}. In \texttt{sim\_telarray}, the PSF can be reproduced by smearing the horizontal and vertical alignment of individual mirror facets (whose optical properties are known from lab measurements), along with ray-tracing of light of a simulated point-like source. The optical PSF was measured for both telescopes in December 2023, providing $D_{80} = 10.4\,\rm{mm}$ and $12.2\,\rm{mm}$ for SST-1M-1 and SST-1M-2, respectively, where $D_{80}$ represents the diameter of a circle containing $80\%$ of the reflected light by the mirrors. This translates to the on-axis PSF of approximately $0.1^\circ$. The stiffness of the telescope structure ensures that the PSF does not depend on the telescope pointing direction and is stable in time. The PSF is measured and optimized by pointing at a bright star and adjusting individual mirror facets to reach the smallest PSF in the PSF screen after any optical system intervention, and periodically twice a year \citep{sst1m_hw_paper}. The current observation did not show any PSF degradation over time.

\subsection{Model of the atmosphere in Ond\v{r}ejov} \label{sec:atmoshpere}
A model of the molecular atmosphere, namely, its density profile as a function of altitude, is crucial for a reliable description of particle interactions in air showers, production of Cherenkov light, and its subsequent attenuation along the path to the telescope.
For the Ond\v rejov Observatory, it was extracted from the ECMWF ERA 5 database\footnote{\url{https://www.ecmwf.int/en/forecasts/datasets/reanalysis-datasets/era5}}.
Moreover, precise estimation of the amount of aerosols present in the atmosphere is important when determining the primary particle's energy, since the incoming Cherenkov light gets attenuated by Mie scattering. The aerosols also contribute to the overall NSB, since the scattering aerosol particles act as a light source. Unfortunately, there is no instrument capable of high-precision measurements of the vertical aerosol optical depth (VAOD) at the Ond\v{r}ejov Observatory. As such, we made use of the open access results from the Sun photometer located in Ko\v{s}etice\footnote{\url{https://aeronet.gsfc.nasa.gov/cgi-bin/data_display_aod_v3?site=Kosetice_NAOK}} \citep{1997ThApC..57...95B}. This site is located 45 km away from Ond\v{r}ejov, but otherwise bears remarkable similarities. It is located at the same height above sea level, in the midst of a hilly landscape and shielded from the nearest village,  at a sufficient distance from the nearest town.

The instrument measures the VAOD at a range of wavelengths. We make use of the filter band at 380 nm. This is justified by the transmissivity of the camera and of the optical system being rather flat in the relevant spectral region, but the Cherenkov spectrum increasing sharply at lower wavelengths. The Sun photometer provides only daytime measurements, so for each nighttime measurement, we calculate the VAOD as the mean of the average values of the preceding and following day. It turns out that the VAOD is remarkably stable for this particular dataset of Crab sample measurements and oscillates around the value of 0.05. We estimated that the systematic uncertainties in this sample related to VAOD estimation are approximately of this order, and so we use this value for MC simulations for the whole period. It is worth noting that the observed VAOD is remarkably low; for example, a similar period in 2024 exhibits large variations with a mean value of approximately 0.2. It should be mentioned, however, that the sun photometer measures an integral VAOD, while the shower reconstruction requires a proper measurement of the height-differential VAOD and this must be properly addressed in the atmospheric calibration strategy of IACTs \citep[e.g.,][]{2017EPJWC.14401003G}. Such a simplification can lead to systematics on the energy scale, as discussed  in Section~\ref{sec:systematic}.

With this VAOD value in hand we then calculate the atmosphere transmissivity using the MODTRAN software \citep{modtran} in 50 layers spaced between 0.5 and 100 km, utilizing an average seasonal vertical atmospheric density profile from the ECMWF ERA 5 dataset. It can then serve as an input to the \texttt{sim\_telarray} simulation.

\section{MC simulations} \label{sec:mc}

The MC simulations were produced in two steps. First, particle interactions and Cherenkov light from air showers in \texttt{CORSIKA v7.7402} \citep{1998cmcc.book.....H}. Second, the attenuation of the light and the response of the SST-1M telescopes in \texttt{sim\_telarray v2021-12-25} \citep{BERNLOHR2008149}.
To cover different zenith angles at which the Crab Nebula had been observed, five fixed zenith angle regions of $20^{\circ}$, $30^{\circ}$, $40^{\circ}$, $50^{\circ}$, and $60^{\circ}$ were simulated.
In each of the regions, three sets were simulated, namely the diffuse samples of gamma rays and protons to establish the shower reconstruction, and a point-like sample of gamma rays to directly calculate the SST-1M response to a point-like source.
To keep the number of simulations manageable, we chose to simulate for a fixed azimuth at which the sources culminate, that is, the telescopes were looking to the South in all simulations.
Incoming directions of primaries for the diffuse samples were randomly thrown around the telescope-axis direction up to $10^{\circ}$ opening angle and sampled on the ground in a way to emulate an isotropic flux.
The maximum distance up to which the impact points were sampled in the telescope-tilted ground frame was set between $928\,\rm{m}-1273\,\rm{m}$ for gammas and $1032\,\rm{m}-1414\,\rm{m}$ for protons, both evolving with the zenith angle $\theta$ as $\cos^{-0.5}{\theta}$ motivated by \cite{2023ApJ...956...80A} and confirmed by us to cover the full detection volume.
Similarly, for calculation efficiency purposes, the lowest simulated energy evolved with the zenith angle from $200\,\rm{GeV}$ to $2\,\rm{TeV}$ for gammas and from $400\,\rm{GeV}$ to $3\,\rm{TeV}$ for protons to roughly catch the changing energy threshold of the detection.
The maximum simulated energy was $800\,\rm{TeV}$ for gammas and $1300\,\rm{TeV}$ for protons to probe the SST-1M sensitivity at the highest energies, and the simulated energy spectrum was $\frac{\rm{d}N}{\rm{d}E} \propto E^{-2}$.

\begin{figure*}
\centering
\begin{tabular}{cc}
\includegraphics[width=.49\textwidth]{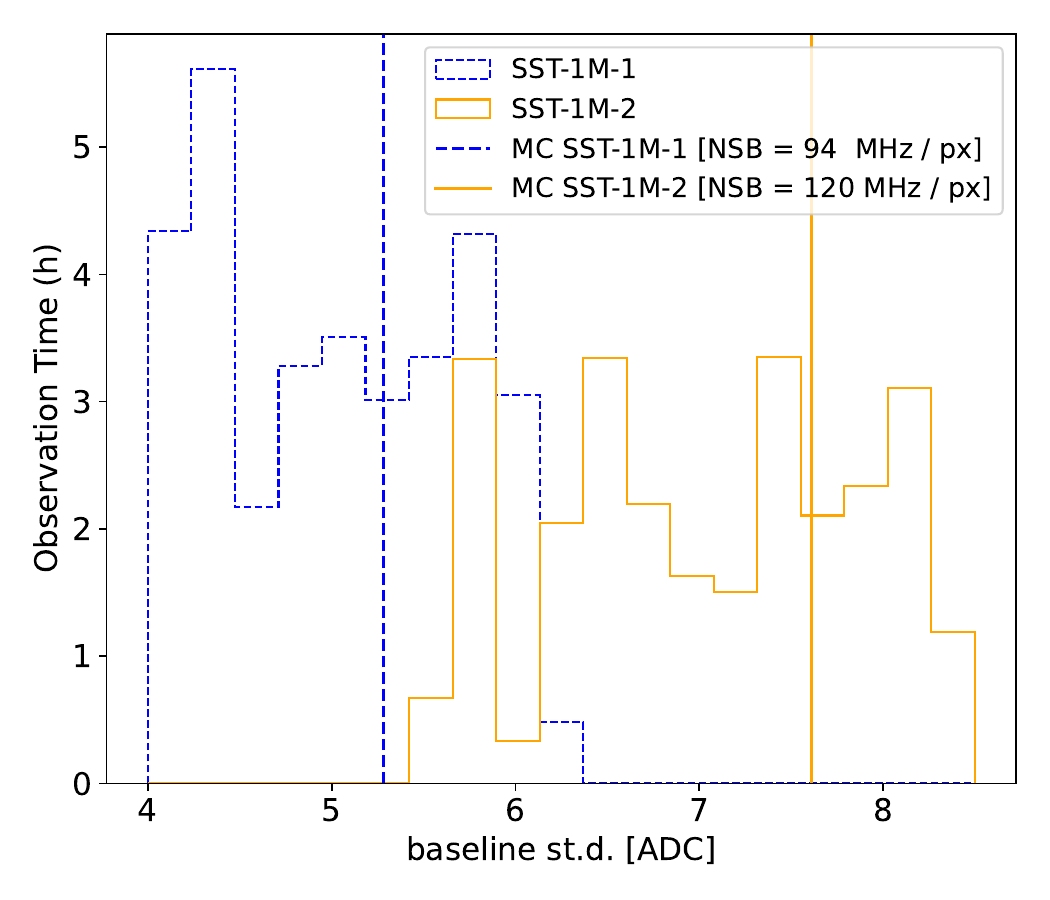} & \includegraphics[width=.49\textwidth]{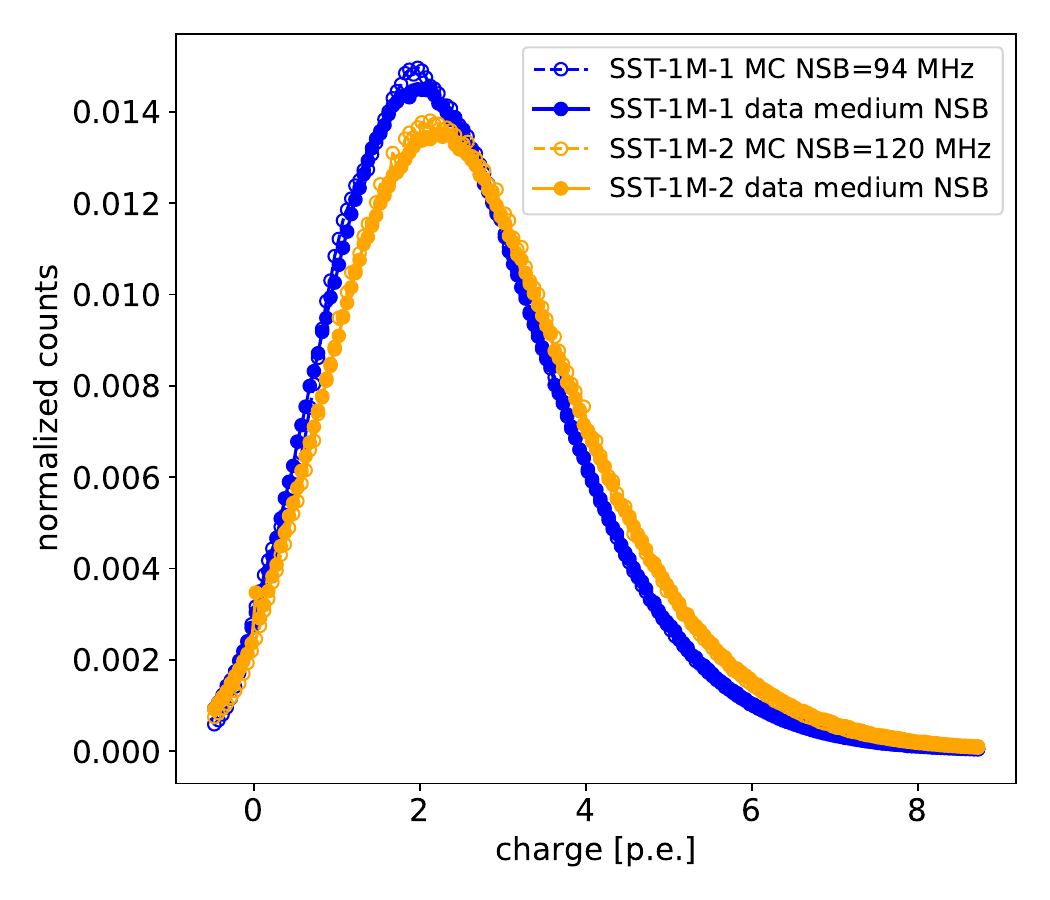}
\end{tabular}
\caption{\textit{Left:} Distribution of the averaged standard deviation of the baselines for SST-1M-1 (dashed line) and SST-1M-2 (solid line). The vertical lines represent the value assumed in the MC model of SST-1M-1 ($94\,\rm{MHz/pixel}$) and SST-1M-2 ($120\,\rm{MHz/pixel}$). The vertical axis represents the number of hours with a baseline standard deviation (on the horizontal axis) in the final Crab Nebula data sample. \textit{Right:} Comparison between the MC-predicted and pedestal events-measured p.e. distributions under the slightly higher than typical NSB conditions during the Crab Nebula observations.}
\label{fig:nsb}
\end{figure*}

While the SST-1M camera is designed to operate under a wide range of NSB conditions, and the telescopes can also collect data under the full moon, the results presented in the following analysis are limited to dark conditions.
The gain, PDE, and optical crosstalk were set following the model described in \cite{nagai_sipm_2019}. The MC events were corrected in the same way as the real data and used to train the energy and angular regression models, as well as the gamma-hadron classification one. 
For typical dark nights at the Ond\v{r}ejov Observatory, the average NSB rate is approximately 100 MHz per pixel. Due to slightly different hardware of the cameras, see Section~\ref{sec.voltage_drop}, the rate of random pulses in pixels caused by NSB was set differently for SST-1M-1 and SST-1M-2 to $94\,\rm{MHz}$ and $120\,\rm{MHz}$, respectively. The left panel of Figure \ref{fig:nsb} shows the distribution of the averaged standard deviation of the baselines, which are directly impacted by the NSB level, compared to the values used in the MC simulations for both cameras.
To stay conservative, the NSB levels in the MC were selected to describe the reconstructed p.e. distributions from pedestal events under the NSB slightly larger than typical during the Crab Nebula observations, depicted in the right part of Figure~\ref{fig:nsb}.
Despite that the Ond\v{r}ejov Observatory is located in an area with a strict public lighting policy, occasional car flashes or the light from nearby villages and towns lead to a substantial NSB variability and a strong dependence on the zenith angle.
To estimate the systematic uncertainty of the telescope sensitivities connected to the discrepancy between the NSB level in the data and MC (Section~\ref{sec:systematic}) we also performed simulations under the lower NSB conditions, representing the darkest nights of observation, with NSB rates of $55\,\rm{MHz}$ and $72\,\rm{MHz}$ for SST-1M-1 and SST-1M-2, respectively.

At the zenith angle of $20^{\circ}$, we simulated 33 millions diffuse gamma-ray showers and 3 millions point-like gammas, each resampled in \texttt{CORSIKA} 20 times, and 26 millions protons, each resampled 100 times.
The amount was reduced for other zenith angles to keep roughly the same number of events above the minimum simulated energy.
In total, 1.86 billion of diffuse gammas, 260 millions of point-like gammas and 7.4 billions of proton induced events were available for the telescope simulation in \texttt{sim\_telarray} and further processing in \texttt{sst1mpipe} (see Section~\ref{sec:pipeline}).

Besides the simulations necessary for the reconstruction setup and Instrument Response Functions (IRFs) determination, a dedicated set of muon simulations was performed to assess the overall optical throughput of the telescopes described in Section~\ref{subsec:eff}.
It consists of 5 millions of muons directed to the telescope with the random scatter around the telescope axis direction up to $6^{\circ}$, impact distances up to $3\,\rm{m}$, and the energy range of $4\,\rm{GeV}-1\,\rm{TeV}$ following the power-law spectrum $\frac{\rm{d}N}{\rm{d}E} \propto E^{-2}$.

The full MC production was split into two independent samples for machine learning-based reconstruction (using random forests, RF) of the energy, direction, and classification -- the training sample used for RF training, and the testing sample for the estimation of the performance and production of IRFs. In order not to limit the RF performance with the training sample size, we conducted a detailed study of mono and stereo performance, varying the number of training events. We carefully monitored energy, angular resolution, and gamma-hadron separation power, especially for high energies, which are most affected by the lack of statistics. It turned out that the best performance can be reached if the number of mono diffuse gammas and protons in the last arbitrary energy bin (500-800 TeV) is higher than $\approx15000$, which we kept roughly constant for all zenith angles (events which survived cleaning; see Sect.~\ref{sec:pipeline}).

\section{Event reconstruction} \label{sec:pipeline}

The data and MC shower reconstruction and analysis are performed in \texttt{sst1mpipe}\footnote{\url{https://github.com/SST-1M-collaboration/sst1mpipe}} \citep{sst1mpipe_073}, which is a standalone \texttt{Python}-based pipeline based on \texttt{ctapipe} \citep{karl_kosack_2021_5720333} functionalities. The basic structure of \texttt{sst1mpipe} is inspired by \texttt{lstchain} \citep{ruben_lopez_coto_2022_6344674} and closely follows the \texttt{ctapipe} data model. For low-level event handling \texttt{sst1mpipe} relies on some functionalities of \texttt{digicampipe}\footnote{\url{https://github.com/cta-sst-1m/digicampipe/tree/master}}, a pipeline developed for camera commissioning, calibration, and low-level data processing. High-level data product of \texttt{sst1mpipe} (photon lists together with IRFs) follows the modern ``gamma astro data formats" (GADF) standards \citep{universe7100374}, making it compatible with contemporary tools for spectral or spatial data analysis like \texttt{gammapy} \citep{axel_donath_2021_5721467}. 

\subsection{Charge extraction} \label{sec:charge_extraction}
Raw data recorded by each telescope contains for each triggered event a 12-bit digitized waveform in all 1296 camera pixels with 50 samples and $4\,\rm{ns}$ intervals. First, the baseline, determined as an average of 1024 samples before the readout, is subtracted. The baseline-subtracted waveforms are then calibrated, namely converted into the number of p.e. using the camera gain described in Section \ref{sec:calib}. Except for this last step, all the following steps are common for data and MC, this last containing already calibrated waveforms, so that both are processed in the same way. Each calibrated waveform is integrated with an eight-sample window around the sample with the maximum signal using the LocalPeakWindowSum algorithm of \texttt{ctapipe}. An integration correction based on the pulse template is applied to correct the remaining charge outside the integration window. In addition to the integrated charge, the pulse time, defined as the amplitude-weighted average of the time samples within the window, is extracted. Faulty pixels, for which the determination of the camera gain failed (typically $\sim5/20$ out of 1296 pixels for SST-1M-1/2), are flagged and the integrated charge, together with the pulse time are replaced by the average value of their neighboring pixels using NeighborAverage method of \texttt{ctapipe}. Constant monitoring of the pixels is done during observation,s and faulty detected pixels are treated the same. We note that in the future, the dead pixels can be corrected at the trigger level, exploiting the fully digital readout and trigger system of the SST-1M camera \citep{CameraPaperHeller2017}.

\subsection{Image cleaning and parametrization} \label{sec:cleaning}
To remove noisy pixels containing only the NSB, we adopted the two-stage modification of the standard tailcut image cleaning \citep{2023ApJ...956...80A}. The default tailcuts were set to $(8\, \mathrm{p.e.},4\, \mathrm{p.e.})$ for both telescopes. These tail cuts were determined empirically as a trade-off between the fraction of true Cherenkov pixels that survived cleaning on one side, and too many noisy pixels left in the cleaned image on the other. First, in the MC sample and the data we optimized for the reference NSB level by demanding about $1\%$ survival rate of pedestal events (which contain only noise). Then we considered the high variability of NSB conditions in Ond\v{r}ejov, and set the tailcuts higher (by $1\,\rm{p.e.}$ to $1.5\,\rm{p.e.}$) to prevent a higher NSB from degrading the overall performance.

To further suppress noisy pixels due to stars and planets in the FoV or to a significant difference between the NSB in the MC and the data, we increased the default tailcuts in each pixel to $\max(8\, \mathrm{p.e.}, \langle Q_\mathrm{PED} \rangle + 2.5 \sigma_{Q_\mathrm{PED}}$), where $\langle Q_\mathrm{PED} \rangle$ is a mean charge in each pixel over 1000 interleaved pedestal events corresponding to 10 seconds, and $\sigma_{Q_\mathrm{PED}}$ stands for its standard deviation \citep{2023ApJ...956...80A}. The multiplicative factor on standard deviation was set to keep the number of altered pixel thresholds at the level of $1-2\%$ so that its impact on the MC and data agreement is negligible. Further improvements of the cleaning performance was achieved by demanding that the arrival time of the neighbor pixel is within $8\,\rm{ns}$ of the arrival time of the pixel, which exploits the natural correlation of Cherenkov pulse times. In the next step, cleaned pixels containing signal from the Cherenkov photons were parameterized with an extended set of Hillas parameters \citep{1985ICRC....3..445H} to describe the shape, orientation, and position of the shower image (for a full list of image parameters used for the event reconstruction, see Figure~\ref{fig.feature_importance}).

\subsection{ RF reconstruction} 
\subsubsection{Mono reconstruction} 
To reconstruct the properties of the primary gamma-ray photon and the shower geometry, the image parameters are fed into RFs \citep{10.1023/A:1010933404324} trained on the training set of MC events using \texttt{scikit-learn} framework \citep{scikit-learn}. The RF classifier is used for gamma-hadron separation (resulting in so-called "gammaness"\footnote{A numerical value between 0 and 1 representing how much a given shower image is gamma-like.} parameter for each reconstructed event), and the RF regressor is used for energy reconstruction. To determine the arrival direction of each shower in mono analysis, we adopt the so-called DISP method \citep{2001APh....15....1L}, assuming that the source lies on the main axis of the shower image. An RF regressor is trained to reconstruct the distance of the source from the image centroid (disp norm), together with an RF classifier to determine on which side of the centroid along the image axis the source lies (disp sign).

\subsubsection{Stereo events} 
The Crab Nebula observations were conducted with both telescopes triggered independently. Therefore, in the stereo reconstruction of such a data sample, the first step is to match events resulting from the same shower seen stereoscopically with both telescopes. With the typical mono trigger rate of around 200~Hz and White Rabbit sampling precision of 16~ns, the event matching can be based only on event timestamps with negligible probability of random coincidences. Moreover, event matching is performed on the subset of events that survived cleaning, further suppressing the likelihood of random coincidences (below 0.01\% of events). In the subsequent analysis, we consider the event stereoscopic, if, for a given shower seen with SST-1M-1, there is a counterpart seen with SST-1M-2 within $10 \, \mathrm{\mu s}$ window. Another possibility is exploiting coincidence event tagging by the Software Array Trigger \citep{sst1m_hw_paper}, which is currently being tested.

In the reconstruction of stereo events, we proceed similarly to \citet{2023A&A...680A..66A}. First, the geometrical reconstruction of the shower impact distance ($\mathrm{tel\_impact\_distance}$) and the height of shower maximum ($\mathrm{h\_max}$) is calculated using the axis crossing method \citep{Hofmann:1999rh}. The stereo parameters are then used as additional features for the RF training. The energy regressor and gamma-hadron classifier are trained independently for each telescope and their outputs are averaged using the image intensity as a weighting factor. For the shower direction, the stereo reconstruction follows the approach of the MAGIC analysis and reconstruction software \citep{ALEKSIC201676}. First, disp norm was reconstructed for both telescopes with the RF regressor. Then, four tentative source positions are calculated, averaging the reconstructed positions for both telescopes for all possible combinations of the head-tail shower orientations. Finally, the combination providing the closest angular distance between the reconstructed coordinates from the two telescopes is adopted.

\subsubsection{RF performance} 

Hyperparameters of the RFs were set to reach a trade-off between a good performance and manageable size and training time for each RF. The RF regressors used for energy and disp norm reconstruction use 150 estimators, maximum tree depth 30, minimal number of samples to split a node 10, and the squared error criterion to measure the quality of each split. The gamma-hadron classifier, and disp sign classifier in the case of the mono reconstruction, employs 100 estimators, a maximum depth of 100, and 10 samples needed to split a node. The Gini impurity is adopted as a measure of the quality of each split, but only the square root of the total number of features is randomly selected to evaluate the Gini index in each step.

\begin{figure*}[!t]
\centering
\begin{tabular}{cc}
\includegraphics[width=.49\textwidth]{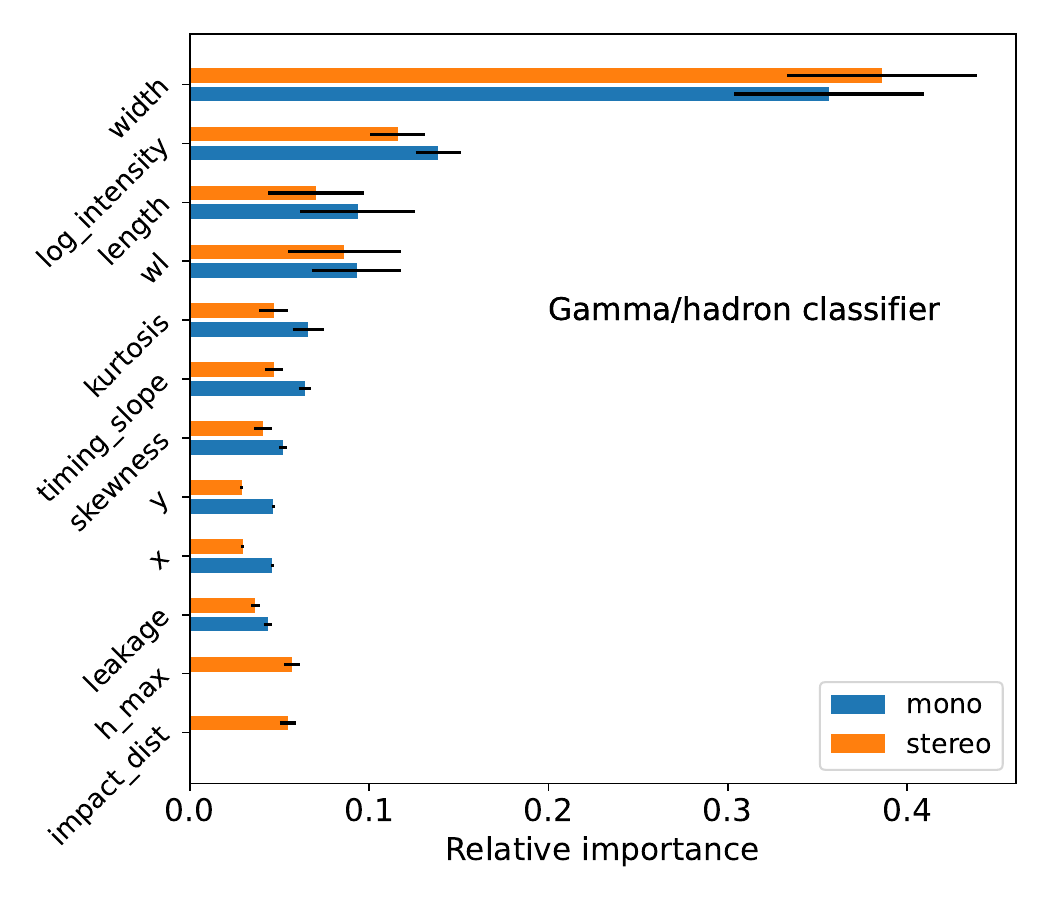} & \includegraphics[width=.49\textwidth]{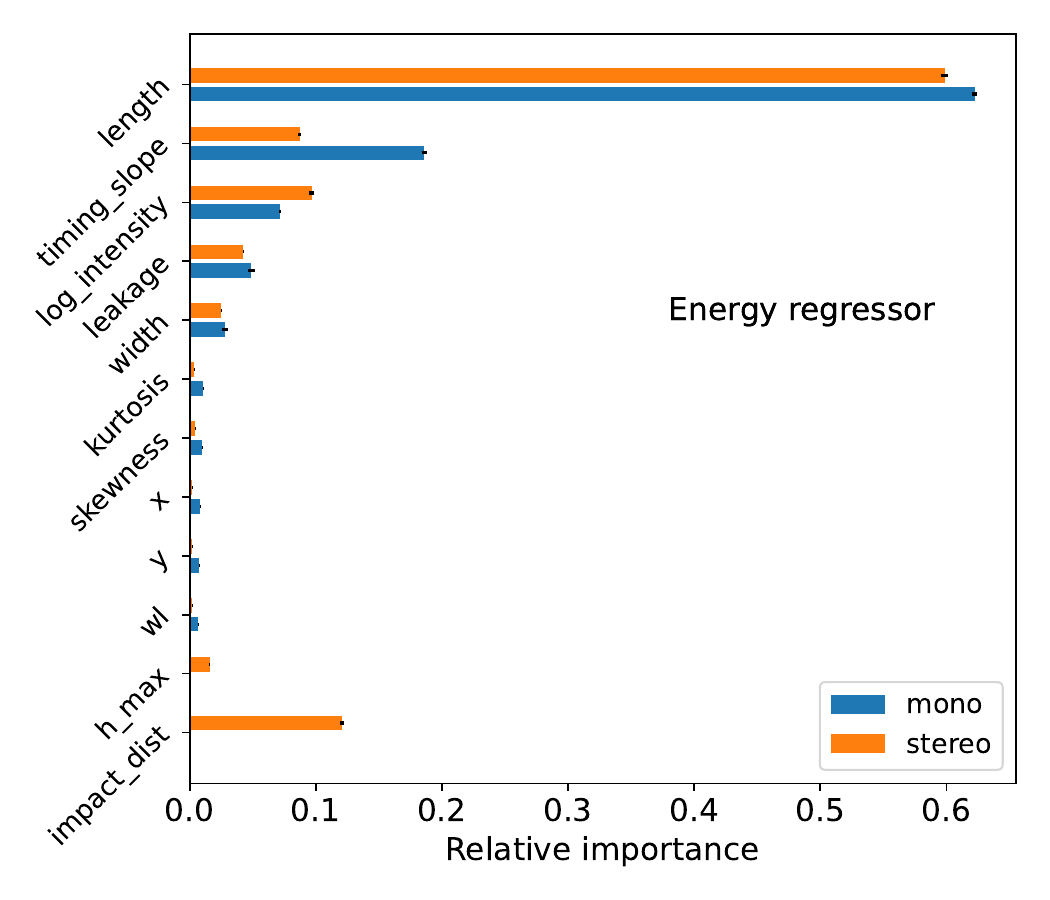} \\
\includegraphics[width=.49\textwidth]{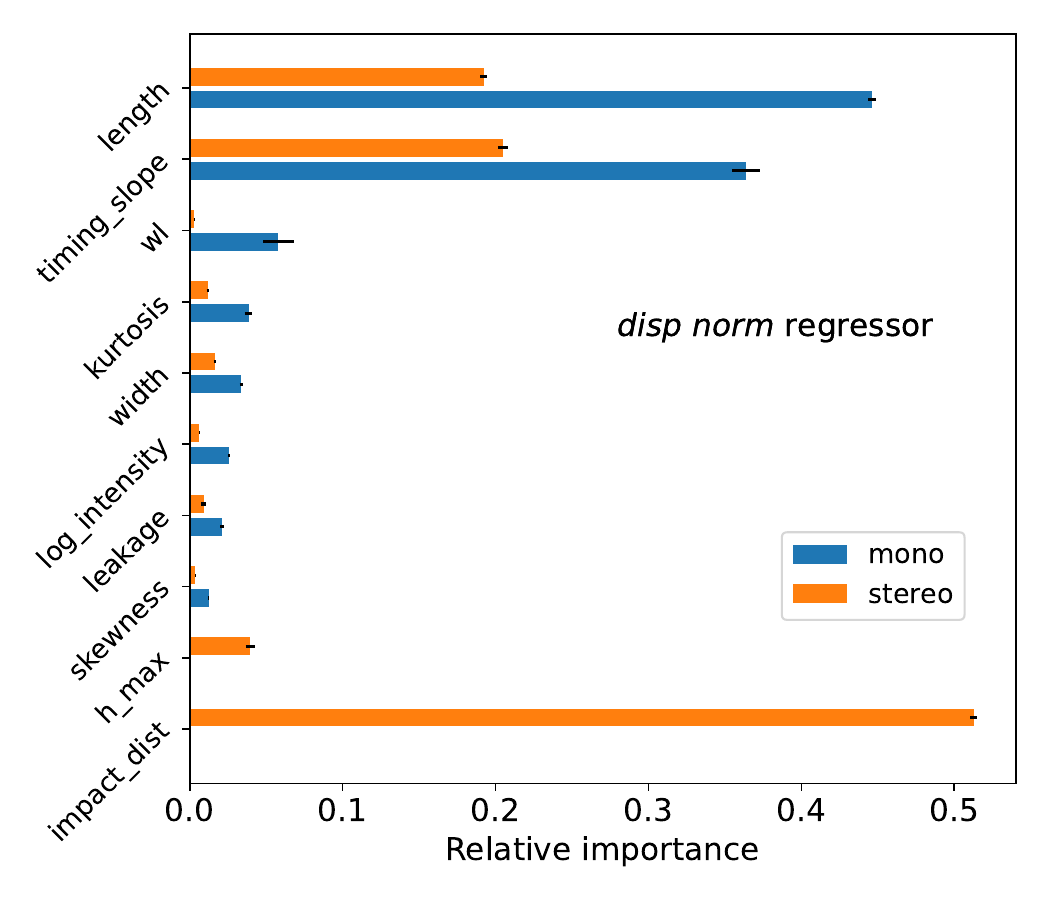} & \includegraphics[width=.49\textwidth]{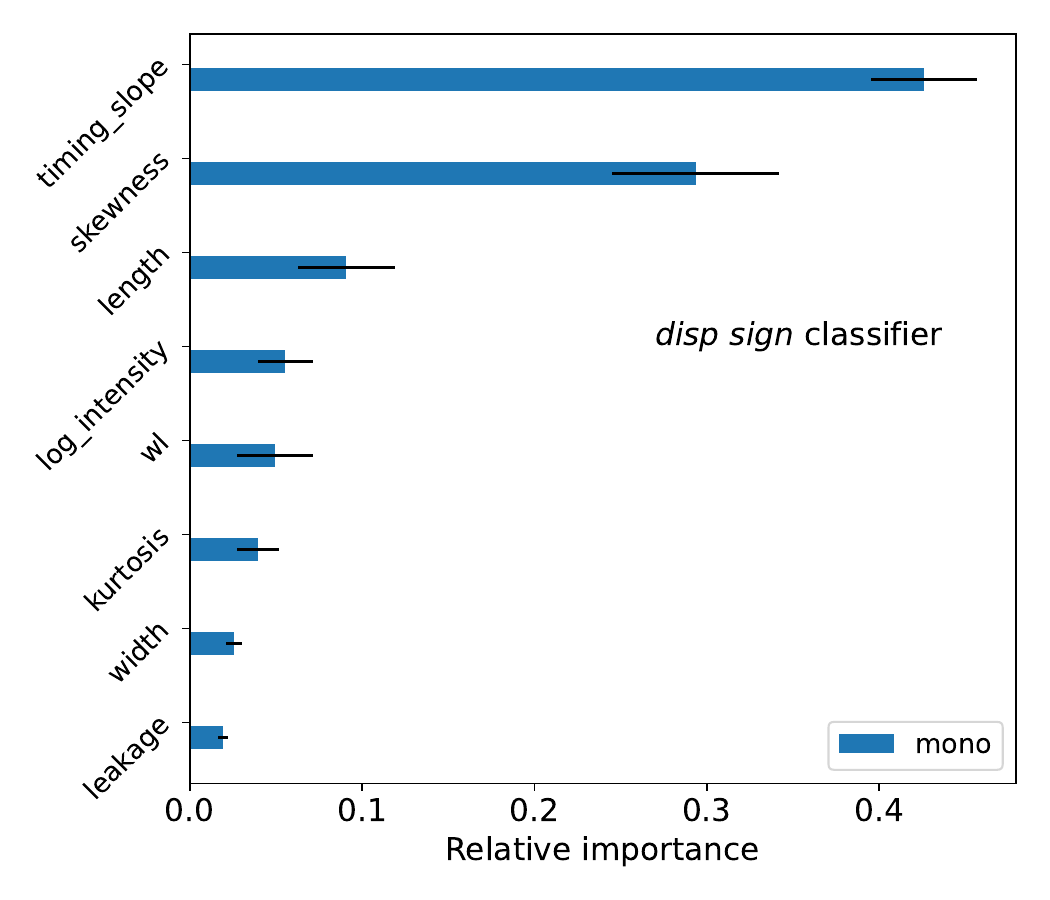} \\
\end{tabular}
\caption{Gini importance for different RFs for the  SST-1M-1. The features used for the reconstruction: log of the intensity, width, length, width/length ratio (wl), timing slope, skewness, kurtosis, leakage, and coordinates of the shower center of gravity in the FoV (x,y). Height of the shower maximum (h max) and distance of the impact point from the telescope (impact dist) are used only in the stereo reconstruction. \textit{Upper left:} Gamma-hadron classifier. \textit{Upper right:} Energy regressor. \textit{Bottom left:} \textit{disp norm} regressor. \textit{Bottom right:} disp sign classifier (Only used in mono reconstruction).}
\label{fig.feature_importance}
\end{figure*}

The training parameters and their relative importance\footnote{Defined as the total reduction of Gini impurity due to each training parameter over all trees in the RF.} for each model and both mono and stereo reconstructions are shown in Figure~\ref{fig.feature_importance}. All RFs, including those for stereo reconstruction, are trained per-telescope, but we only show feature importance for SST-1M-1, as the difference between the telescopes is very small. The most important of the RF mono and stereo parameters of the gamma-hadron classifiers is the Hillas' ellipse width, which is related to the higher lateral momentum of secondary particles in the hadronic showers. The energy reconstruction is surprisingly dominated by the length of the ellipse, which is due to its correlation with its intensity and impact parameter. We tested the importance of the parameters when the length is not included between the RF parameters, so that the impact and intensity became the most relevant parameters, while the performance degraded by a few percent. We also note that the training sample is dominated by low-energy showers close to the energy threshold with relatively small images, where the correlation of Hillas intensity and Hillas length is even more profound. Interestingly, the stereoscopic parameters ($\mathrm{h\_max}$ and $\mathrm{impact\_dist}$, see Fig.~\ref{fig.feature_importance}) are not taking over the importance for stereo reconstruction, which may suggest that there is room for further performance improvement in stereo. In the case of direction reconstruction, quite expectedly, the Hillas length and time gradient play an important role in mono reconstruction, while in the stereo reconstruction, the impact point distance to the telescope is the most relevant parameter.

\begin{figure}[!t]
\includegraphics[width=.49\textwidth]{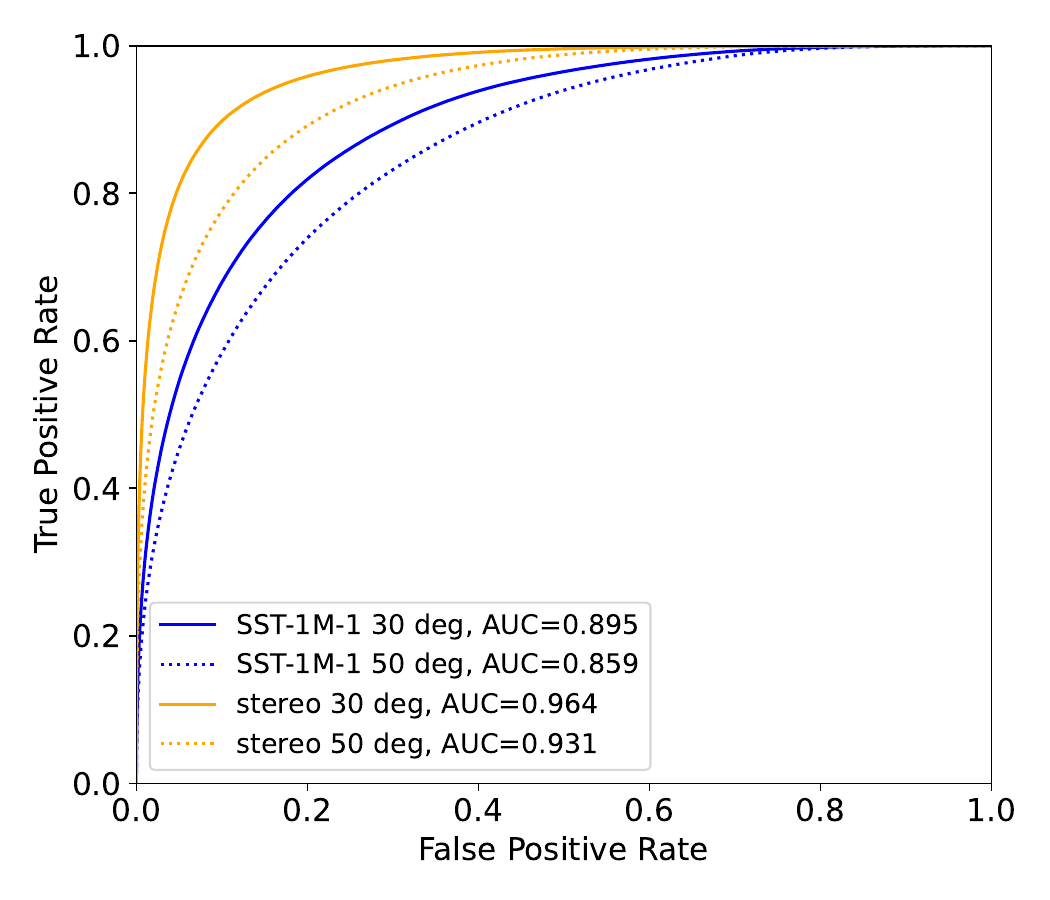}
\caption{ROC curve of the RF gamma-hadron classifier for $30^\circ$ (solid line) and $50^\circ$ (dotted line) zenith angles. Line colors represent different regimes of observations.}
\label{fig.roc}
\end{figure}

The performance of RFs for hadronic background suppression can be assessed with the receiver operating characteristic (ROC) curve. For each gammaness cut, a position along the ROC curve represents the fraction of true gammas and false gammas (i.e., protons mistakenly reconstructed as gammas) left in the sample. Maximizing the former and minimizing the latter, the performance of the gamma-hadron classifier can be expressed as the integral of the ROC curve -- area under the curve (AUC). Figure~\ref{fig.roc} shows the ROC and AUC for both mono and stereo observation regimes at two different zenith angles, with $30^\circ$ being the typical zenith angle of the Crab Nebula observation from Ond\v{r}ejov, and $50^\circ$ representing the high zenith angle performance for comparison. Including stereo parameters in the training features and a combination of gammaness from both telescopes improves the performance of the classifier as expected. It can be noted that the degradation of the performance with increasing zenith angle is due to the increasing telescope(s)-shower distance, leading to smaller shower images and lower precision of the reconstructed shower geometry.

\section{Mono and stereo performance at $510\,\rm{m\,a.s.l.}$} \label{sec:performance}

The RFs trained on the training MC dataset were used to reconstruct the direction, the energy, and to classify the particle of the point-like and diffuse gammas and diffuse protons at all simulated zenith angles. The resulting reconstructed events were then used to calculate the IRFs (effective area, PSF, energy migration matrix, and background model), applying the same quality and selection cuts as in the real data analysis. The IRFs were computed from the MC testing sample using the \texttt{pyirf} Python package \citep{2023arXiv230916488D, maximilian_linhoff_2024_11190775}. A global cut on the image "intensity" of $> 45\, \mathrm{p.e.}$ was applied to bring the data and MC on the same intensity threshold (see Section~\ref{sec:crab}). A cut on the "leakage2"\footnote{leakage2 represents a fraction of shower pixels in the two outermost rings of the camera pixels.} $< 0.7$ was also introduced to remove significantly incomplete shower images from the sample. 
The gamma-hadron separation performance based on parametrized shower images depends on the energy of the primary gamma ray. Low-energy gamma ray and proton showers produce less Cherenkov light and so smaller images, so that the gamma-hadron discrimination capability is challenged by the limited resolution of the optical system and by the pixel size of the camera.
Similarly to the approach in \citet{2023ApJ...956...80A},  we introduced an energy-dependent gammaness cut to keep the same fraction (i.e., $60\%$) of gammas in the final sample in all energy bins.

\subsection{Energy threshold and effective area} 

\begin{figure*}[!t]
\centering
\begin{tabular}{cc}
\includegraphics[width=.49\textwidth]{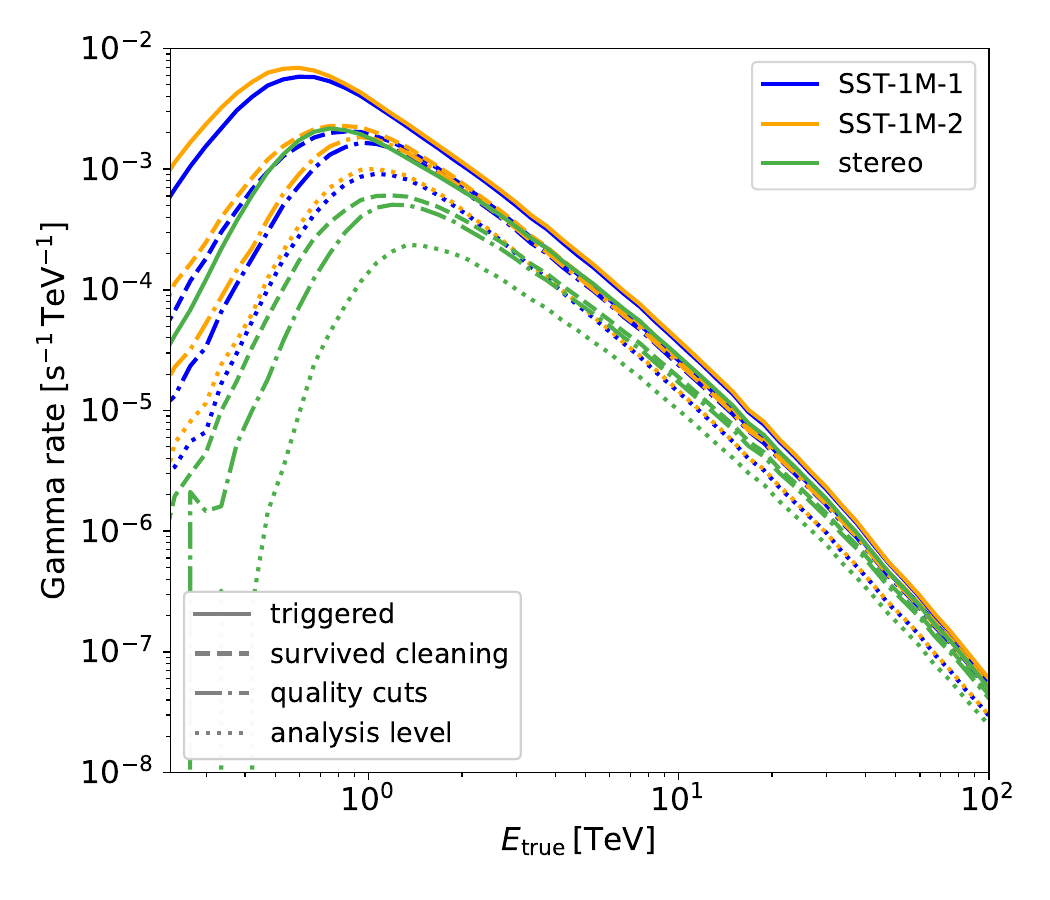} & \includegraphics[width=.49\textwidth]{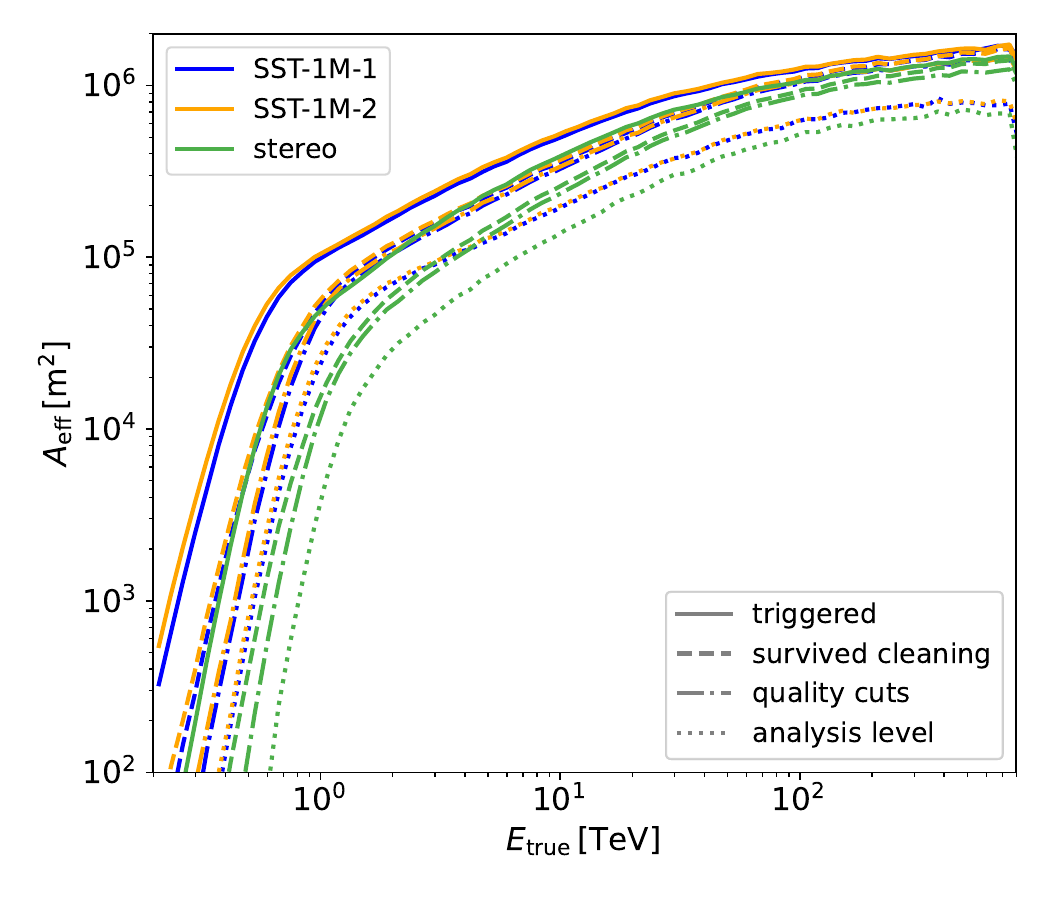}
\end{tabular}
\caption{\textit{Left:} Differential rate of point-like gamma rays with Crab Nebula spectrum in mono and stereo for $20^\circ$ zenith angle. Event rates for different stages of the analysis are shown: All triggered events (solid line), all events that survived cleaning (dashed line), events that survived quality cuts (dash-dotted line), and events that survived $\theta^2$ and gammaness cuts used in the analysis (dotted line). \textit{Right:} Effective area for $20^\circ$ zenith angle.}
\label{fig.effective_area}
\end{figure*}

In IACT, the energy threshold ($E_\mathrm{T}$) is the true energy of primary gamma rays, for which the differential event rate of gamma rays with a given spectral energy distribution reaches its maximum. The left part of Figure~\ref{fig.effective_area} shows the rate of point-like gamma rays at $20^\circ$ zenith angle, weighted on the Crab Nebula spectrum \citep{ALEKSIC201676} for different stages of the analysis. As a consequence of a low altitude of the Ond\v{r}ejov observatory, compared to the other IACT sites, the trigger threshold of SST-1M is relatively high. This is because the amount of emitted Cherenkov light depends on the energy of the primary gamma ray, while the amount of collected light also depends on its attenuation between the emission point and the telescopes. Thus, with decreasing altitude of the telescopes, the distance to the shower maximum at typically 10 km a.s.l. increases, and the trigger efficiency drops. At the trigger level, for mono observation $E_\mathrm{T}\approx0.6\, \mathrm{TeV}$, and it raises to $E_\mathrm{T}\approx1.0\, \mathrm{TeV}$ at the analysis level, where the energy-dependent cut on gammaness and global cut on $\theta$ are applied. Also, $\theta$ represents the angular distance between the true and reconstructed gamma-ray directions. The low-energy events are naturally more affected by the quality cuts on image intensity. The energy threshold is higher in the stereo regime, namely $E_\mathrm{T}\approx0.7\, \mathrm{TeV}$ and $E_\mathrm{T}\approx1.3\, \mathrm{TeV}$ at the trigger and analysis level, respectively, as the low-energy showers emitting fewer Cherenkov photons have a lower probability of triggering both telescopes for significantly different impact parameters. The total trigger rate is also lower in stereo because the probability that the same shower triggers both telescopes is lower than just one telescope being triggered. The energy threshold also strongly depends on the zenith angle of observation as the distance between the shower and the telescopes increases with the zenith angle. For $60^\circ$ zenith, the analysis-level energy threshold is 6~TeV and 10~TeV for mono and stereo, respectively.

The effective area is the ratio of reconstructed point-like gamma rays over the number of simulated ones, multiplied by the area on which the simulated showers have been thrown in the frame orthogonal to the telescope pointing. It is shown in bins of true energy in the right panel of Figure~\ref{fig.effective_area}.

\subsection{Energy and angular resolution} 

\begin{figure*}[!t]
\centering
\begin{tabular}{cc}
\includegraphics[width=.49\textwidth]{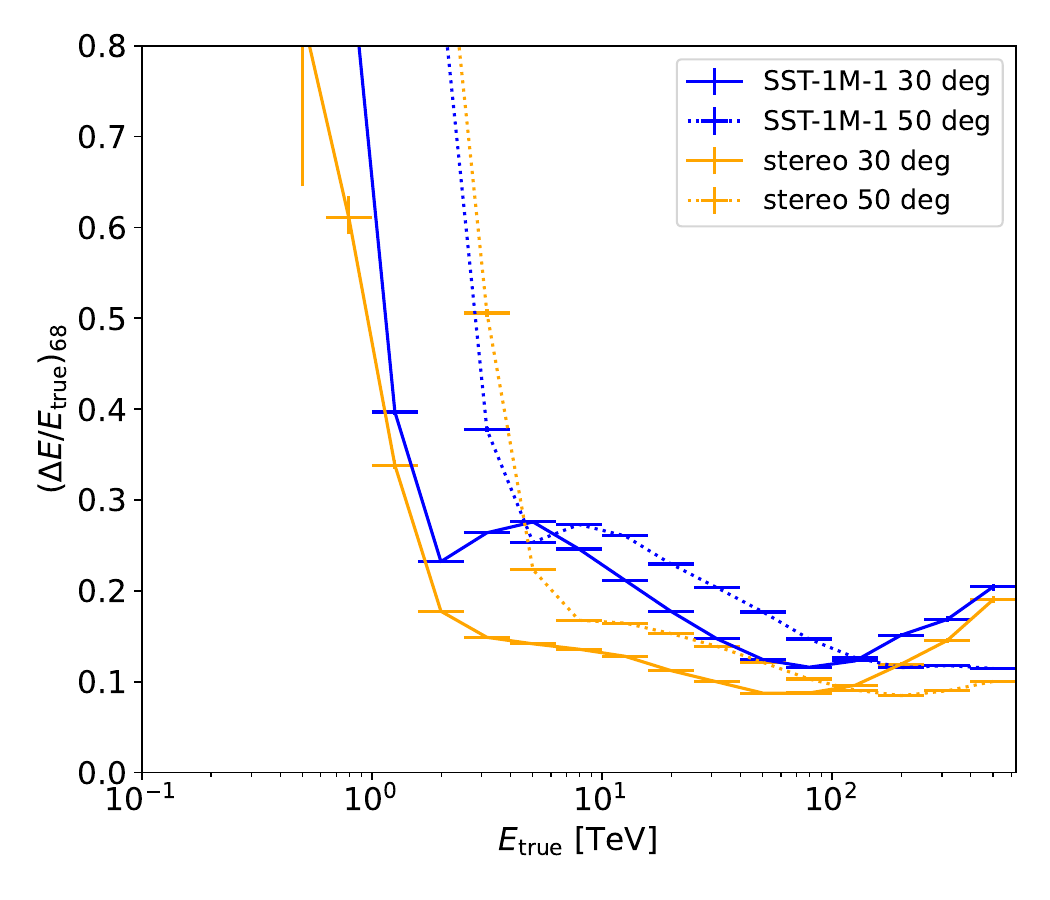} & \includegraphics[width=.49\textwidth]{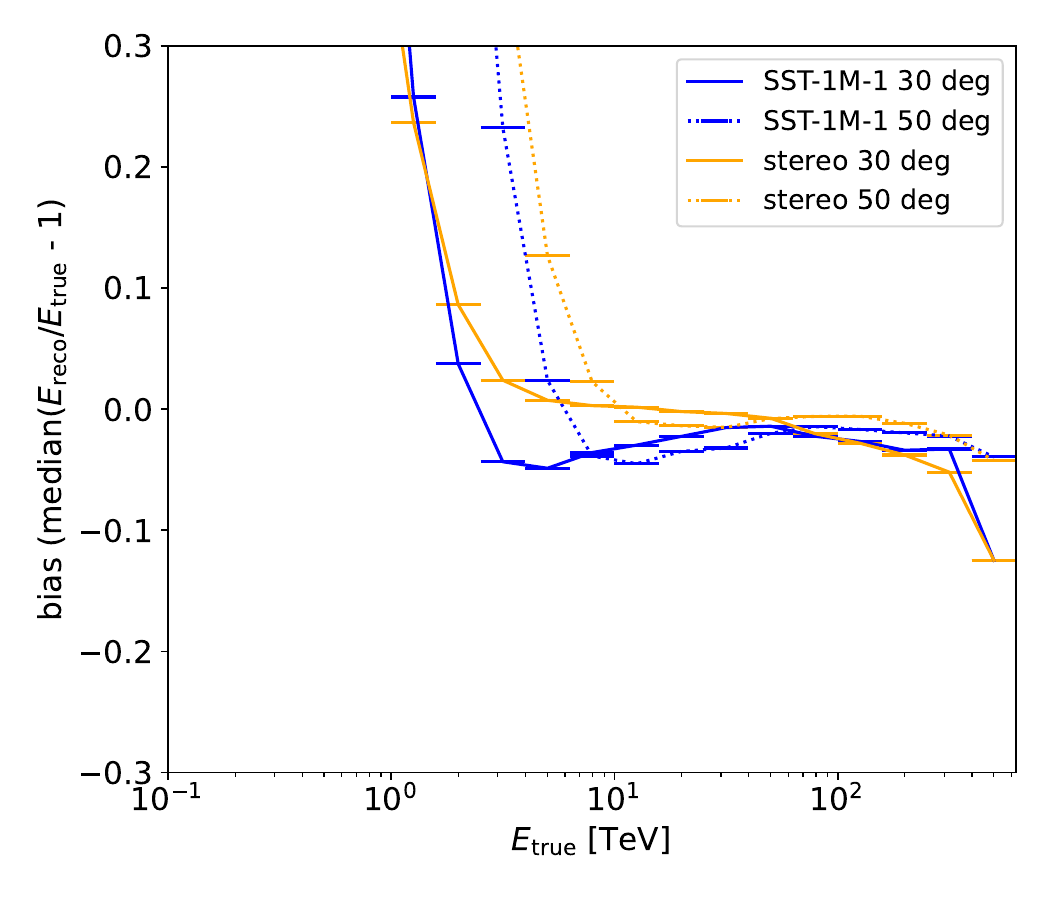}
\end{tabular}
\caption{Energy resolution (\textit{left}) and energy bias (\textit{right}) for $30^\circ$ (solid line) and $50^\circ$ (dotted line) zenith angles evaluated using on-axis point-like gammas for testing. Line colors represent different regimes of observations. For the sake of clarity, only results for SST-1M-1 are shown in mono as both telescopes are very similar.}
\label{fig.energy_resolution}
\end{figure*}

The energy resolution is defined as $68\%$ containment around the median value of $\Delta E / E_\mathrm{True} = (E_\mathrm{Reco}-E_\mathrm{True}) / E_\mathrm{True}$ for reconstructed gammas, while the median itself is the so-called energy bias. Both are shown in Figure~\ref{fig.energy_resolution} as a function of $E_\mathrm{True}$ for mono and stereo at two different zenith angles. As expected, the energy resolution in Figure~\ref{fig.energy_resolution} improves with energy, reaching $\Delta E / E_\mathrm{True, mono} \approx15\%$ and $\Delta E / E_\mathrm{True, stereo} \approx10\%$. At very high energies, the event reconstruction starts to be limited by the containment of the shower images within the camera FoV, degrading the performance. High zenith angle observations, increasing the average shower-to-telescope distance, partially compensate for this effect at the expense of a higher energy threshold. Stereo reconstruction improves the energy resolution significantly at energies below $10\,\rm{TeV}$, where the impact parameter is not well reconstructed in the mono analysis.

\begin{figure}[!t]
\centering
\includegraphics[width=.49\textwidth]{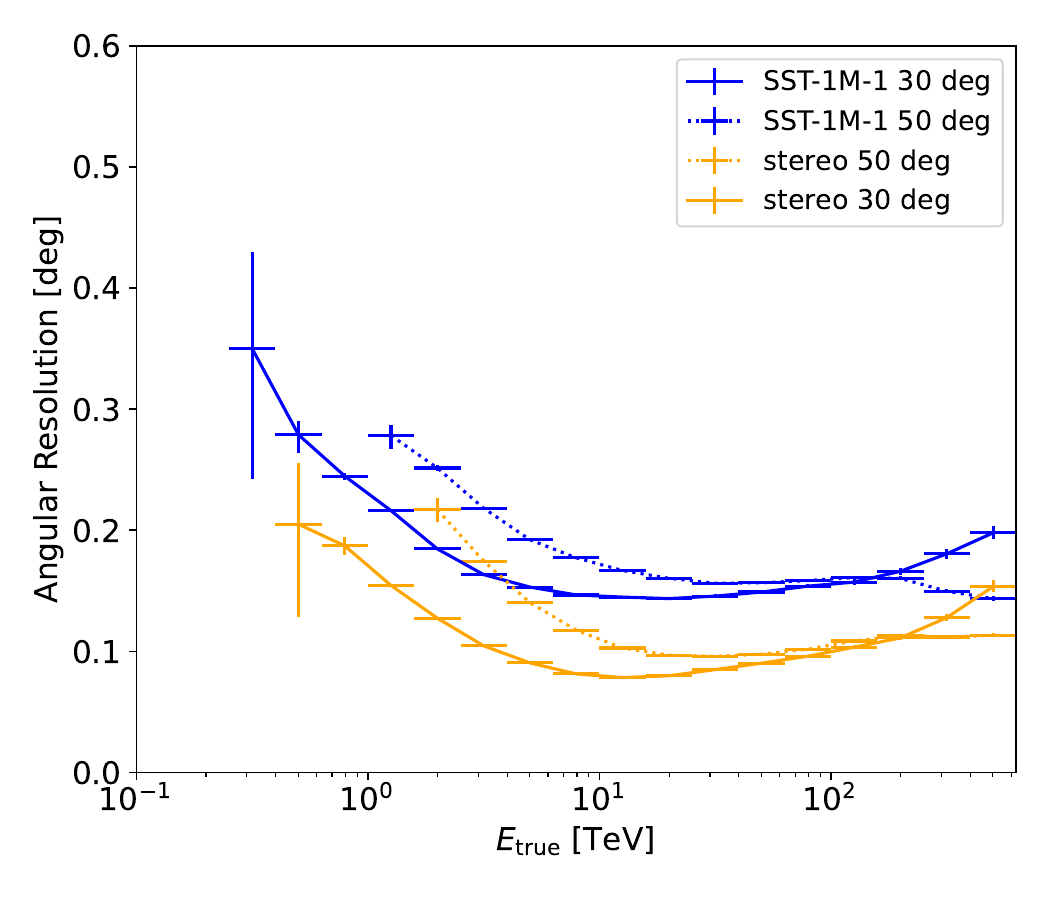}
\caption{Angular resolution for $30^\circ$ (solid line) and $50^\circ$ (dotted line) zenith angles evaluated on testing on-axis point-like gammas. Line colors represent different regimes of observations. For the sake of clarity, only results for SST-1M-1 are shown in mono, as both telescopes are very similar.}
\label{fig.angular_res}
\end{figure}

The angular resolution is defined as the $68\%$ quantile of the distribution of $\theta$. As pointed out by \citet{2023ApJ...956...80A}, in mono reconstruction, the PSF consists of two components: the central component is made of events with properly reconstructed shower orientation (the disp sign parameter). It is surrounded by a ring of events with a wrongly reconstructed disp sign, which becomes the dominant fraction of the PSF near the energy threshold. This feature is not present in stereo by the nature of the stereoscopic reconstruction. The radius of the ring is about $1.5^\circ$, which means that these events do not affect the capability of the instrument to resolve two point-like sources and thus we consider only the sample of properly reconstructed gamma rays to characterize the angular resolution in Figure \ref{fig.angular_res}. We note that the IRFs used for skymaps and spectral analysis contain the full PSF. The angular resolution in mono is below $\approx0.20^\circ$ above the energy threshold for both telescopes. In stereo reconstruction, the angular resolution reaches  $\approx0.07^\circ$. Figure \ref{fig.angular_res} shows the angular resolution for different zenith angles. The effect of the increasing energy threshold with the zenith angle is clearly visible. At very high energies, on the other hand, the increasing zenith and thus increasing telescope-to-shower distance helps to lower the fraction of events not fully contained in the camera, which leads to better angular resolution.

\subsection{Flux sensitivity}\label{sec:sensitivity}

The flux sensitivity is defined as the minimum flux of a source that can be detected in a given time, usually 50 hours in the IACT community. We followed the usual definition of the differential sensitivity \citep[e.g.,][]{2023ApJ...956...80A, 2023A&A...680A..66A}, where the detection with $5\sigma$ statistical significance \citep{1983ApJ...272..317L} should be achieved in individual energy bins dividing each energy decade into five equal logarithmic intervals, assuming the ratio between the signal and background region sizes, namely, the alpha parameter \citep{1983ApJ...272..317L}, of 0.2. We also required at least ten excess events surviving cuts in each energy bin and the signal to background ratio of at least $5\%$. The former condition affects mostly the high-energy sensitivity bins, where the number of detected gamma rays is typically low due to the naturally decreasing spectrum of all gamma-ray sources, while the latter is the most important at low energies due to low gamma-hadron separation performance, leading to a  low signal-to-background ratio.

\begin{figure}[!t]
\centering
\includegraphics[width=.49\textwidth]{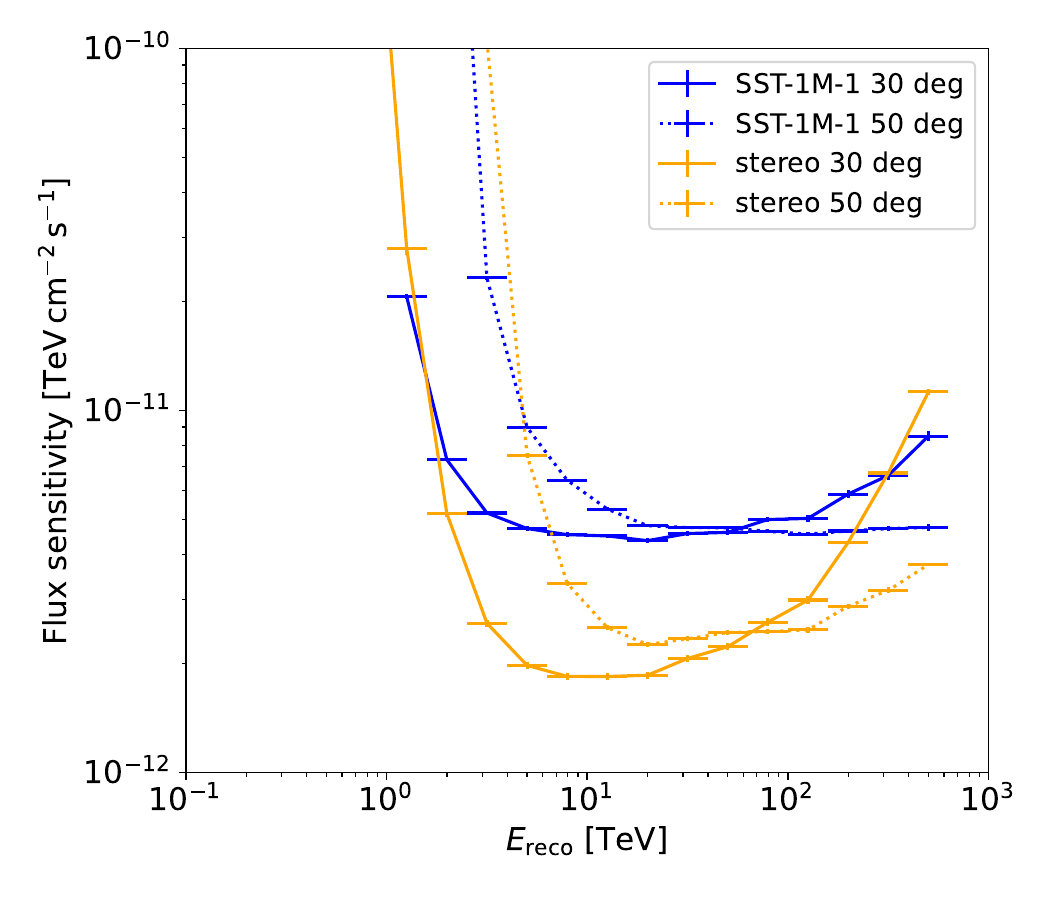}
\caption{Differential sensitivity at $5\sigma$ confidence level for $30^\circ$ (solid line) and $50^\circ$ (dotted line) zenith angles evaluated on testing on-axis point-like gammas and diffuse protons. Line colors represent different regimes of observations. For the sake of clarity, only results for SST-1M-1 are shown in mono, as both telescopes are very similar. We note that in contrast to stereo, the additional condition requiring at least ten excess events does not affect the last energy bins for the $50^\circ$ zenith angle in mono, as the number of excess events is satisfied by the $5\sigma$ condition only (see Section~\ref{sec:sensitivity} for more details).}
\label{fig.diff_sensitivity}
\end{figure}

The differential sensitivity of SST-1M mono/stereo for different zenith angles shown in Figure~\ref{fig.diff_sensitivity} was evaluated for on-axis point-like gammas re-weighted using the Crab Nebula spectrum \citep{ALEKSIC201676}, and diffuse protons were re-weighted on protons + Helium spectrum \citep{2024PhRvD.109l1101A}. To select the region of interest, we applied a global $\theta$ cut, keeping $\approx 80\%$ of reconstructed point-like gamma-rays. Both cuts on gamma efficiency\footnote{A percentage of the MC gammas surviving the cut.} and $\theta$ were optimized on MC, to reach the best detection significance for a source with the Crab Nebula spectrum. The stereo observation improves the sensitivity by a factor of $\approx2$. It should be noted, however, that the sensitivity at low energies reflects the current triggering scheme of the telescopes, and can be further improved using a shorter coincidence window of the acceptance of the two mono-triggers. This would further suppress NSB and enable  lower energy events to be included than is currently possible. As expected, the energy threshold increases at high zenith angles as the layer of the atmosphere is larger, requiring higher energy events to generate detectable showers. On the other hand, at energies above $\approx 50\,\mathrm{TeV}$, the sensitivity improves with the zenith angle due to the effect of the Cherenkov cone geometry, leading to an increased effective area.

\subsection{Off-axis performance}\label{sec.offaxis_performance}

Observations of the SST-1M telescopes are taken in the wobble mode \citep{1994APh.....2..137F}, where the telescope pointing has a small offset to the true position of the source (see Section~\ref{sec:crab}). However, in the case of observation of extended sources, the offset may be larger to avoid contamination of the signal region with the background. Moreover, in the case of follow-up of a poorly localized alert (e.g., gamma-ray bursts, gravitational waves, or neutrino events), the source may end up located anywhere in the telescope FoV. We therefore study the performance of the SST-1M mono and stereo for different source angular distances from the camera center, using MC simulation of point-like gammas and diffuse protons at $20^\circ$ zenith angle. 

\begin{figure}[!t]
\centering
\includegraphics[width=.49\textwidth]{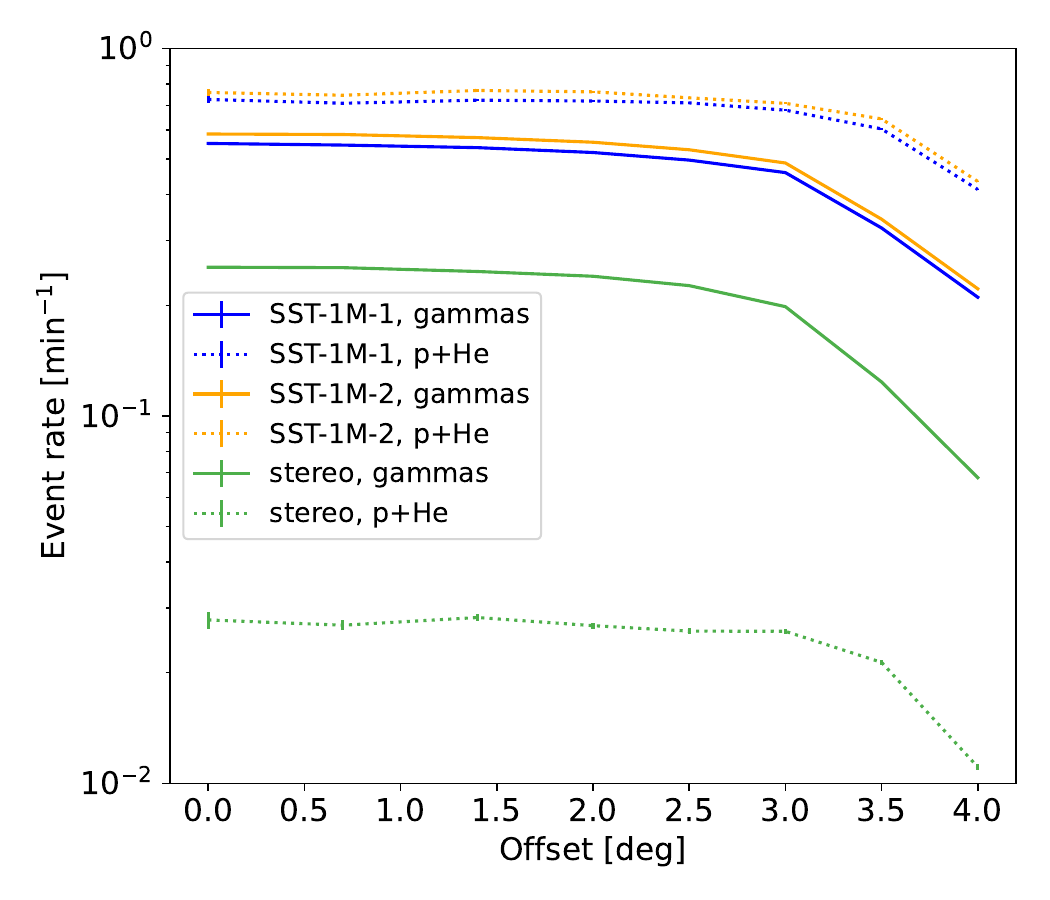}
\caption{Simulated gamma-ray rate from the Crab Nebula and simulated protons weighted by the CR spectrum in the signal region after analysis cuts as a function of the offset at $20^\circ$ zenith angle. Only events with energies above the $E_\mathrm{T}$ are taken into account.}
\label{fig.offaxis_rate}
\end{figure}

Figure~\ref{fig.offaxis_rate} shows the integral rate of simulated gammas and protons above the energy threshold (re-weighted on the Crab Nebula, and CR spectrum, respectively) in the signal region after the analysis cuts for both telescopes. We note that in this analysis, we apply an energy-dependent gamma efficiency cut of $60\%$, optimized for each offset. Due to the large camera optical FoV ($9^\circ$) and having all pixels part of the trigger, the acceptance is almost flat up to $~2.5^\circ$, where it drops by $\approx10\%$, which makes the SST-1M an ideal instrument for observation of extended gamma-ray sources.

\begin{figure*}[!t]
\centering
\begin{tabular}{cc}
\includegraphics[width=.49\textwidth]{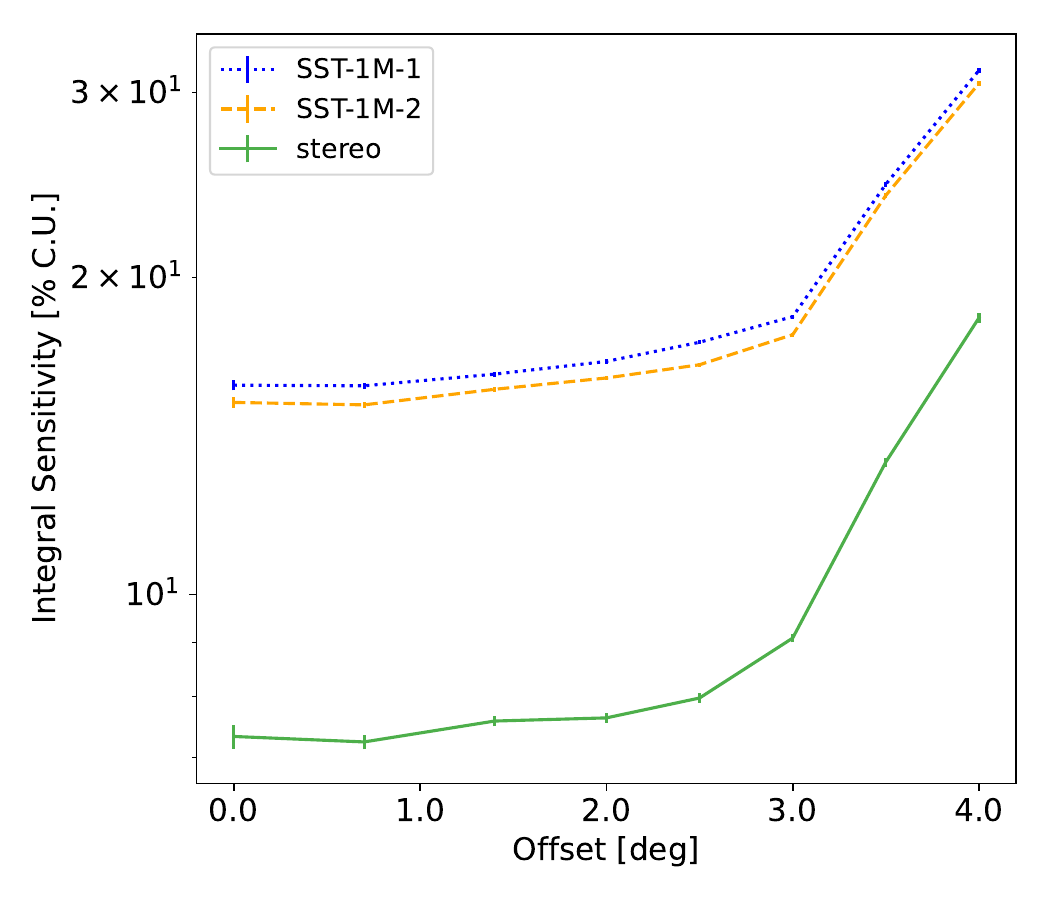} & \includegraphics[width=.49\textwidth]{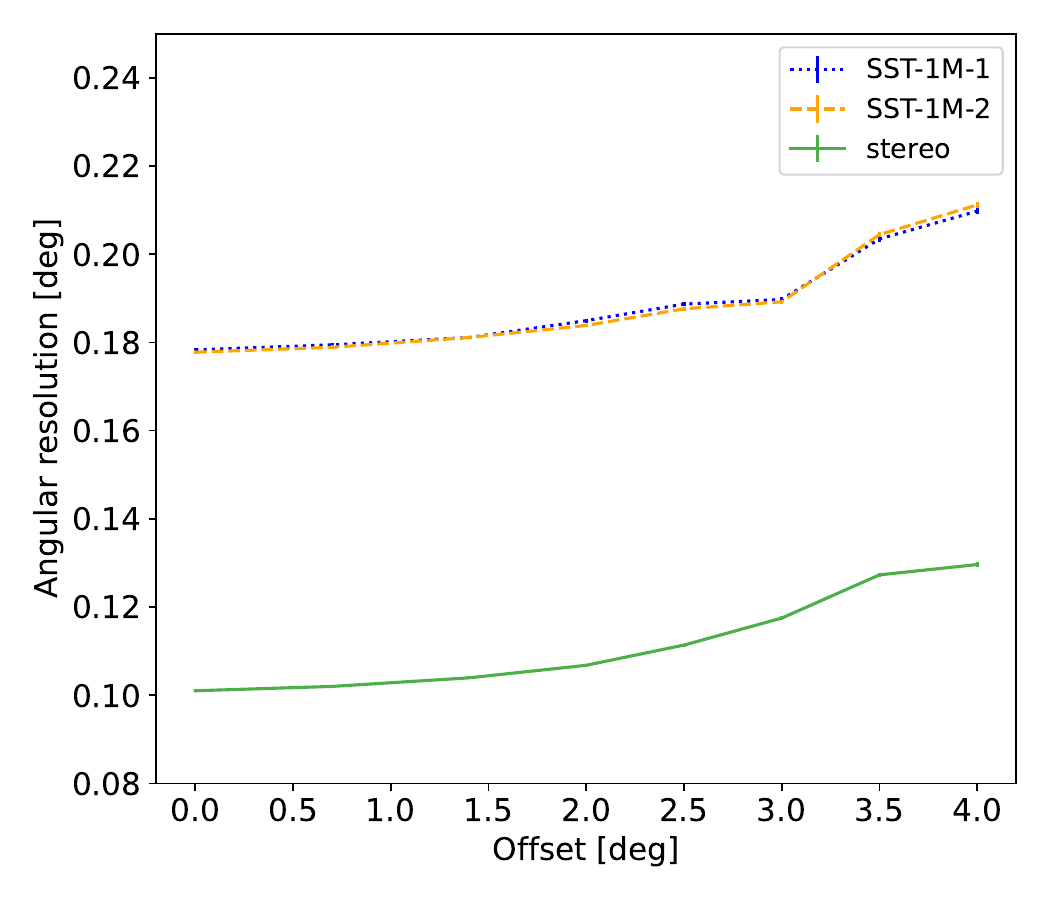}
\end{tabular}
\caption{Integral sensitivity for 50 hours of observation (\textit{left}), and angular resolution for a source with the Crab Nebula spectrum (\textit{right}) as a function of the offset at $20^\circ$ zenith angle. All events with energies above the $E_\mathrm{T}$ are integrated.}
\label{fig.offaxis_performance}
\end{figure*}

The integral sensitivity shown in Figure~\ref{fig.offaxis_performance} is given in a fraction of the Crab Nebula flux (Crab units, C.U.). It shows a faster degradation than the acceptance above $\sim3^\circ$, as the background acceptance drops slower than the acceptance for point-like gammas. The angular resolution integrated over all energies above the energy threshold (Fig.~\ref{fig.offaxis_performance}, right) slowly degrades with increasing offset, reaching about $5\%$ and $11\%$ degradation at $2.5^\circ$, for mono and stereo, respectively.

\section{Crab Nebula observation} \label{sec:crab}

The data sample used in this study was collected between September 2023 and March 2024. The White Rabbit timing network was installed in October 2023, and all data collected before this date are mono only. Due to the maintenance campaign between September 2023 and January 2024, the September 2023 data were all taken with SST-1M-1 only, while a part of the January 2024 data were taken only with SST-1M-2. The observations were performed in wobble mode. It allows for an estimation of the rate of the background events from the regions with the same offset as the signal region, reflected with respect to the center of the FoV. Utilizing the wobble observation strategy, the background can be determined directly from the same dataset as the signal, and no dedicated observations of the background are needed. Being in the commissioning phase, the observation strategy was not yet optimized during the Crab Nebula data acquisition, and thus part of the data sample was taken with $0.7^\circ$ wobble offset, while all data taken after December 2023 were taken with $1.4^\circ$ offset to take advantage of having more background regions and therefore better statistics of the background events. 

Each observation run, in which telescopes were tracking fixed equatorial coordinates $\alpha$ and $\delta$ (one wobble offset), was typically 20 minutes long. In total 46 hours and 52 hours of data were collected with SST-1M-1 and SST-1M-2, respectively. After event matching, we end up with 33 hours of stereo data in total. The zenith angle in the dataset ranges from $27^\circ$ up to $53^\circ$.

\subsection{Run selection}

\begin{figure*}
\centering
\begin{tabular}{ccc}
\includegraphics[width=.32\textwidth]{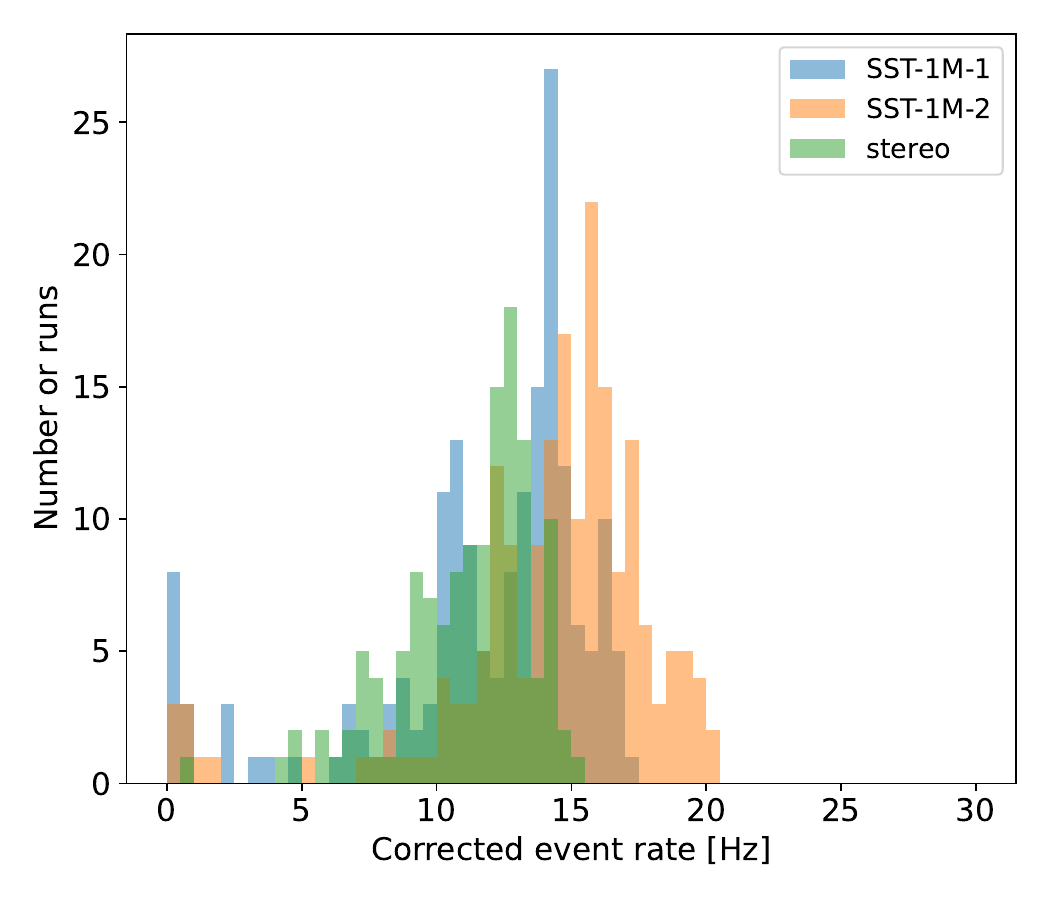} & \includegraphics[width=.32\textwidth]{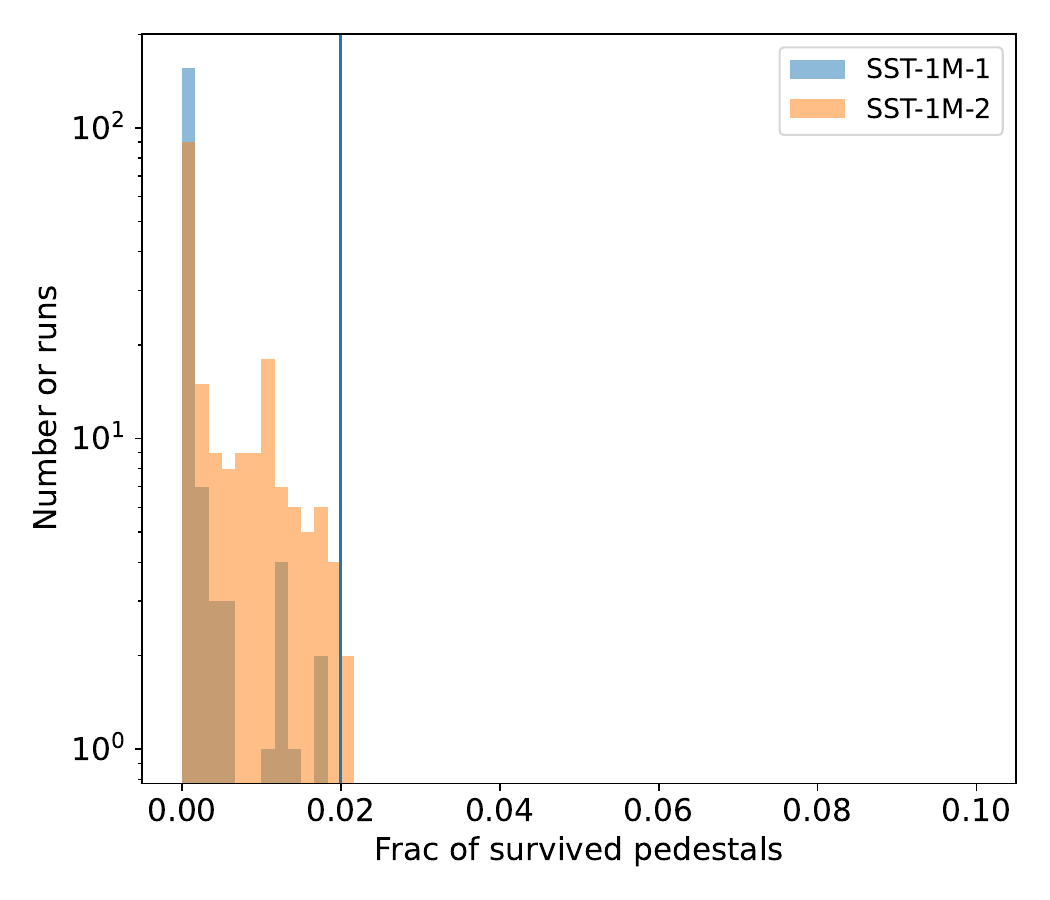} & \includegraphics[width=.32\textwidth]{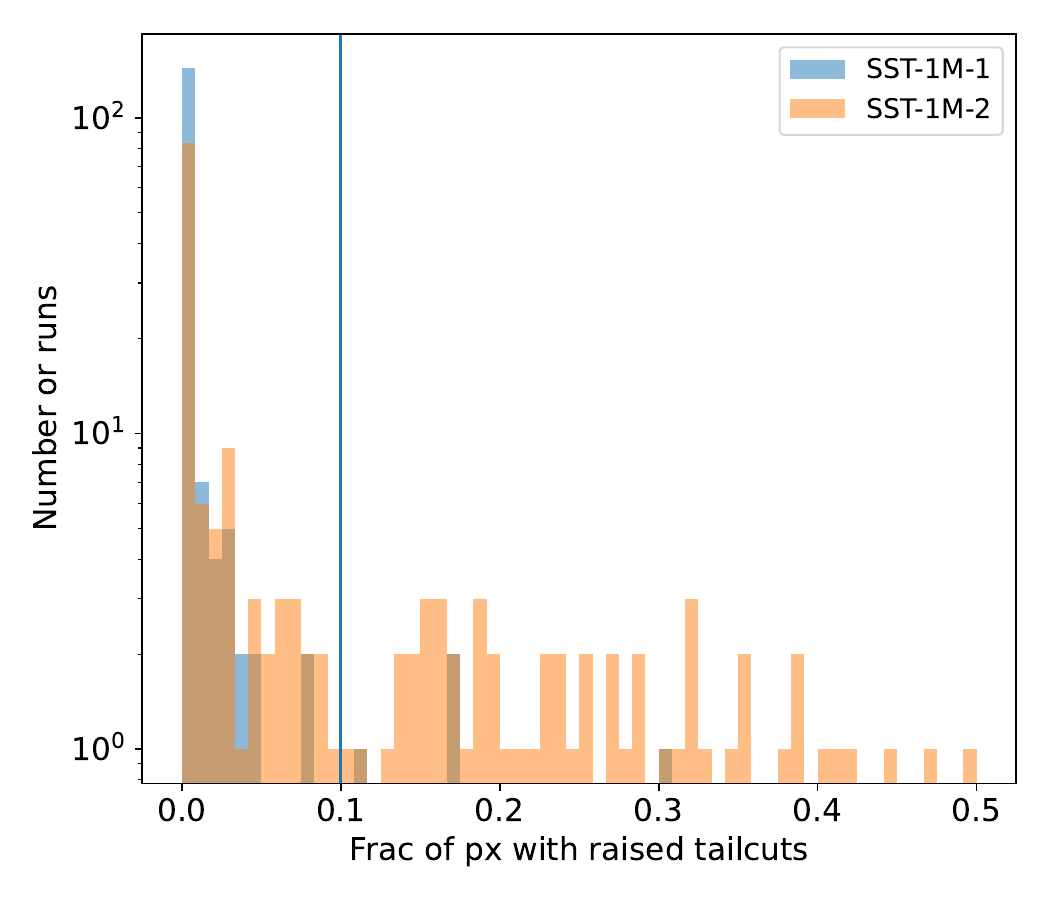}
\end{tabular}
\caption{\textit{Left:} Run-averaged rate of mono events with intensity above $200\,\rm{p.e.}$, corrected for $\cos(\mathrm{zenith\; angle})$. \textit{Center:} Run-averaged fraction of pedestal events that survived the tailcut cleaning. \textit{Right:} Run-averaged fraction of pixels with increased tailcuts. The solid blue line marks the cuts used in the run selection.}
\label{fig.event_rate}
\end{figure*}

At first, we identified a high-quality dataset taken under good atmospheric conditions, also taking into account the highly variable NSB level at Ond\v{r}ejov. As atmospheric transparency modulates the event rate, the main selection criterion is the stability of the average event rate per run (after the cleaning). The event rate is also affected by the trigger threshold, which can be changed during the observation to account for NSB variability \citep{sst1m_hw_paper}. To avoid this affecting the run selection, we require the event rate to be stable for events with intensity above 200 p.e., which is far enough from the intensity threshold to be unaffected by the trigger threshold in low to moderate NSB conditions. The event rate also typically decreases with increasing zenith angle of the source, which may bias the selection. This has been accounted for by correcting the event rate by a factor of $1/\cos(\mathrm{zenith\; angle})$, which yields a good parameterisation of the airmass as a function of the zenith angle over the range of zenith angles in our data sample. Figure~\ref{fig.event_rate} (left panel) shows the airmass-corrected, run-averaged event rate for both telescopes. To ensure stable atmospheric conditions over the entire data set, we require the airmass-corrected event rate to be $> 10 \, \mathrm{Hz}$ in mono and $> 9 \, \mathrm{Hz}$ in stereo to account for a slightly lower rate of stereoscopic events of lower intensities. We also rejected very short runs with a live time $< 200$ s, where the statistics of the cleaned events is insufficient to reliably estimate the averaged event rates.

With increasing NSB level, the capability of cleaning to remove noisy pixels is degraded. This degradation can be observed by monitoring the fraction of pedestal events that survive the cleaning cuts (center of Figure~\ref{fig.event_rate}), for which we require a survival probability lower than 2\% in the final sample. Local NSB fluctuations, for instance, due to bright stars in the FoV, are handled with the adaptive cleaning, which raises the tailcuts locally in the affected pixels. To avoid deviating too much from the MC sample with fixed homogeneous NSB across the full FoV, we require that the average fraction of affected pixels in the entire run to stay below $10\%$. We also checked for a stricter cut on the fraction of affected pixels of $1\%$, resulting in the ratio of the Hillas parameter distributions (for shower images after the quality cuts) staying within $(0.95, 1.05)$.

\begin{figure*}
\centering
\begin{tabular}{ccc}
\includegraphics[width=.32\textwidth]{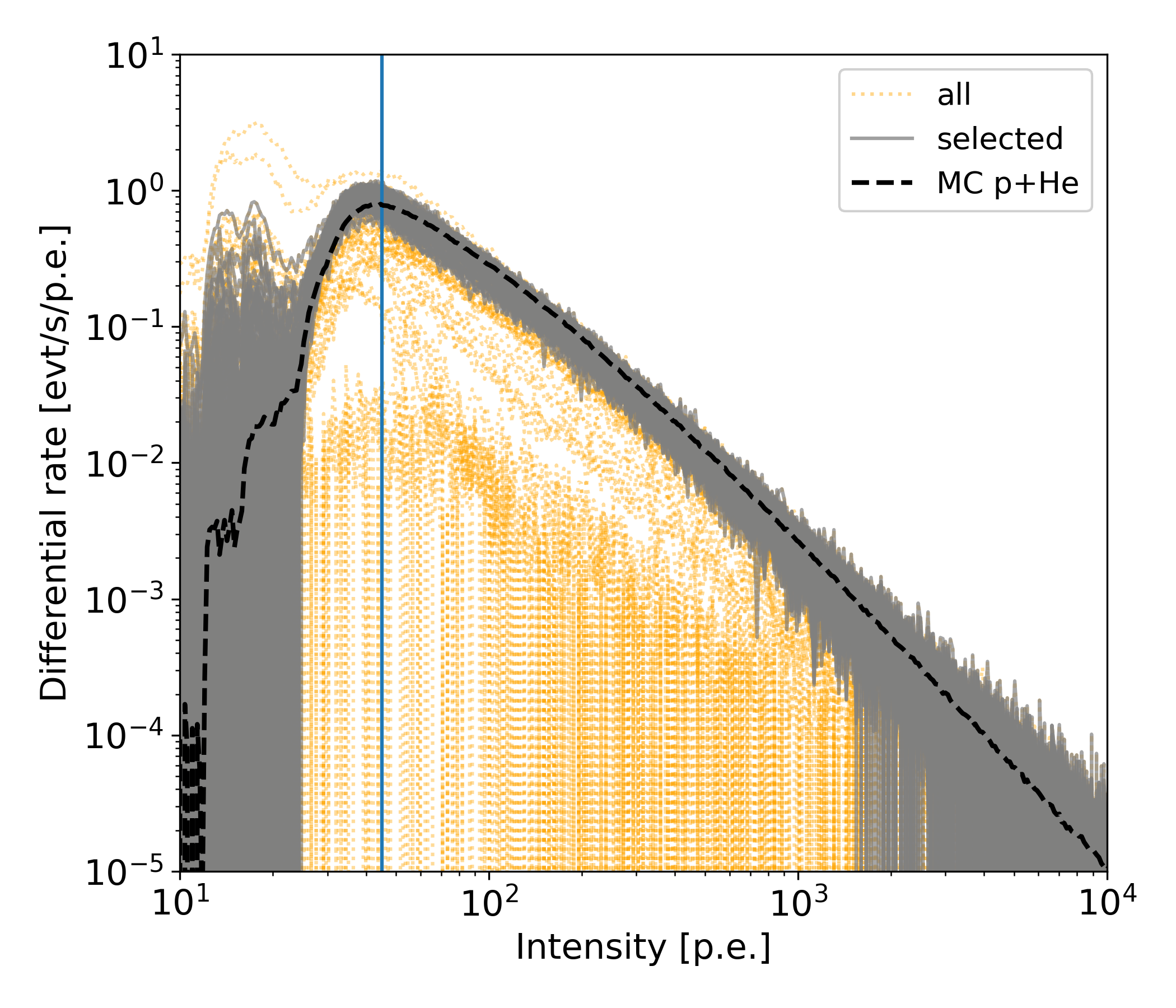} & \includegraphics[width=.32\textwidth]{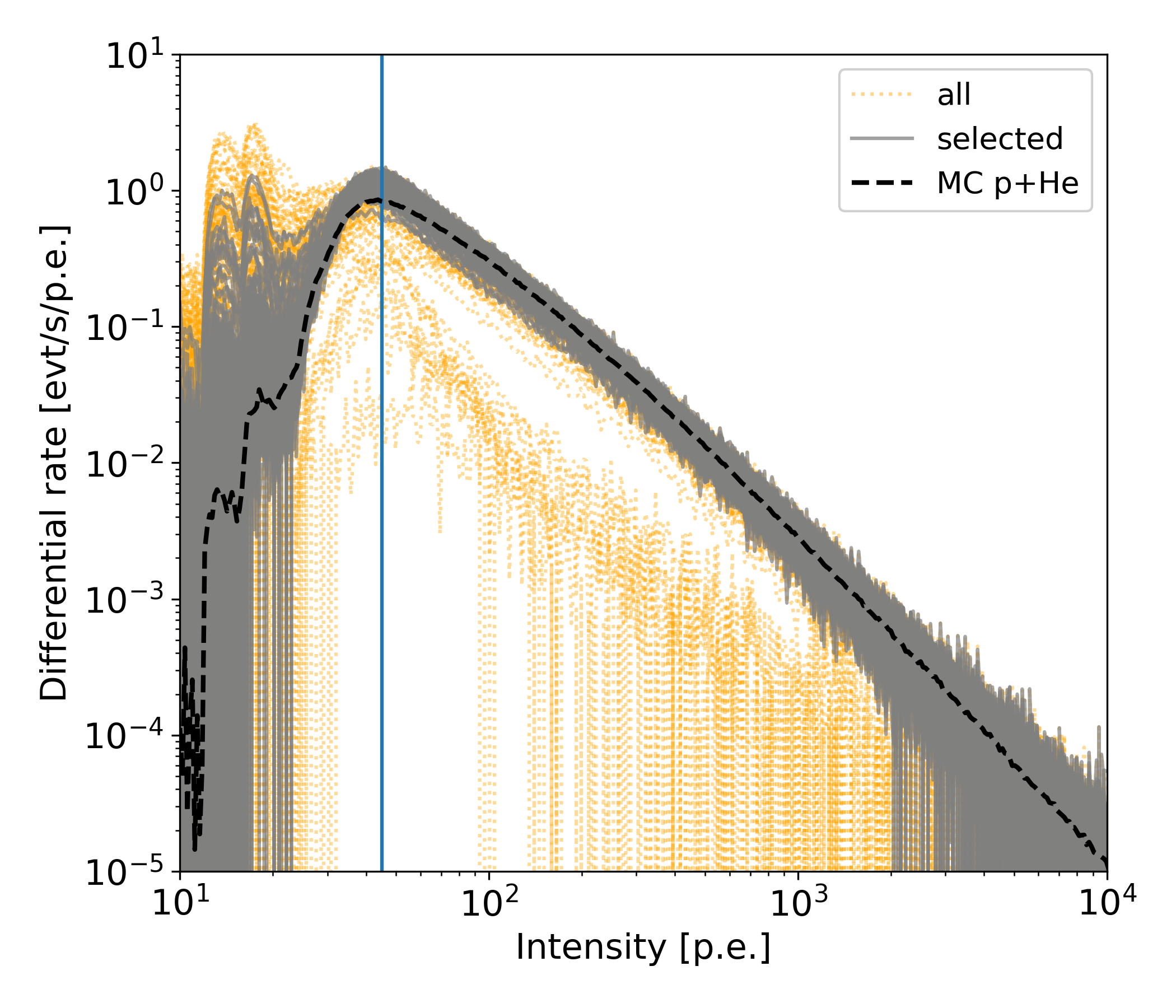} & \includegraphics[width=.32\textwidth]{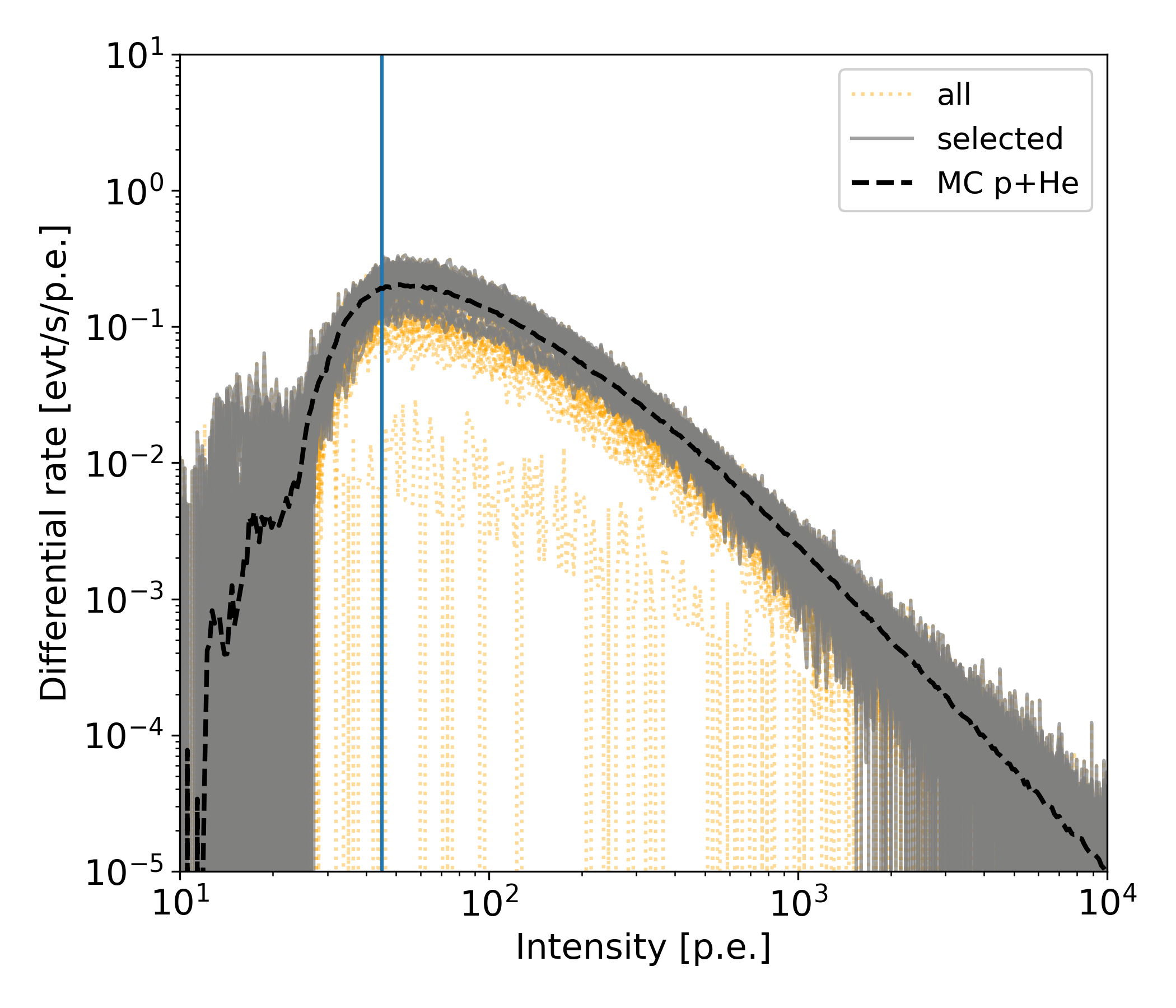}
\end{tabular}
\caption{Per-run rates of the event intensities (i.e., only events which survived cleaning contribute to the distributions) for SST-1M-1 mono (\textit{left}) and SST-1M-2 mono (\textit{center}), and SST-1M-1 stereo (\textit{right}). The distributions shown are corrected for $\cos(\mathrm{zenith})$. Dotted lines mark all runs taken, solid lines mark runs after selection cuts. The dashed line shows the distribution of MC protons re-weighted on the proton + helium spectrum. The solid blue line marks the intensity threshold of $45\,\rm{p.e.}$ introduced in the higher-level analysis.}
\label{fig.diff_rate}
\end{figure*}

Figure~\ref{fig.diff_rate} shows the run-averaged distributions of event intensities before and after the run selection compared with MC simulations. Above the intensity threshold (which we define as the intensity where the differential distribution reaches the maximum), the differential event rates of selected runs agree very well. Below the intensity threshold, the differences due to variable trigger settings and fluctuations of the NSB spoil the MC-data agreement. Therefore, in further analysis, we introduce a common cut on the intensity $> 45\,\rm{p.e.}$, bringing both telescope data and MC to a common analysis threshold.

After the selection cuts, the final sample used for the analysis is 33 hours and 27 hours for SST-1M-1 and SST-1M-2, respectively. The sample of stereo data after the quality cuts is 25 hours. Due to the negligible dead time of the detector \citep{2017EPJC...77...47H}, we consider the final amount of collected hours equal to the effective observation time.

\subsection{Crab Nebula spectrum}\label{sec:crab_sed}

\begin{table*}[!ht]
\centering
\begin{tabular}{ccc}
\hline
& $\phi_0$ & $\Gamma$ \\ 
Telescope & $\times 10^{-13} \,\mathrm{cm^{-2} s^{-1} TeV^{-1}}$ & \\ 
\hline
SST-1M-1 & $1.83\pm0.12_\mathrm{stat}\pm 0.40_\mathrm{syst}$ & $2.70\pm0.08_\mathrm{stat}\pm 0.11_\mathrm{syst}$ \\ 
SST-1M-2 & $2.02\pm0.14_\mathrm{stat}\pm 0.44_\mathrm{syst}$ & $2.68\pm0.09_\mathrm{stat}\pm 0.16_\mathrm{syst}$ \\ 
stereo & $1.76\pm0.12_\mathrm{stat}\pm 0.40_\mathrm{syst}$ & $2.78\pm0.10_\mathrm{stat}\pm 0.08_\mathrm{syst}$ \\ 
\hline
\end{tabular}
\caption{Best-fit parameters for the spectral analysis performed on the SST-1M data using a PL model of the spectrum, together with their statistical and systematic uncertainties (see Section~\ref{sec:systematic}).}
\label{tab:pl_fit}
\end{table*}

\begin{figure*}[!ht]
\centering
\begin{tabular}{cc}
\includegraphics[width=.49\textwidth]{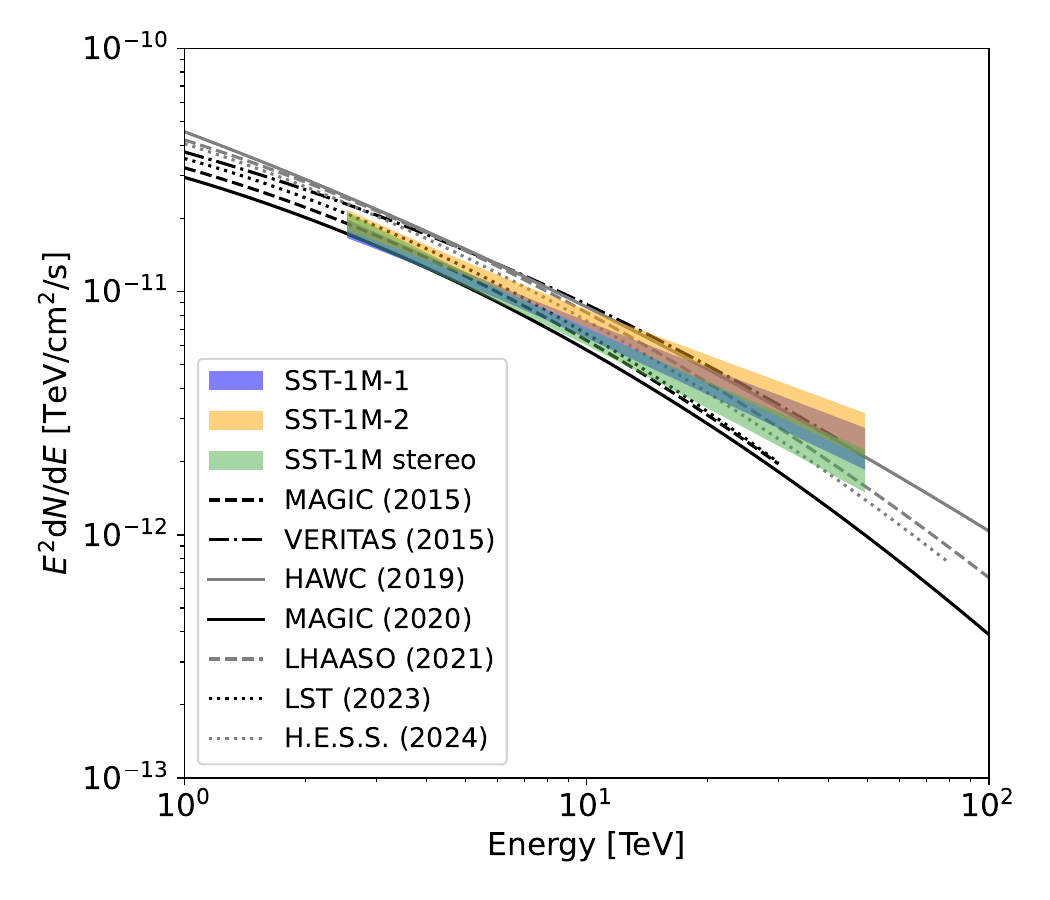} & \includegraphics[width=.49\textwidth]{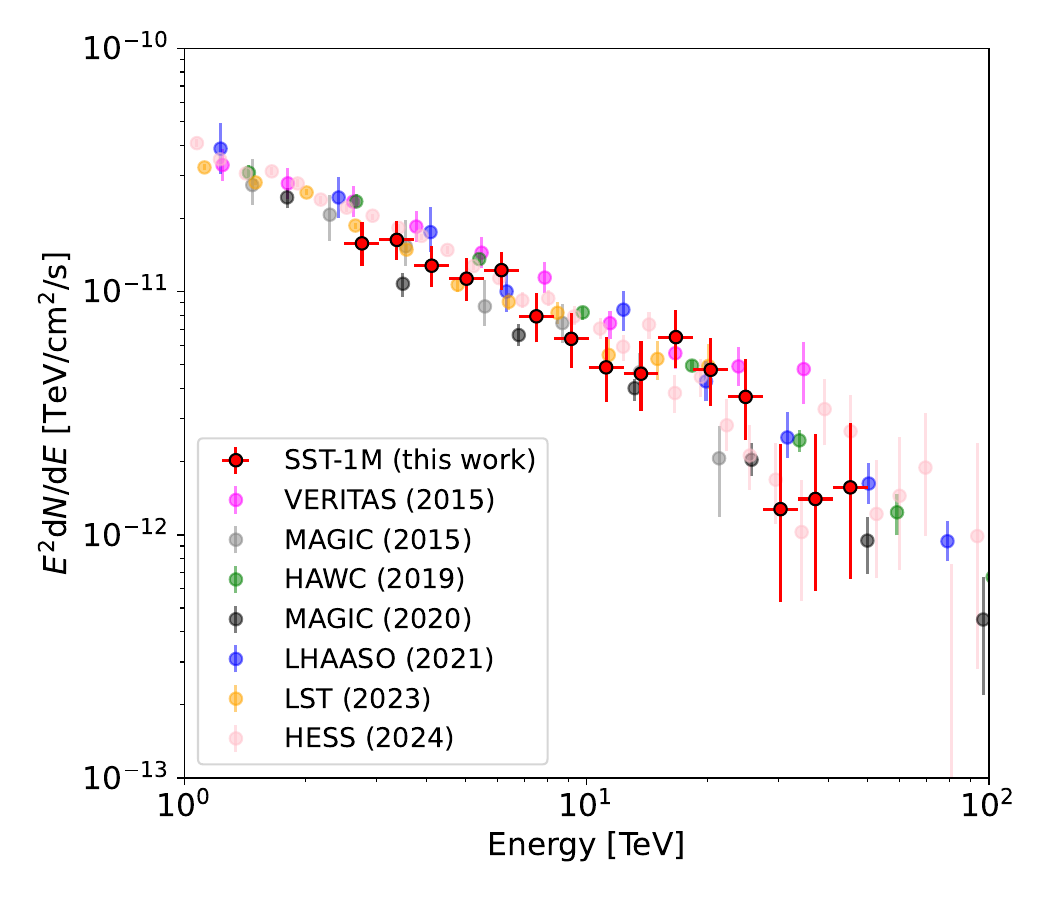} \\
\end{tabular}
\caption{\textit{Left:} SED of the Crab Nebula measured with SST-1M telescopes. Different color bands represent the best-fitting spectra for the SST-1M-1 mono (blue), SST-1M-2 mono (yellow), and SST-1M stereo (green) datasets. \textit{Right:} Stereo flux points derived from the PL spectral model. SEDs from other experiments are shown for comparison in the energy range of their validity (\textit{left}), and corresponding flux points (\textit{right}) - MAGIC \citep{2015JHEAp...5...30A,refId0}, H.E.S.S. \citep{2024A&A...686A.308A}, VERITAS \citep{2015ICRC...34..792M}, LST-1 \citep{2023ApJ...956...80A}, LHAASO \citep{2021Sci...373..425L}, and HAWC \citep{Abeysekara_2019}.}
\label{fig:crab_sed}
\end{figure*}
To assess the capabilities of the SST-1M telescopes to reproduce the spectral energy distribution (SED) of a known source, we performed the spectral analysis on the run-selected Crab data set in \texttt{gammapy v1.0.1} \citep{axel_donath_2021_5721467}. The analysis was performed independently for stereoscopic events, as well as for all mono events detected with both telescopes, to evaluate the consistency of the obtained results. The cuts applied on the dataset were exactly those used for performance estimation (Section~\ref{sec:performance}). The signal region size (cut on $\theta$) was set to the value providing the best significance of the Crab Nebula detection (optimized on MC) - $0.2^\circ$, and $0.12^\circ$ for mono and stereo, respectively, which correspond to about $80\%$ containment of the point-like gamma-ray events. We note that the angular resolution of the instrument is rather stable in the range of zenith angles analyzed (except for the effect of the energy threshold). As the dataset consists of data collected using the wobble method, we adopt the reflected region method for background estimation, where the background is estimated from regions with the same offsets from the center of the FoV as the signal region, assuming the radial symmetry of the acceptance \citep{2007A&A...466.1219B}. As the position of the signal region, we adopted the Crab Nebula coordinates determined by \citet{2020NatAs...4..167H}: $\alpha_{2000}=83.62875^\circ$, $\delta_{2000}=+22.01236^\circ$.

We performed forward-folding likelihood fit in the energy range 2.5-50 TeV, assuming a Power-Law (PL) spectral shape for the differential flux: $d\phi / dE = \phi_0 (E / E_0)^{-\Gamma}$, where the reference energy $E_0 = 7\,\rm{TeV}$ was fixed on the value close to the decorrelation energy. The lower bound of the energy range was selected correspondingly to the stereo energy threshold below $40^\circ$ zenith angle, in which most data were taken. We note that the energy threshold for mono is lower ($\approx 1.5\,\rm{TeV}$), but aiming at comparing the results of both observation modes, we want to avoid any bias, given  by, for instance, a curvature of the intrinsic Crab Nebula spectrum. To justify the use of a single PL spectral model, we tested for a possible curvature in our dataset assuming an alternative Log-parabola (LP) spectral shape in a form of $d\phi / dE = \phi_0 (E / E_0)^{-\alpha - \beta \log{E/E_0}}$, where $\beta$ represents the curvature. As PL and LP are nested models (LP with $\beta=0$ degenerates to PL), we can use likelihood ratio test to determine whether LP is preferred over PL \citep{Wilks:1938dza, 2002ApJ...571..545P}, resulting in $-2\Delta\log\mathcal{L} < 3$ and $\sigma < 2$ for both mono and stereo. We note that on the relatively narrow energy range given by the high energy threshold on one side, and relatively short integration time on the other, the curvature is expected to be relatively small. The best-fit PL model is shown in Figure~\ref{fig:crab_sed}, and the best-fitting parameters are listed in Table~\ref{tab:pl_fit}. The model parameters are consistent for all three datasets within the uncertainties. We also performed a maximum likelihood estimation of the Crab Nebula flux using the \texttt{gammapy} function FluxPointsEstimator in 15 logarithmically spaced energy bins in the entire fitting range. In this procedure, \texttt{gammapy} recalculates the flux normalization in each energy bin, assuming a PL model with spectral index fixed on the best-fitting value for the entire dataset.

\subsection{Crab Nebula skymap}\label{sec:crab_skymap}
\begin{figure*}[!htb]
\minipage{0.32\textwidth}
  \includegraphics[width=\linewidth]{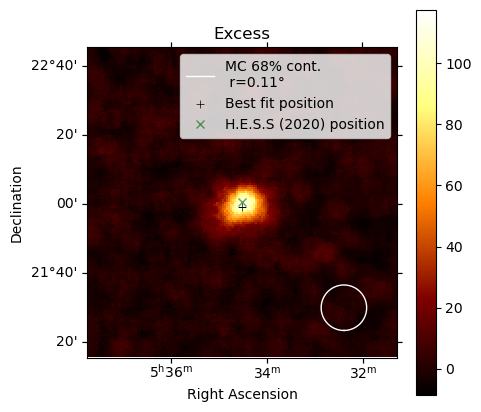}
\endminipage\hfill
\minipage{0.32\textwidth}
  \includegraphics[width=\linewidth]{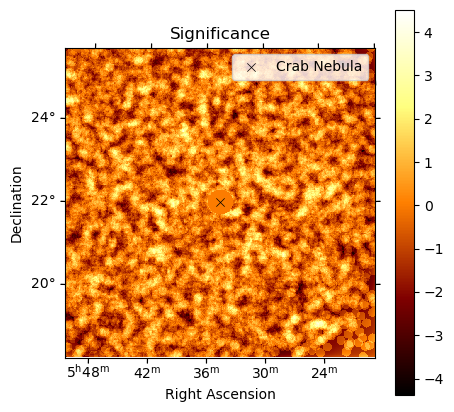}
\endminipage\hfill
\minipage{0.32\textwidth}
  \includegraphics[width=\linewidth]{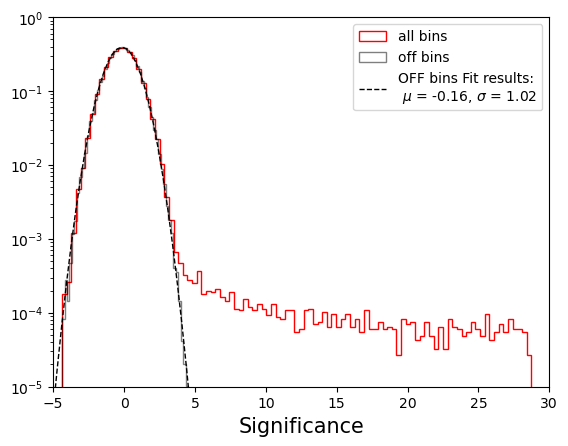}

\endminipage
\caption{\textit{Left}: Excess map of the Crab Nebula region within a region of 1.5$^\circ \times 1.5^\circ$. The best-fit position of the excess is marked with a ``+" symbol, and the expected position of the Crab Nebula, as measured in \cite{2020NatAs...4..167H}, is indicated by the green ``×".  The distribution was convolved with a disk kernel of 0.05$^\circ$. 
\textit{Center}: Significance map across the 7.5$^\circ \times 7.5^\circ$ field of view centered on Crab Nebula position. The map convolved with a disk kernel with a radius of 0.11$^\circ$. 0.3$^\circ$ region around the Crab Nebula is masked. 
\textit{Right}: One-dimensional distribution of the significance values, including all bins (red) and OFF-region bins (gray). The OFF-region significance distribution is fitted by a Gaussian function (dashed black line).
}
\label{fig:sky_dist}
\end{figure*}

The excess and significance distributions of the Crab Nebula observations were computed using the ring background method \citep{puhlhofer_technical_2003} implemented in the \texttt{gammapy} framework \citep{axel_donath_2021_5721467}. A background ring with a radius of 1$^\circ$ and a width of 0.3$^\circ$ was used to estimate background events, $0.3^\circ$ region around the Crab Nebula was excluded from the background estimation. The radial behavior of background events was estimated using MC simulations of diffuse protons. We fitted a two dimensional, symmetrical Gaussian on the excess distribution to estimate the position of the excess. The best-fit coordinates were found to be $((\alpha_{2000} = 83.62^\circ, \delta_{2000} = 21.99^\circ ) \pm 0.02^\circ$ This position is consistent with the expected Crab Nebula location and in line with the pointing precision of the instrument measured to be $0.02^\circ$ \citep{sst1m_hw_paper}. The excess distribution, convolved with a disk kernel of $0.05^\circ$ is shown in Figure \ref{fig:sky_dist}.

The significance map shown in Figure \ref{fig:sky_dist} was calculated using the Li and Ma formula for a wide region and convolved with a disk kernel of 0.11$^\circ$. The Crab Nebula is masked on the figure to highlight the remarkable homogeneity across the entire 7.5$^\circ \times 7.5^\circ$ wide field of view. The one-dimensional distribution of the significance map is shown in Figure~\ref{fig:sky_dist} and compares the significance distribution across all bins and the OFF regions. The OFF distribution is closely described by a Gaussian with the mean $\mu = -0.16$ and standard deviation $\sigma = 1.02$, consistent with theoretical expectations for an unbiased background. This demonstrates the accuracy of our hadronic background modeling using MC simulations and paves the way for observations of bright, large, extended sources. 

\section{MC model validation}

The final sample of the Crab Nebula mono and stereo data was used to validate the MC model of both telescopes. A good agreement between the cosmic-ray-dominated event rate and MC protons (re-weighted on p+He spectrum to account for the CR composition) shown in Figure~\ref{fig.diff_rate} demonstrates that the MC model provides a valid low-level description of the instruments, including calibration, cleaning, and the atmospheric effects discussed in Section~\ref{sec:atmoshpere}. 

To further demonstrate a good agreement at the analysis level, we used the gamma-like events in the signal region in the data and compared them with the point-like MC simulations, re-weighted on the number of events from the Crab Nebula \citep{ALEKSIC201676} detected in the effective observation time. The data sample is restricted to the zenith angles between $25^\circ$ and $35^\circ$ (representing a majority of the dataset), and simulations with the zenith angle fixed at $30^\circ$. We further suppressed the background in the signal region by subtracting the background events, taken from three regions radially symmetrical to the center of the FoV. For all parameter comparisons, we carefully checked the differences between the two wobble offsets ($0.7^\circ$/$1.4^\circ$), each compared with the respective off-axis MC. Due to only a negligible difference ($<2\%$) in the acceptance between the two offsets (see Section~\ref{sec.offaxis_performance}), and to increase the statistics in the data sample, we merge them in the distributions shown. We note, however, that in the spectral (Section~\ref{sec:crab_sed}) and skymap (Section~\ref{sec:crab_skymap}) analysis, each offset is handled properly as the full enclosure IRFs are used. 

\begin{figure*}[!t]
\centering
\begin{tabular}{ccc}
\includegraphics[width=.32\textwidth]{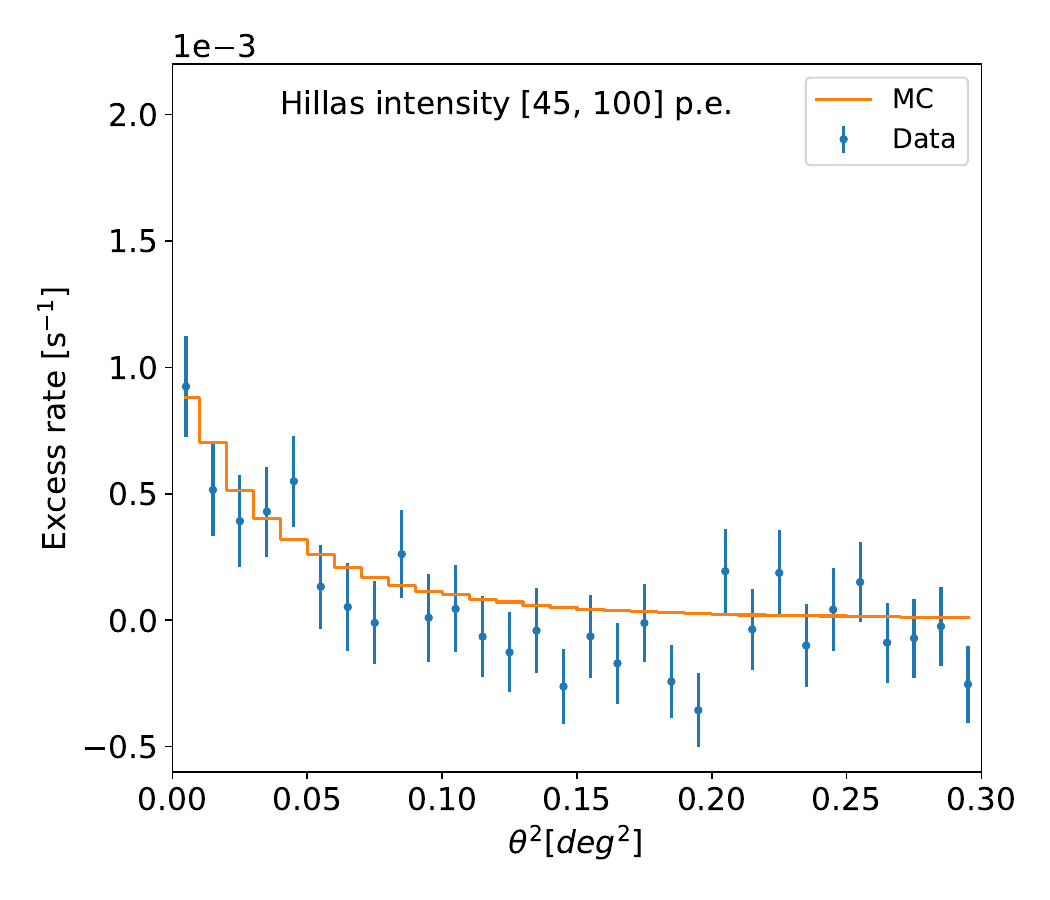} & \includegraphics[width=.32\textwidth]{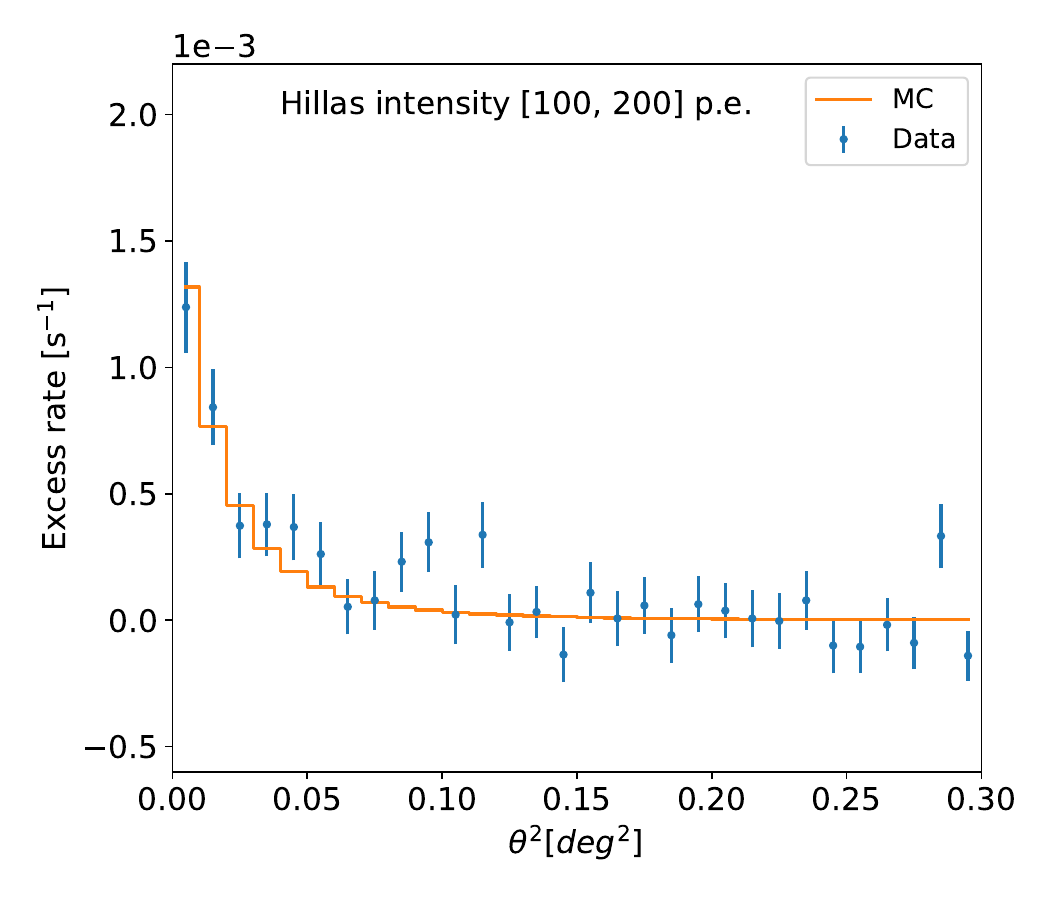} & \includegraphics[width=.32\textwidth]{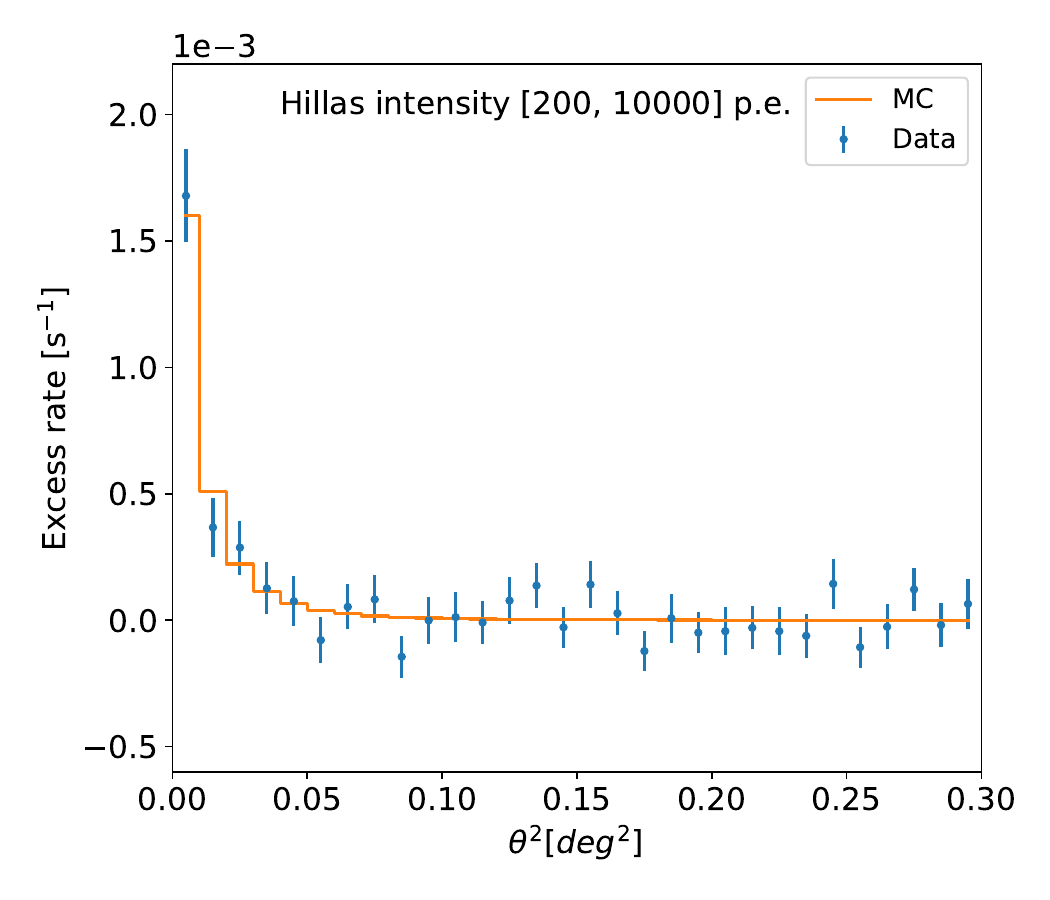} \\ 
\includegraphics[width=.32\textwidth]{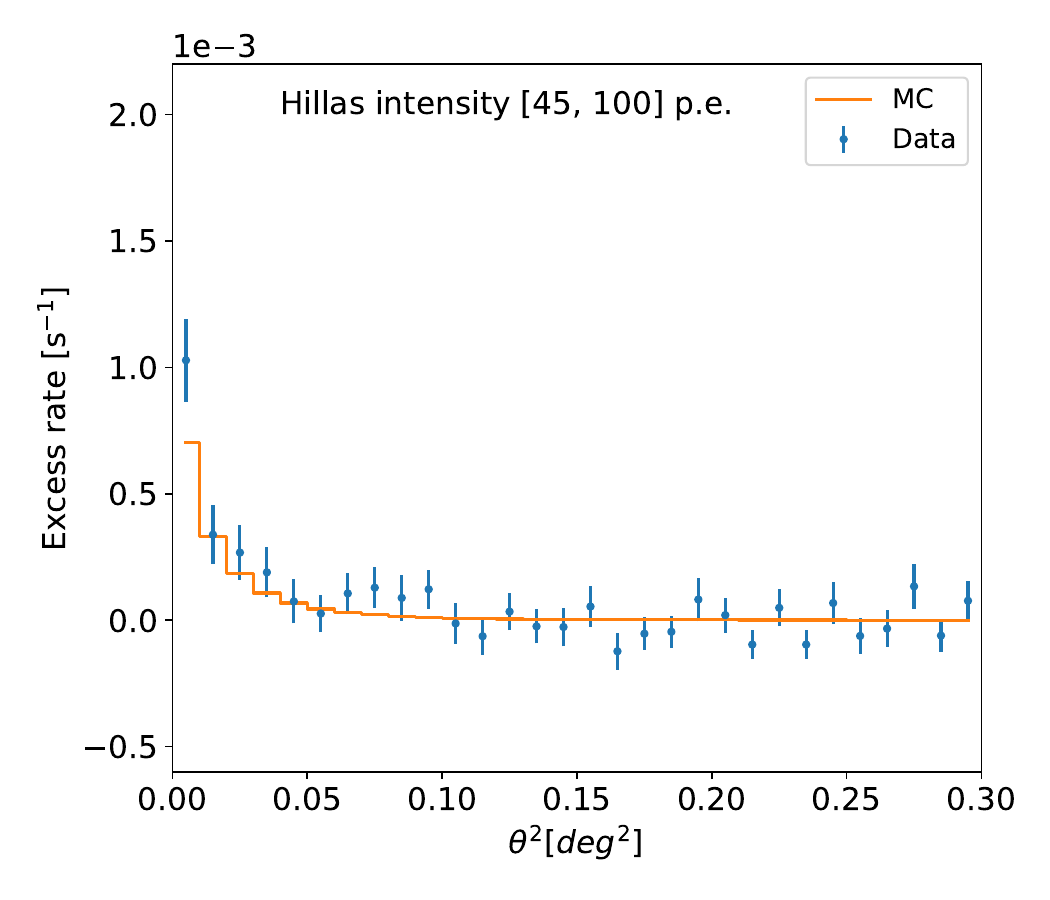} & \includegraphics[width=.32\textwidth]{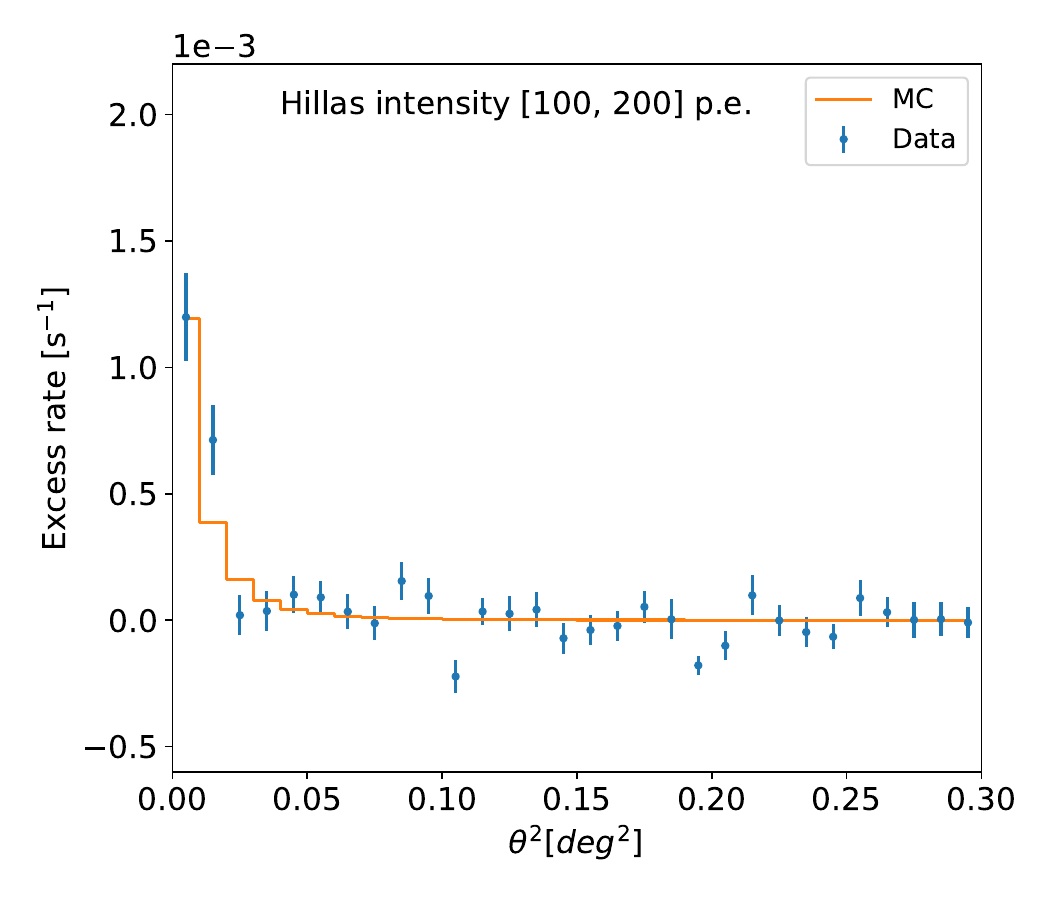} & \includegraphics[width=.32\textwidth]{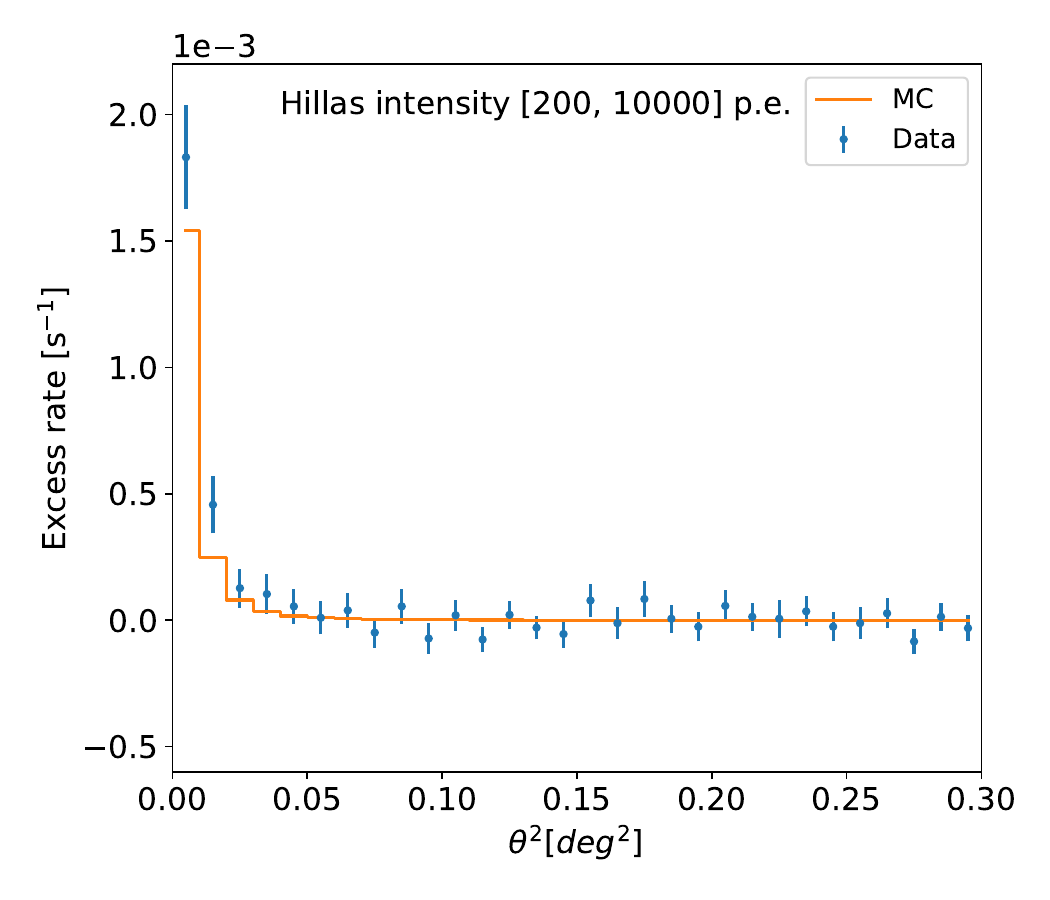}
\end{tabular}
\caption{$\theta^2$ distributions for Crab Nebula excess event rates compared with point-like gamma MC in different bins of Hillas intensities. \textit{Top row:} SST-1M-1 mono, \textit{Bottom row:} stereo.}
\label{fig.mc_data_excess_theta2}
\end{figure*}

\begin{figure*}[!t]
\centering
\begin{tabular}{cc}
\includegraphics[width=.45\textwidth]{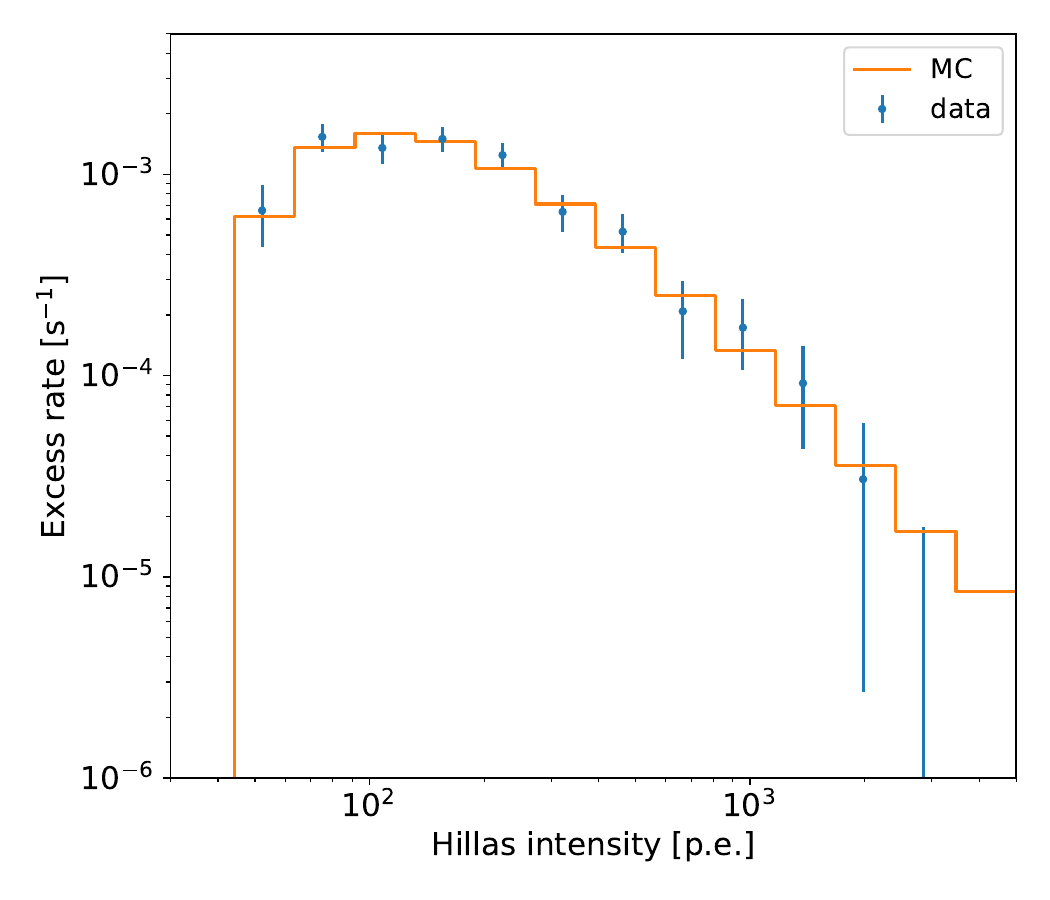} & \includegraphics[width=.45\textwidth]{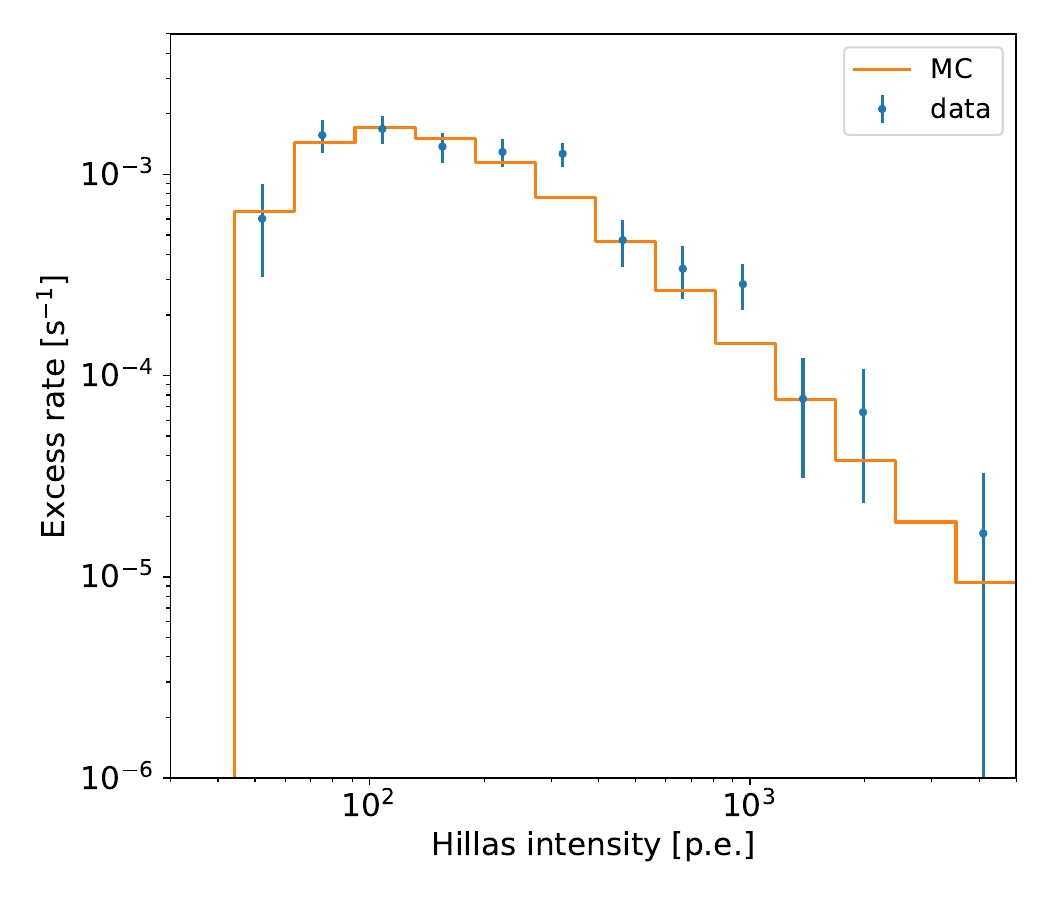}
\end{tabular}
\caption{Excess event rates binned in Hillas intensity for SST-1M-1 (\textit{left}) and SST-1M-2 (\textit{right}).}
\label{fig.mc_data_excess_int}
\end{figure*}

Figure~\ref{fig.mc_data_excess_theta2} shows the distribution of the squared angular distances of the reconstructed event directions from the center of the Crab Nebula TeV gamma-ray emission \citep{2020NatAs...4..167H}, the so-called $\theta^2$ distribution. The same energy-dependent gammaness cut on $60\%$ gamma efficiency as in the final analysis in Section~\ref{sec:crab} is applied (it was optimized on MC and turns out to yield the best signal-to-noise ratio for a source with the Crab Nebula spectrum). The comparison is shown in three bins of intensity, selected so that the rate of excess mono events in each intensity bin is comparable, demonstrating a good agreement for small shower images close to the intensity threshold as well as the extended shower images with very high intensities. Naturally, the distributions become narrower with increasing intensity, following the expected performance of the arrival direction reconstruction. For the sake of clarity in the case of mono, only results for SST-1M-1 are shown, as they are very similar for SST-1M-2. One can notice a clear improvement in the angular resolution for the stereo analysis. Small discrepancies between MC and data may be due to small mispointing (so far no pointing correction has been applied on the analysis level), or due to a small contamination of the background region with the signal events. We note that Figure~\ref{fig.mc_data_excess_theta2} compares the absolute rates as no normalization between the data and MC is applied, proving overall agreement of the excess event rates between Crab Nebula data and MC. This is further demonstrated in Figure~\ref{fig.mc_data_excess_int}, showing the rate of excess events at the analysis level in different intensity bins integrated in the signal region with a radius of $0.2^\circ$, which was carefully selected to prevent the background regions of the same size from overlapping.

\begin{figure*}[!t]
\centering
\begin{tabular}{ccc}
\includegraphics[width=.32\textwidth]{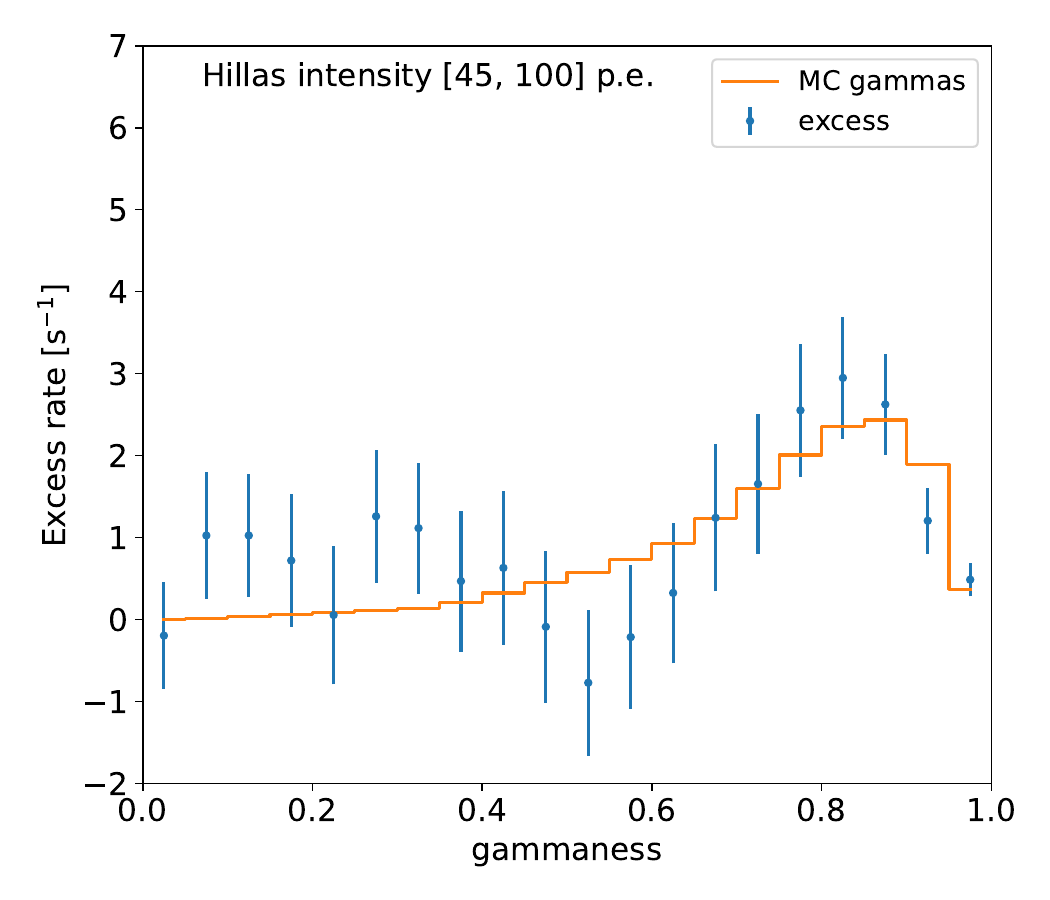} & \includegraphics[width=.32\textwidth]{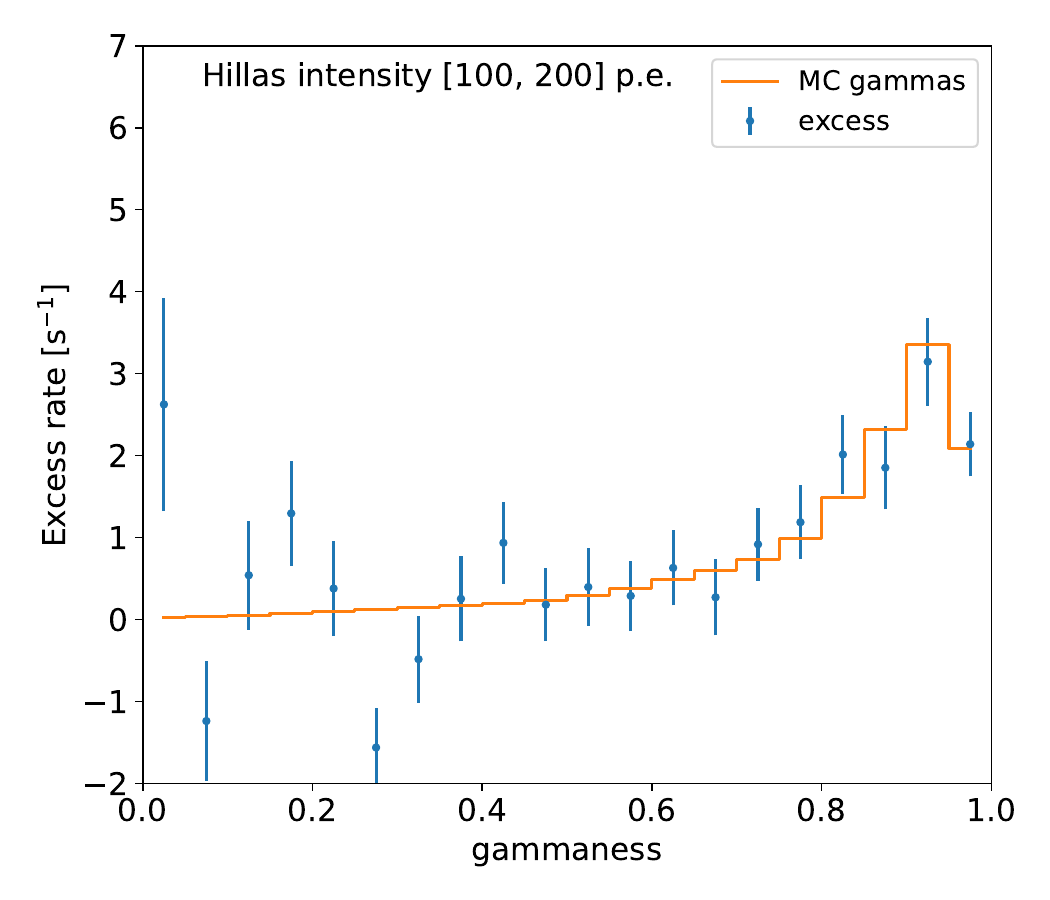} & \includegraphics[width=.32\textwidth]{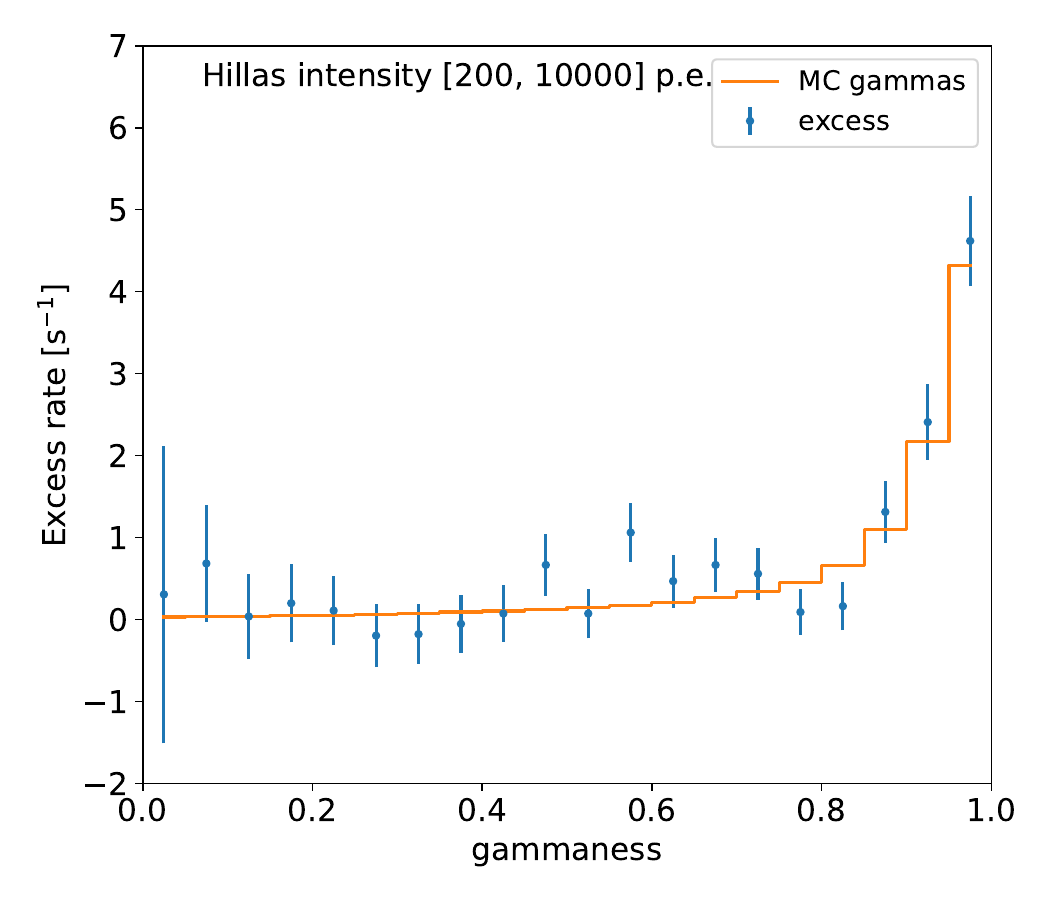} \\
\includegraphics[width=.32\textwidth]{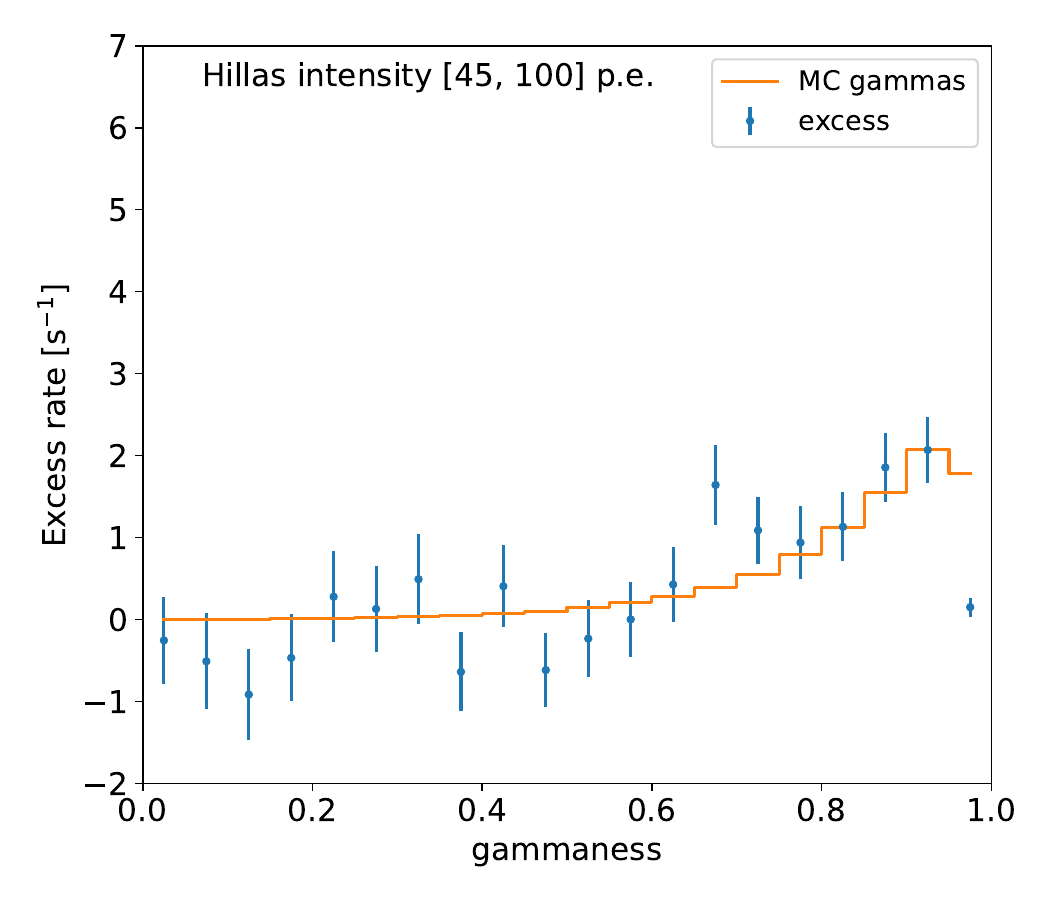} & \includegraphics[width=.32\textwidth]{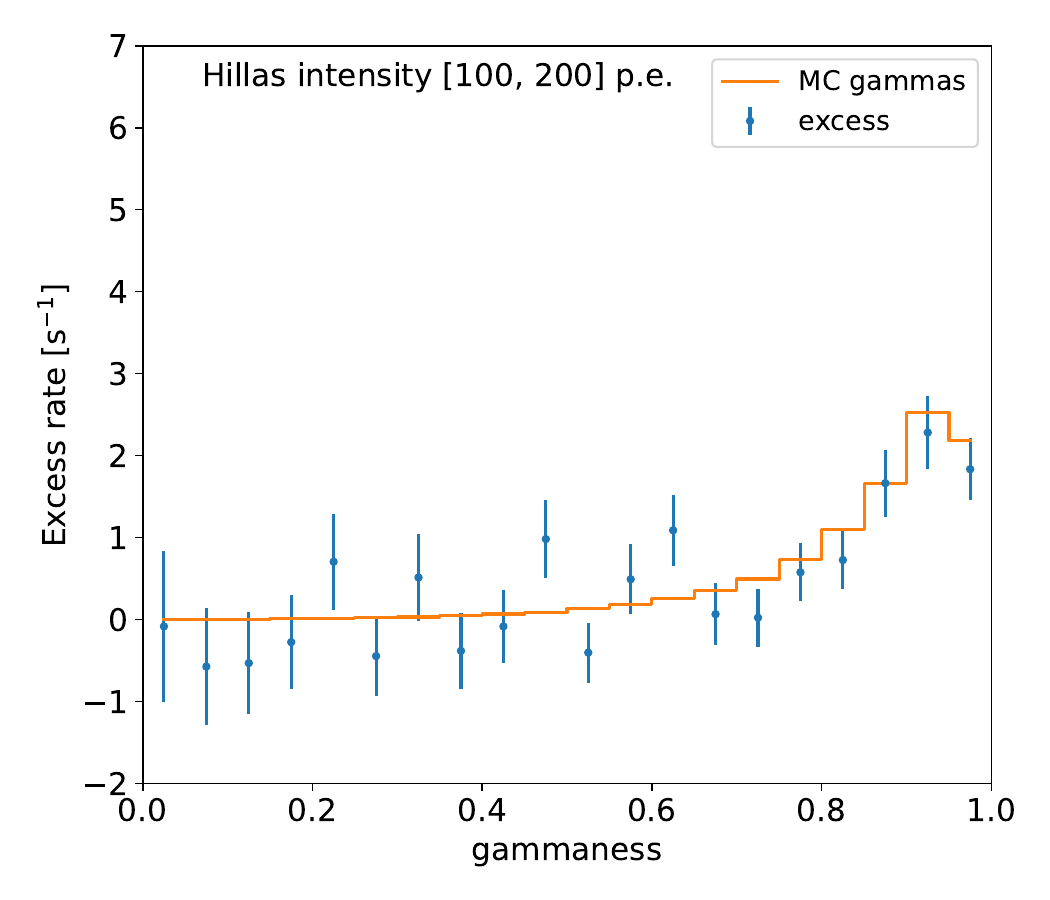} & \includegraphics[width=.32\textwidth]{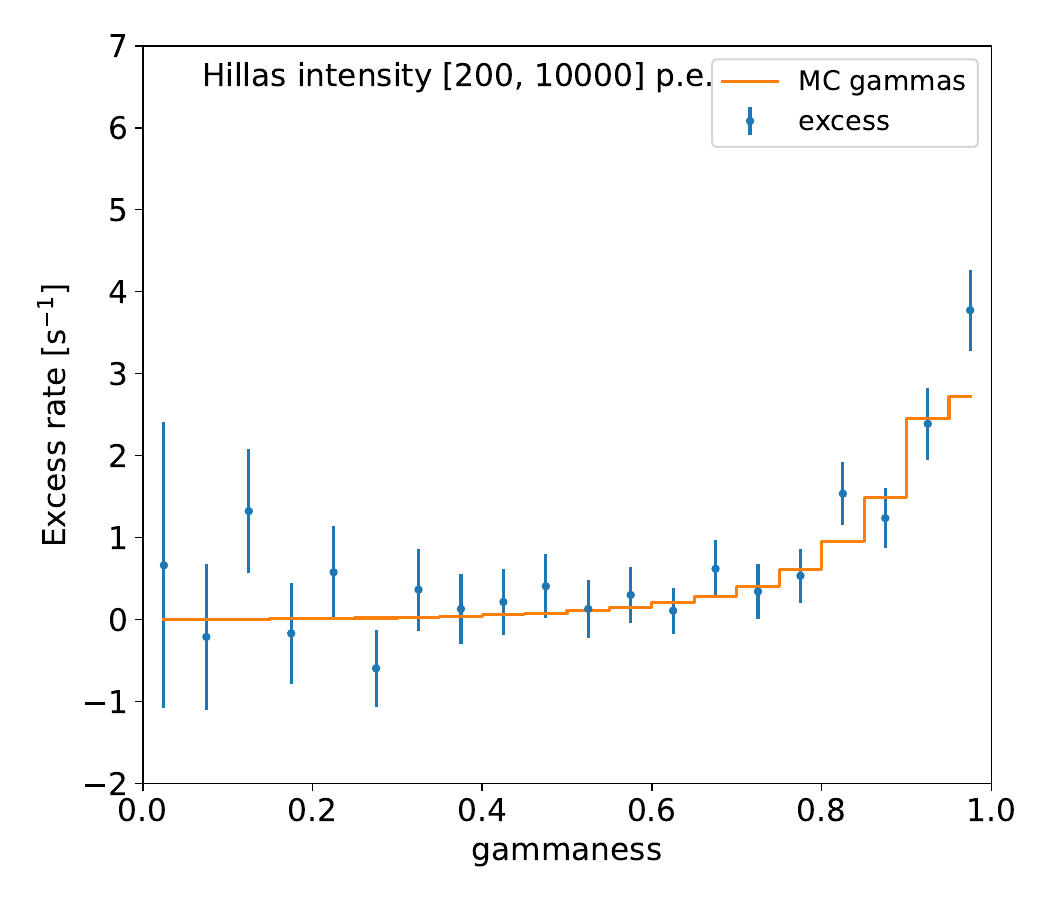}
\end{tabular}
\caption{Comparison of gammaness distribution for MC simulations and Crab Nebula excess events in different intensity bins. \textit{Top row:} Mono SST-1M-1, \textit{Bottom row:} Stereo (binned in the SST-1M-1 intensity).}
\label{fig.mc_data_excess_gammaness}
\end{figure*}

A good agreement is reached also in the distribution of gammaness shown in Figure~\ref{fig.mc_data_excess_gammaness}, proving that the Hillas parameters in the data are well reproduced in MC used for RF training. In this analysis, we consider all events in the signal and background regions (no gammaness cut applied) to avoid cropped distributions, which results in both signal and background regions being background-dominated. Distributions of gammaness in both are then subtracted, leaving only the gammaness distribution for the excess event rate. The disagreement for high gammaness close to the intensity threshold in stereo may be due to small NSB differences between MC and data, leading to different fractions of image pixels surviving the tailcuts, and it is a subject of further investigation.

\section{Systematic uncertainties} \label{sec:systematic}

Event reconstruction in the IACT technique relies on MC simulations, which describe the particle interactions in the atmosphere, the production and subsequent propagation of the Cherenkov light, and finally, the response of the telescope optics and camera electronics. All these parameters are known with only limited precision and are often variable in time, and therefore contribute to the total systematic uncertainties on the reconstructed quantities. In this section, we estimate three classes of systematic uncertainties on the energy scale, flux normalization, and spectral index. Results of this study are summarized in Table~\ref{tab:systematics_summary}. We note that as the statistical uncertanties of the resulting Crab Nebula SED are relatively high due to the limited size of the data sample, some of the systematic estimates are not conclusive and will be subject to a future study.

\begin{table}[h]
\centering
\begin{tabular}{lr}
\hline
\multicolumn{2}{c}{Energy scale} \\ 
\hline
Atm. (VAOD) & $5\%$ \\
Atm. (seasonal) & $<3\%$ \\
Optical eff. & $5\%$ (SST-1M-1)\\
& $6\%$ (SST-1M-2)\\
Reco. charge & $<4\%$ \\
NSB (voltage-drop) & $<2\%$ \\
NSB (RFs) & $<2\%$ \\
\hline
\textit{Total} & $<10\%$ \\
\hline
\hline
\multicolumn{2}{c}{Flux normalization ($\phi_0$)} \\
\hline
Energy scale & $19\%$ (stereo)\\
& $18\%$ (SST-1M-1/2)\\
Bkg. norm & $\approx5\%$\\
NSB & $10\%$\\
MC and data diff. & $6\%$ (stereo)\\
& $5\%$ (SST-1M-1)\\
& $7\%$ (SST-1M-2)\\
\hline
\textit{Total} & $23\%$ (stereo) \\
& $22\%$ (SST-1M-1/2) \\
\hline
\hline
\multicolumn{2}{c}{Spectral index ($\Gamma$)} \\
\hline
Energy scale & $<1\%$\\
Bkg. norm & $<1\%$ (stereo), $2\%$ (SST-1M-1/2)\\
MC and data diff. & $3\%$ (stereo, SST-1M-1) \\
& $6\%$ (SST-1M-2)\\
\hline
\textit{Total} & $3\%$ (stereo)\\
& $4\%$ (SST-1M-1)\\
& $6\%$ (SST-1M-2)\\
\hline
\hline
\end{tabular}
\caption{Estimated values of the main sources of systematic uncertainties. See the text for a detailed explanation.}
\label{tab:systematics_summary}
\end{table}

\subsection{Energy scale}\label{sec.energy_scale}

The absolute energy scale of the instrument is affected by any effect that contributes to the uncertainty of the image "intensity." For example, if the reconstructed Hillas intensities in MC are over-estimated (i.e., for a given true energy, the Hillas intensity in data is lower than in MC), some of the real events would not survive the intensity cut, leading to an underestimation of the flux, especially at low energies. The overall effect of the miscalibrated energy scale is given by the shape of the source spectrum and the effective area, and thus is quite complex to determine \citep[see, e.g.,][]{ALEKSIC201676}.

The Earth's atmosphere is a critical part of the IACT technique, enabling particle showers to develop, and presenting an attenuation medium for emitted Cherenkov photons. Using MC simulated with a fixed model of the atmosphere inevitably leads to systematic uncertainty if used on data taken during variable conditions. In Section~\ref{sec:atmoshpere}, we describe how the average VAOD used in MC was determined. To estimate the effect of its variability on the light scale, we trained an RF regressor with atmospheric transmissivity corresponding to VAOD = 0.2, and then reconstructed MC with different simulated VAOD. It turned out that a change of 0.1 roughly corresponds to a $10\%$ change in the light scale, and correspondingly, a $10\%$ bias in reconstructed energy. This is in line with what is naively expected since the Mie attenuation causes a $I/I_{0} \varpropto e^{\mathrm{-VAOD}} = e^{-0.1} \approx 0.9$ shift in detected flux of Cherenkov photons. Our estimated systematic uncertainty of 0.05 in the measured VAOD (for the Crab Nebula dataset with exceptionally low VAOD and its variations, see Section~\ref{sec:atmoshpere}) therefore corresponds approximately to a $5\%$ uncertainty on the light scale.

Seasonal variations of atmospheric molecular profiles (and thus variations of the index of refraction and other relevant quantities) can have a sizable impact on the density of Cherenkov photons detected at the ground \citep{BERNLOHR2000255}. Since we use only a single representative profile for simulations covering the whole period of the Crab measurements, this represents another potential source of systematic error. To estimate this effect, we constructed three average seasonal profiles out of weekly ECMWF data, corresponding to winter, summer, and an intermediate period. We then produce three sets of simulations, one for each profile. We let the RF regressor train on the intermediate profile and then let it reconstruct the MC for all three cases. The systematic shift in reconstructed energy and the light scale between a fixed intermediate atmosphere and a winter or summer atmosphere is roughly 1--3\%, slightly depending on the energy.

The amount of detected light is also affected by the total optical efficiency of the telescope, considering mirror reflectivity, transmissivity of the camera window, and shadowing of the telescope structure. This effect is taken into account by analysis of the muon rings (Section~\ref{subsec:eff}), which allows for the tuning of the optical efficiency in MC.
The systematic uncertainties induced by the discrepancies between the instrument and MC optical efficiencies are evaluated through the precision of the fit between the muon radius and the muon image charge (see Fig.~\ref{fig:muon_RvsI}), which is used to tune the MC optical efficiency. The selection criteria of muon events do not significantly affect the agreement between MC and data. However, for SST-1M-2, the slope of the linear fit differs between MC and data, resulting in an additional 5\% uncertainty. The systematic uncertainty on the shower charge due to discrepancies between MC and data is estimated to be 4\% for SST-1M-1 and 6\% for SST-1M-2.
 
Moreover, slow degradation of the optical components of the telescopes leads to about $5\%$ loss of optical efficiency in a year, while the MC are tuned to the average value. During the course of the Crab observation presented in this study, the optical efficiency of both telescopes decreased by about $2\%$, and thus we conclude the total systematics on the optical efficiency to be 5/6\% for SST-1M-1/2.

The NSB-dependent effects described in Section~\ref{sec.voltage_drop} can introduce significant systematic uncertainties in the energy scale, reaching up to 20\% for SST-1M-1 and 5\% for SST-1M-2 if not corrected. We note that such a large difference between the two telescopes is driven by their different susceptibility to the NSB (see Section~\ref{sec.voltage_drop}). However, the dependence of the electronic response on the baseline shift is accounted for and corrected in calibration. The correction factor is evaluated on the analysis level using the baseline shift determined from pedestal events taken with a frequency of 100 Hz. The precision of the correction is assessed using the muon data, where the dispersion of the average muon charge across different NSB bins after correction remains below 2\%, consistent with statistical uncertainties.

Digicam readout is characterized with a negligible dead time \citep{2017EPJC...77...47H}. However, in case of extreme trigger rates on the level of a few GHz, which may briefly occur due to a flash from a car passing along a road near the observation site, or under extreme NSB conditions (Moon in the FoV, very Large Zenith Angle observations in Ond\v{r}ejov), the camera server event buffer may reach saturation. This can lead to a reduction in the rate of stored events. The Crab Nebula sample used in this study was cleaned for these effects, and only runs with low-to-moderate NSB conditions were used in the analysis.

The reconstructed charge is given by the integral of the waveform on a predefined integration window width. We test the difference in the number of reconstructed p.e. varying the size of the integration window, even though this effect is mitigated by the integration correction (Section~\ref{sec:charge_extraction}). We find the difference in the reconstructed pixel charge $< 4\%$ for both telescopes.

Hillas intensity is one of the most important features in RF reconstruction of the energy of the primary gamma-ray photon. The number of pixels, that survive cleaning depends on the NSB level. Moreover, the amount of reconstructed charge in the individual pixels is affected by the NSB-dependent PDE, gain, and optical crosstalk, and thus any discrepancy in NSB between MC and data may lead to an offset in reconstructed energy.
Although the impact of these effects is reduced by the adaptive cleaning thresholds, see Section~\ref{sec:cleaning}, and voltage drop correction based on muon-ring analysis, Section~\ref{sec.voltage_drop}, 
we tested for this by running a dedicated MC production with a very low NSB level on which we trained a set of RFs used to reconstruct testing data from our main MC production. We compared the results with those obtained if the RFs trained on MC with the same NSB level were used, resulting in $2\%$ systematics in the energy scale.

The total systematic error on the energy scale of the Crab Nebula dataset is below $10\%$. We note that for periods with higher VAOD variations when no VAOD monitoring device is present on site, the energy scale systematics are dominated by its uncertainty, and thus can be higher. To evaluate the effect of the energy scale systematics on the measured SED, we perform a set of spectral analyses of the Crab Nebula data sample, each with a true energy axis in the IRFs scaled in the range of $\pm10\%$ for both telescopes. We found that the spectral indices are not significantly affected by the energy scaling ($\Delta\Gamma < 0.3\%$), while the maximum variation of the flux normalization is $\Delta\phi_0 = \pm18\%$ for mono and $\Delta\phi_0 = \pm19\%$ for stereo, effectively due to the shift of the SED along the energy axis. We note that using the scale invariance property of PL results in similar systematics in $\phi_0$, while no systematic in $\Gamma$ is expected.

\subsection{Flux normalization}\label{sec.flux_systematics}

The number of excess detected counts, and thus the source flux, depend on the estimated background counts. In the wobble observation mode, the number of background events in the source region is estimated from one or more regions of the same size at the same distance from the camera center as the source, assuming radially symmetrical acceptance. This method is therefore naturally affected by any inhomogeneity in the distribution of the background events in the FoV, leading to a systematic error in the estimated flux. Small inhomogeneities are usually caused by stars present in the FoV, locally increasing event rate, or by dead pixels, which decrease the acceptance in some parts of the camera. On top of that, small asymmetry of acceptance is expected in the case of stereo observations with only two telescopes \citep[e.g.,][]{ALEKSIC201676}. The latter is mitigated by wobbling, where the source and the background region positions are swapped every run. To evaluate the systematics related to background inhomogeneity, we compared the background estimated in two off-source regions for the full Crab Nebula sample, at a distance of $1.4^\circ$ from the center (and equidistant to the Crab Nebula). Unfortunately, having a relatively low event rate due to a high energy threshold, the total number of background events in our sample is not large enough to make a conclusive statement about the background systematics. After applying relatively loose cuts (gamma efficiency = $90\%$ and $\theta < 0.3^\circ$), we end up with $3.5\pm2.0\%$, $2.3\pm2.0\%$, and $2.1\pm5.0\%$ difference in the number of events between the two regions, for SST-1M-1, SST-1M-2, and in stereo, respectively\footnote{The number of background events was $\approx3500\pm59$, $\approx4500\pm67$, $\approx750\pm27$, for SST-1M-1, SST-1M-2, and in stereo, respectively.}. For stereo the difference is consistent with statistical uncertainty. We note that the effect of background uncertainty on the flux normalization depends on the signal-to-noise ratio, and while it may not be important for the Crab Nebula data, it becomes significant for faint sources. 

Variable NSB in the data effectively affects the waveform variances in individual pixels, leading to differences in extracted Hillas parameters, which may result in variable acceptance for gamma rays. One may expect a general performance degradation with increasing NSB level. Even though this effect is partially mitigated with adaptive cleaning (see Section~\ref{sec:cleaning}), we performed a dedicated MC study, carefully testing what is the effective area of gamma rays for both telescopes at different zenith angles and under the variable NSB conditions in the Crab Nebula sample. We found rather energy independent $10\%$ change of the effective area. We note that this level of systematics is only valid for the presented Crab Nebula sample in this paper and may differ for different atmospheric conditions. For analysis of a different data sample spanning over a large range of NSB, one may consider using dedicated MC tuned on data in several NSB bins. The effect of simulated NSB on the intensity threshold in our data sample turns out to be at the level of a few p.e., below the intensity threshold adopted in the final analysis (45 p.e.).

To evaluate what is the effect of using RFs trained on MC with fixed level of NSB on the gamma-ray acceptance, we extended the study described in Section~\ref{sec.energy_scale}, and calculated the relative change in the effective area, if RFs trained on low NSB MC are used to reconstruct the baseline testing MC. We found the difference lower than $3\%$ (energy independent) for all bins in the zenith angle, for both mono and stereo.

We combined both NSB-related factors that affect the effective area and performed a spectral analysis of the Crab Nebula sample with the effective area changed by $\pm10\%$. We confirmed that the relative difference in flux normalization is $\Delta\phi_0 = \pm10\%$ (for both mono and stereo), not changing the spectral index as expected for energy-independent effective area variation.

Due to small remaining discrepancies between MC and data, the reconstructed flux depends on the cuts applied in the analysis (tighter cuts result in underestimated fluxes). We test this effect on the final Crab Nebula SED, varying the gammaness and direction cuts in the range of $\pm10\%$ the fraction of gamma-like events left in the sample after the cut. We varied the gamma efficiency (energy-dependent gammaness cut) applied on the data and corresponding IRFs between 50--70\%, and $\theta$ cut between $0.17^\circ-0.30^\circ$ and $0.10^\circ-0.18^\circ$, for mono and stereo, respectively, which correspond to the 70--90\% containment, averaged over all energy bins considered in the analysis. All other parameters of the spectral analysis were left the same as described in Section~\ref{sec:crab_sed}. We found that the effect on flux normalization is $\Delta\phi_\mathrm{0, stereo} = \pm6\%$,  $\Delta\phi_\mathrm{0, 1} = \pm5\%$, and $\Delta\phi_\mathrm{0, 2} = \pm7\%$.

\subsection{Spectral index}

Uncertainty on the background counts is the largest in low energy bins, where the number of excess gamma rays is usually smaller than the background, and thus the signal-to-noise ratio is quite low. For high energies, the situation is usually the opposite, and therefore the background systematics discussed in Section~\ref{sec.flux_systematics} results also in an uncertainty in the spectral index $\Delta \Gamma$. Following \citet{ALEKSIC201676}, $\Delta \Gamma$ can be approximately estimated as:
\begin{equation}
\Delta\Gamma = 2 \frac{\sqrt{(5\%/\mathrm{SBR_{LE}})^2 + 5\%/\mathrm{SBR_{HE}})^2}}{\log(E_\mathrm{max}/E_\mathrm{min})},
\end{equation}
where we consider reconstruction of a spectrum in the energy range $(E_\mathrm{min}, E_\mathrm{max})$, with two energy bins only, having signal to background ratio $\mathrm{SBR_{LE/HE}}$. 5\% in the formula comes from a relatively conservative estimate of systematics in the background for both mono and stereo modes of operation (Section~\ref{sec.flux_systematics}). For the Crab Nebula data sample, with high SBR in both energy bins, $\Delta \Gamma_\mathrm{mono} = 0.04$, $\Delta \Gamma_\mathrm{stereo} = 0.01$. We note that this is an estimate relevant only for the Crab Nebula data sample presented in this study, as $5\%$ uncertainty on the background counts is only a limit given by relatively small statistics.

Using the results of the analysis performed in Section~\ref{sec.flux_systematics}, we can evaluate how the spectral index is affected by the analysis cuts. Varying the gamma efficiency and the size of the signal region, we found that for stereo and SST-1M-1 in mono, the spectral index change within $3\%$ ($\Delta \Gamma_\mathrm{1} = +0.08/-0.05$, $\Delta \Gamma_\mathrm{stereo} = +0.03/-0.08$); whereas for SST-1M-2 in mono, it is within $6\%$ ($\Delta \Gamma_\mathrm{2} = +0.0/-0.2$). 

\section{Summary and conclusions} \label{sec:summary}

We evaluated the low-altitude performance of the SST-1M telescopes in mono and stereo modes using MC simulations and Crab Nebula data taken during commissioning from September 2023 to March 2024. The MC model of the telescopes was carefully tuned to accurately represent both telescopes, and the atmospheric conditions at the Ond\v{r}ejov observatory. Due to the dependence of the gain, PDE, and the optical crosstalk on the NSB level, special care was taken with the telescope calibration and tuning of the simulated NSB level to reflect low to moderate NSB conditions in Ond\v{r}ejov.

We obtained the trigger threshold (for a source with the spectrum of the Crab Nebula) at low zenith angles to be $E_\mathrm{T, mono} = 0.6\,\rm{TeV}$, and $E_\mathrm{T, stereo} = 0.7\,\rm{TeV}$, rising up to $E_\mathrm{T, mono} = 1\,\rm{TeV}$, and $E_\mathrm{T, stereo} = 1.3\,\rm{TeV}$ at the analysis level. At the low zenith angle, the energy resolution in mono is about $20\%$ at low energies near the energy threshold and reaches about $15\%$ at higher energies. In stereo, the energy resolution is between $10-15\%$. The angular resolution is better than $0.18^\circ$ above the energy threshold in mono and reaches $0.10^\circ$ in stereo. The flux sensitivity in stereo shows an improvement of about a factor of 2 compared to the mono reconstruction. The integral sensitivity above the energy threshold in stereo reaches about $7\%$ C.U. in 50 hours. We also show the zenith angle dependence of these quantities, showing an improvement at large energies with increasing zenith angle at the expense of increasing energy threshold.

Additionally, we performed an MC study of the off-axis performance, which shows a remarkably flat acceptance up to about $2.5^\circ$ offset, where it drops by $10\%$. The angular resolution and the integral sensitivity show similar behavior, making the SST-1M an ideal instrument for multi-TeV observations of extended gamma-ray sources or for the follow-up of poorly localized transients.

We used the low-zenith-angle Crab Nebula data to demonstrate a good MC-data agreement, showing that the real observation matches the expected performance. The derived Crab Nebula SED is in good agreement with the results of other observatories within the reported uncertainties. We carefully studied the main sources of systematic uncertainties and evaluated their impact on the measured SED using dedicated MC simulations. The total systematic uncertainties were found to be $<10\%$ in the energy scale, $23\%$ in the flux normalization, and about 3--6\% in the spectral index. We note that the most important effect on the flux normalization systematics is the uncertainty in the energy scale, dominated by the uncertainty in VAOD and optical efficiency. While the former can be mitigated with the adoption of on-site atmospheric monitoring of VAOD or the so-called Cherenkov transparency correction \citep{HAHN201425, STEFANIK201912}, improvements to the muon analysis  have to be investigated to make  improvements to the latter. The second largest contribution comes from the variable NSB in the data, leading to an effective area mismatch with the MC simulated with a fixed NSB level. Improvements in the analysis methods to reduce these effects will be the subject of future studies.

\begin{acknowledgements}
\textit{Acknowledgements.} This publication was created as part of the projects funded in Poland by the Minister of Science based on agreements number 2024/WK/03 and DIR/\-WK/2017/12. The construction, calibration, software control and support for operation of the SST-1M cameras is supported by SNF (grants CRSII2\_141877, 20FL21\_154221, CRSII2\_160830, \_166913), by the Boninchi Foundation, and by the Université de Genève, Faculté de Sciences, Département de Physique Nucléaire et Corpusculaire. The Czech partner institutions acknowledge support of the infrastructure and research projects by Ministry of Education, Youth and Sports of the Czech Republic (MEYS) and the European Union funds (EU), MEYS LM2023047, EU/MEYS CZ.02.01.01/00/22\_008/0004632, CZ.02.01.01/00/22\_010/0008598, Horizon Europe MSCA COFUND Physics for Future 101081515, and Czech Science Foundation, GACR 23-05827S. \textit{Author contribution.} J. Jury\v{s}ek: paper project coordination, MC performance, and validation, Crab Nebula SED, study of the systematics uncertainties. T. Tavernier: Calibration, muon analysis and voltage drop correction, optical efficiency, Crab Nebula skymap. V. Novotn\'y: MC production, low-level MC-data tuning. J. Bla\v{z}ek: VAOD and atmospheric effect. D. Mand\'at, M. Pech: telescope operation and data acquisition. All these authors contributed to the paper drafting. The rest of the authors contributed to one or more of the following: design, construction, maintenance, SW development, and discussion and approval of the draft.
\end{acknowledgements}

\bibliographystyle{aa}
\bibliography{aa55292-25}

\end{document}